\documentclass[a4paper, 11pt, oneside, onecolumn]{article}

\usepackage[english]{babel} 
\usepackage{fontawesome}
\usepackage{microtype} 
\usepackage{xspace} 
\usepackage[utf8]{inputenc}	
\usepackage[T1]{fontenc} 
\usepackage{float}
\usepackage{dsfont}
\usepackage{soul}
\usepackage{setspace} 
\usepackage[normalem]{ulem}
\usepackage{orcidlink}
\usepackage{cprotect}

\usepackage{amsmath,amssymb,bbm}
\usepackage{sansmath}

\usepackage[compat=1.1.0]{tikz-feynman}
\usepackage{tikz}
\usetikzlibrary{shapes.geometric, arrows.meta, positioning, matrix}

\DeclareMathVersion{sans}
\SetSymbolFont{operators}{sans}{OT1}{cmbr}{m}{n}
\SetSymbolFont{letters}{sans}{OML}{cmbrm}{m}{it}
\SetSymbolFont{symbols}{sans}{OMS}{cmbrs}{m}{n}
\SetMathAlphabet{\mathit}{sans}{OT1}{cmbr}{m}{sl}
\SetMathAlphabet{\mathbf}{sans}{OT1}{cmbr}{bx}{n}
\SetMathAlphabet{\mathtt}{sans}{OT1}{cmtl}{m}{n}
\SetSymbolFont{largesymbols}{sans}{OMX}{iwona}{m}{n}

\DeclareMathVersion{boldsans}
\SetSymbolFont{operators}{boldsans}{OT1}{cmbr}{b}{n}
\SetSymbolFont{letters}{boldsans}{OML}{cmbrm}{b}{it}
\SetSymbolFont{symbols}{boldsans}{OMS}{cmbrs}{b}{n}
\SetMathAlphabet{\mathit}{boldsans}{OT1}{cmbr}{b}{sl}
\SetMathAlphabet{\mathbf}{boldsans}{OT1}{cmbr}{bx}{n}
\SetMathAlphabet{\mathtt}{boldsans}{OT1}{cmtl}{b}{n}
\SetSymbolFont{largesymbols}{boldsans}{OMX}{iwona}{bx}{n}

\usepackage{afterpage}

\usepackage{amsfonts,amsmath,amssymb,bm,mathtools,slashed} 
\usepackage{mathrsfs}

\usepackage{multicol} 
\usepackage[dvipsnames,table]{xcolor}	

\definecolor{indigo}{rgb}{0.0, 0.25, 0.42}

\definecolor{rwth}{RGB}{0,84,159}
\definecolor{rwth2}{RGB}{64,127,183}

\definecolor{rmp}{RGB}{41, 43, 133}
\newcommand{\LinkColor}[0]{rwth}

\colorlet{darkBlue}{blue!45!black}
\colorlet{linkColor}{blue!80!black}
\usepackage{hyperref} 
\hypersetup{
	colorlinks, 
	bookmarksopen, 
	bookmarksnumbered,
	pdftoolbar=false, 
	pdfmenubar=false, 
	pdffitwindow=false, 
	pdfstartview={FitH},	 
	citecolor=\LinkColor, 
	linkcolor=\LinkColor,	
	urlcolor=\LinkColor, 
	pdftitle={UFO Sightings with Matchete}, 
	pdfauthor={Hourtz, Ketheeswaran, Krämer, Thomsen, Weber, and Wilsch}, 
	pdfnewwindow=true, 
	linktoc=all
}

\usepackage{lipsum}
\setlipsum{%
  par-before = \begingroup\color{gray},
  par-after = \endgroup
}

\usepackage{geometry}
\usepackage{pdflscape}

\usepackage{fancyhdr}
\fancypagestyle{main}{%
    \fancyhf{}
    \setcounter{page}{2}
    \pagenumbering{arabic}
	\cfoot{--~\thepage~--}
	
}

\usepackage{titlesec}
\titleformat{\section}{\LARGE\sffamily\bfseries}{\thesection}{1.0em}{}
\titleformat{\subsection}{\Large\sffamily\bfseries}{\thesubsection}{1.0em}{}
\titleformat{\subsubsection}{\large\sffamily\bfseries}{\thesubsubsection}{1.0em}{}
\titleformat{\paragraph}[runin]{\sffamily\bfseries}{}{1.0em}{}

\makeatletter
\g@addto@macro\bfseries{\boldmath} 
\makeatother

\usepackage[sort&compress,numbers]{natbib}
\addto\captionsenglish{}

\usepackage{graphicx}
\usepackage{epstopdf}
\graphicspath{{./Figures/}{./figures/}}
\usepackage{caption,subcaption}
\DeclareCaptionFormat{custom}
{%
    \textbf{#1#2} #3
}
\DeclareCaptionLabelSeparator{custom}{ }
\numberwithin{equation}{section}

\usepackage{booktabs} 
\usepackage{longtable} 
\usepackage{multirow} 
\renewcommand{\arraystretch}{1.2}

\usepackage[shortlabels]{enumitem}
\setlist{itemsep=.1em,topsep=.5em}
\SetEnumerateShortLabel{i}{\textit{\roman*}}

\usepackage{listings} 
\lstdefinelanguage{json}{
  morestring=[b]",
  morecomment=[l]{//},
  stringstyle=\color{red!70!black},
  commentstyle=\color{gray}\ttfamily,
  basicstyle=\ttfamily\small,
  literate=
   *{0}{{{\color{blue}0}}}{1}
    {1}{{{\color{blue}1}}}{1}
    {2}{{{\color{blue}2}}}{1}
    {3}{{{\color{blue}3}}}{1}
    {4}{{{\color{blue}4}}}{1}
    {5}{{{\color{blue}5}}}{1}
    {6}{{{\color{blue}6}}}{1}
    {7}{{{\color{blue}7}}}{1}
    {8}{{{\color{blue}8}}}{1}
    {9}{{{\color{blue}9}}}{1},
} 
\allowdisplaybreaks

\usepackage{bbold,bbm}
\usepackage{slashed}

\newcommand{\dd}{\mathop{}\!\mathrm{d}}

\newcommand{\unit}{\text{\usefont{U}{dsrom}{m}{n}1}}

\newcommand{\cA}{\mathcal{A}}

\newcommand{\cD}{\mathcal{D}}

\newcommand{\cO}{\mathcal{O}}

\newcommand{\cW}{\mathcal{W}}

\newcommand{\cY}{\mathcal{Y}}
\newcommand{\cZ}{\mathcal{Z}}

\newcommand{\LL}{\mathrm{L}}

\newcommand{\U}{\mathrm{U}}
\newcommand{\SU}{\mathrm{SU}}

\newcommand{\matchete}{{\textsc{Matchete}}\xspace}

\newcommand{\eminus}{\vcenter{\hbox{\scalebox{0.6}[1]{$ - $}}}}	

\newcommand{\transpose}{^{\intercal}}

\newcommand{\sscript}[1]{{\scriptscriptstyle \mathrm{#1}}}

\DeclareUnicodeCharacter{202F}{\,} 

\newlength{\mymailwidth}

\newcommand{\email}[1]{%
    \rlap{\thanks{\href{mailto:#1}{\sffamily #1}}}%
    \settowidth{\mymailwidth}{\textsuperscript{\thefootnote}}%
}
\newcommand{\emailorcid}[2]{%
    \rlap{\thanks{\orcidlink{#2}\,\href{mailto:#1}{\sffamily #1}}}%
    \settowidth{\mymailwidth}{\textsuperscript{\thefootnote}}%
}

\usepackage{titling}       
\usepackage{authblk}       
\usepackage{lipsum}        

\pretitle{%
    \begin{center}
    \bfseries  \sffamily \fontsize{26}{30} \selectfont \mathversion{boldsans}
}
\posttitle{%
    \par\end{center}\vskip 1em%
    {\color{rwth}
    \rule{\textwidth}{1.5pt}
    }
}

\preauthor{%
    \begin{center}\begin{spacing}{1.5}\Large
}

\postauthor{%
    \end{spacing}\end{center}
}

\makeatletter
\renewcommand\AB@affilsepx{\\[1.1ex]} 
\makeatother

\newsavebox{\mysuperscriptbox}
\newlength{\mysuperscriptwidth}

\makeatletter
\renewcommand\AB@authnote[1]{%
    \hspace{\mymailwidth}\rlap{\textsuperscript{\normalfont#1}}%
    \settowidth{\mysuperscriptwidth}{\textsuperscript{\normalfont#1}}%
}

\makeatother

\predate{}
\postdate{}

\renewenvironment{abstract}{
    \begin{center}
        \bfseries\sffamily Abstract \vspace{-.5em}\vspace{0pt}
    \end{center}
    \quotation
}{\endquotation}

\makeatletter
\let\MyIntOrig\int
\def\MyIntSpace{\hspace{-.35em}} 
\def\int{\MyInt}
\def\MyInt{\MyIntOrig\MyIntSkipMaybe}
\def\MyIntSkipMaybe{
	\@ifnextchar_{\MyIntSkipScript}{%
		\@ifnextchar^{\MyIntSkipScript}{%
			\@ifnextchar\limits{\MyIntSkipTok}{%
				\@ifnextchar\nolimits{\MyIntSkipTok}{%
					\MyIntSpace}}}}%
}
\def\MyIntSkipScript#1#2{#1{#2}\MyIntSkipMaybe}
\def\MyIntSkipTok#1{#1\MyIntSkipMaybe}
\makeatother

\definecolor{blue-violet}{rgb}{0.54, 0.17, 0.89}

\newcommand{\mmanobreak}[1]{\mbox{#1}\xspace} 
\usepackage{mmacells}
\mmaDefineMathReplacement{**}{\cdot}
\mmaDefineMathReplacement[→]{->}{\to}[2]
\mmaDefineMathReplacement{II}{\mathbb{i}}
\mmaDefineMathReplacement{NCMdot}{\(\,\dot\,\)}
\mmaDefineMathReplacement{SCA}{\mathcal{A}}
\mmaDefineMathReplacement{SCM}{\mathcal{M}}
\mmaDefineMathReplacement{SCW}{\mathcal{W}}
\mmaDefineMathReplacement{SCZ}{\mathcal{Z}}
\mmaDefineMathReplacement{gou}{\mathfrak{u}}
\mmaDefineMathReplacement{god}{\mathfrak{d}}
\mmaDefineMathReplacement{goe}{\mathfrak{e}}
\mmaDefineMathReplacement{gob}{\mathfrak{b}}
\mmaDefineMathReplacement{SCe}{e}
\mmaDefineMathReplacement{SCWbar}{\overline{\mathcal{W}}}
\mmaDefineMathReplacement{aghostWbar}{\overline{\hat{\omega}_{\mathcal{W}}}}
\mmaDefineMathReplacement{ghostWbar}{\overline{\omega_{\mathcal{W}}}}

\mmaSet{
  morefv={gobble=2},
  linklocaluri=mma/symbol/definition:#1,
  morecellgraphics={yoffset=1.9ex},
  moredefined={DefineGaugeGroup, DefineField, DefineCoupling,
    Fermion, Scalar, Charges, Mass, SelfConjugate, Heavy, 
    H, U1, SU, fund, adj, eps, gen, CG, Delta, l, EFTOrder, Indices, LoopOrder, ModelParameters,
    FreeLag, NiceForm, PlusHc, Bar, PR, PL, Match, GreensSimplify, EOMSimplify, LoadModel, Flavor,
    Match, CovariantLoop, HcSimplify, Contract, Light, m, SelectOperatorClass,
    FundAlphabet, AdjAlphabet, IndexAlphabet, DefineFlavorIndex, Chiral, LeftHanded, RightHanded, RelabelIndices, CheckLagrangian,
    FeynmanRules, EXT, SetSymmetryBreakingPattern, SU2L, U1Y, U1em,
    RepresentationDecomposition, Singlet,
    CGDecomposition, 
    ExportUFO, UFO, DefaultParamCard,
    DefineCouplingOrder,
    OutputDirectory, InputFile,
    FieldDecomposition, ImplementVacuumConditions, ToBrokenPhase, GaugeFixLagrangian, SetSSBReplacements, GetVacuumConditions,
    ImposeFlavorSymmetry, DefineGlobalGroup, SubstituteMasses, GetFlavorStructures, FileNameJoin,
    Gauge, Unitary
  }
}

\title{
    UFO Sightings with Matchete:\\
    {\LARGE Connecting BSM Lagrangians with Event Generators}
}

\author[1,2]{Luis~Hourtz\email{luis.hourtz@psi.ch}}
\author[1]{Tamilarasan~Ketheeswaran\email{tamilarasan.ketheeswaran@rwth-aachen.de}}
\author[1]{Michael~Kr\"amer\email{mkraemer@physik.rwth-aachen.de}}
\author[3]{Anders~Eller~Thomsen\emailorcid{anders.thomsen@unibe.ch}{0000-0001-9614-8381}}
\author[1]{Philip~Weber\email{philip.weber@rwth-aachen.de}}
\author[1]{Felix~Wilsch\emailorcid{felix.wilsch@physik.rwth-aachen.de}{0000-0003-2409-1579}}
\affil[1]{%
    Institute for Theoretical Particle Physics and Cosmology,
    RWTH Aachen University, 
    \linebreak 
    Sommerfeldstr.~16, 
    D-52056 Aachen, 
    Germany
}

\affil[2]{%
    PSI Center for Neutron and Muon Sciences, 
    CH-5232 Villigen PSI, 
    Switzerland
}

\affil[3]{%
     Albert Einstein Center for Fundamental Physics, 
     Institute for Theoretical Physics,
     \linebreak 
     University of Bern, 
     Sidlerstrasse 5, 
     CH-3012 Bern, 
     Switzerland
}

\date{}

\fancypagestyle{mypreprint}{
    \fancyhf{} 
    \fancyhead[R]{\footnotesize \ttfamily
        P3H-26-071
        \\
        TTK-26-34%
    } 
    \fancyhead[L]{
        \href{https://matchete.gitlab.io/}{\includegraphics[width=.35\textwidth]{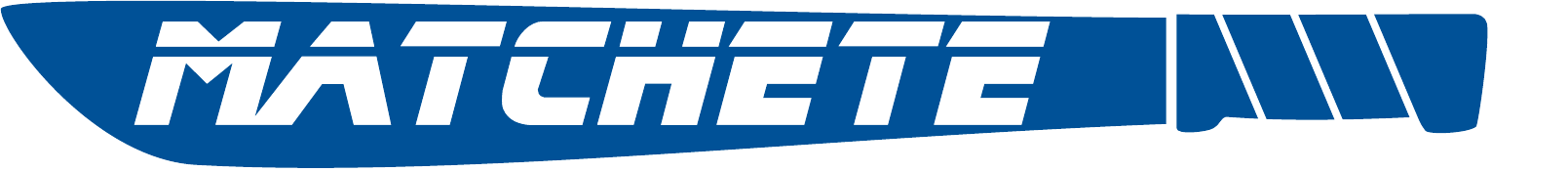}}
    }
}

\begin{document}

\newgeometry{bindingoffset=0cm, inner=2.5 cm, 
	outer=2.5cm, 
	top=3.5cm,
    bottom=2.5cm}

\renewcommand*{\thefootnote}{\fnsymbol{footnote}} 

\maketitle
\thispagestyle{mypreprint} 
\suppressfloats

\vspace{-0.8cm}
\begin{abstract}\noindent
We present a tool for the automated generation of \textsc{Universal Feynman Output}~(UFO) files, used as input by many event generators, for generic Beyond the Standard Model~(BSM) theories and their low-energy Effective Field Theory~(EFT) descriptions. 
Building on the \textsc{Matchete} package, which provides a convenient interface for defining BSM models and deriving their EFTs via matching, our extension obtains the Lagrangian in the broken phase of the electroweak symmetry, performs automatic gauge fixing, computes all Feynman rules, imposes flavor symmetries, and exports the model in the UFO format. 
This significantly streamlines BSM and EFT analyses, as we illustrate with an EFT validity study for a multi-channel leptoquark search. 
As an application, we provide a complete UFO implementation of the Standard Model EFT up to dimension eight using the $M_W$ electroweak input scheme.
\end{abstract}

\newpage
\thispagestyle{main}
\setcounter{footnote}{0}
\renewcommand*{\thefootnote}{\arabic{footnote}}%
\linespread{1.05}
\restoregeometry

{
    \sffamily
    \hypersetup{linkcolor=black}
	\addtocontents{toc}{\protect\hypertarget{toc}{}} 
    \hrule\vspace*{-0.25cm}
    \setcounter{tocdepth}{2} 
	\tableofcontents
    \vspace*{0.5cm}\hrule
}




\section{Introduction}
\label{sec:intro}

The Standard Model~(SM) of particle physics offers a remarkably good description of the interactions among all observed elementary particles. Yet it is known to be incomplete:
Various cosmological observations, such as the existence of Dark Matter or the baryon asymmetry of the Universe, require extensions Beyond the SM~(BSM).
Moreover, the SM comes with various theoretical shortcomings, like the electroweak hierarchy problem.
Nor does the model provide an explanation of neutrino masses or of the observed mass hierarchies and flavor patterns in the fermion sector.
While no conclusive deviations from the SM have been observed in particle physics experiments thus far, scrutinizing the wide range of proposed New-Physics~(NP) extensions of the SM remains one of the principal tasks for experimental and theoretical high-energy physics. 

To that end, one can either compare the predictions of individual BSM theories against experimental data or employ a more model-agnostic approach based on Effective Field Theory~(EFT).
In the latter case, the effects of generic heavy NP states are encoded in a tower of higher-dimensional effective operators, providing a valid description when the masses of the heavy BSM states are well above the energies probed experimentally. 
While light NP is not excluded, the non-observation of signals for NP at the high energies probed at the Large Hadron Collider~(LHC) suggests that there is a mass gap between the electroweak scale of the SM and the scale of~NP, allowing for the reinterpretation of the SM as an EFT.
The resulting theory is the extension of the SM by a minimal set of gauge-invariant higher-dimensional operators and is dubbed the Standard Model Effective Field Theory~(SMEFT)~\cite{Buchmuller:1985jz,Grzadkowski:2010es}; see also~\cite{Brivio:2017vri,Isidori:2023pyp,Aebischer:2025qhh} for reviews.

While the SMEFT covers a broad class of possible BSM models, it comes at the cost of a proliferation of free parameters, i.e., the couplings of the large number of higher-dimensional operators and their unknown correlations within the generic EFT setup.
In addition, the validity of the EFT approach becomes questionable if the unknown NP scale is sufficiently low.
Particularly for studies of NP scenarios at the TeV~scale at the LHC, it is often unclear whether EFTs can be consistently applied~\cite{Contino:2016jqw,Brivio:2021alv,Brivio:2022pyi,Allwicher:2022gkm,Corbett:2024evt,Allwicher:2024mzw,Chang:2025ohh}.
One option for studying the EFT validity is to select a few representative BSM theories and match them onto the SMEFT~\cite{Carmona:2021xtq,Fuentes-Martin:2022jrf,Guedes:2023azv}. 
Then one can perform the same analysis in both the concrete models and the EFT and examine to what extent the resulting limits agree.

The search for BSM physics relies heavily on numerical simulations of all kinds of processes that are potentially modified by~NP.
In order to perform these simulations---often of the Monte Carlo~(MC) variety---the relevant Feynman rules of the model are needed.
While determining Feynman rules is in principle a straightforward algebraic exercise, manual extraction is very tedious, error-prone, and repetitive across different models.
Automation is therefore essential.
One such automated tool is the \textsc{FeynRules} package~\cite{Christensen:2008py,Alloul:2013bka}, greatly simplifying the study of different NP scenarios.
It provides a complete toolbox for tree-level phenomenology~\cite{Alloul:2013bka} and has been extended to one loop as well~\cite{Degrande:2014vpa}.
The \textsc{FeynRules} code not only provides an automatic derivation of the Feynman rules but also offers interfaces to various other codes for subsequent calculations.
Most importantly, it supports the export of Feynman rules to the \textsc{Universal Feynman Output}~(UFO) format~\cite{Degrande:2011ua,Darme:2023jdn}, which consists of a collection of \textsc{Python} files enabling the numerical evaluation of amplitudes. 
This interface has, for instance, been used to derive several SMEFT implementations in the UFO format, namely 
\textsc{SMEFTsim}~\cite{Brivio:2017btx,Brivio:2020onw}, 
\textsc{SmeftFR}~\cite{Dedes:2019uzs,Dedes:2023zws}, 
and \textsc{SMEFT@NLO}~\cite{Degrande:2020evl}.

UFO files are a central ingredient when performing MC simulations for BSM theories and EFTs, and the format is widely supported. 
An incomplete list of software tools supporting the standard includes
\textsc{Achilles}~\cite{Isaacson:2022cwh}, 
\textsc{Comix}~\cite{Gleisberg:2008fv}, 
\textsc{Contur}~\cite{Butterworth:2016sqg}, 
\textsc{GoSam}~\cite{Braun:2025afl},
\textsc{Herwig~7}~\cite{Bellm:2015jjp},
\textsc{MadAnalysis~5}~\cite{Conte:2012fm}, 
\textsc{MadDM}~\cite{Ambrogi:2018jqj}, 
\textsc{MadGraph5\_aMC@NLO}~\cite{Alwall:2011uj,Alwall:2014hca},
\textsc{Recola~2}~\cite{Denner:2017wsf},
\textsc{Sherpa}~\cite{Hoche:2014kca}, 
and \textsc{Whizard}~\cite{Kilian:2007gr}.
While \textsc{FeynRules} is extremely successful and widely used among particle physicists, the implementation of complex models, in particular those involving higher-dimensional operators, spontaneous symmetry breaking, gauge fixing, or non-trivial flavor structures, can become technically involved. 
Given the central position of the UFO in BSM phenomenology, the community benefits from having multiple independent tools to generate UFO files. 
There are already alternatives in the form of the \textsc{Sarah}~\cite{Staub:2015kfa} and \textsc{LanHEP}~\cite{Semenov:2014rea} packages, each with its own strengths. Nevertheless, we think there are compelling reasons to introduce one more. 

In this work, we introduce a complementary UFO-generation tool in the \matchete package~\cite{Fuentes-Martin:2022jrf} for \textsc{Mathematica}.
Originally developed for automated one-loop matching of generic BSM theories onto their EFT descriptions, \matchete has evolved into a growing toolbox for EFT studies.
Its extensions include mapping between different operator bases, simplification of Lagrangians in the presence of evanescent operators~\cite{Fuentes-Martin:2022vvu}, an interface~\cite{Belfatto:2025ids} to the Wilson coefficient exchange format~(\textsc{WCxf})~\cite{Aebischer:2017ugx} for phenomenological studies, and dimensional reduction for thermal field theory~\cite{Fuentes-Martin:2026bhr}.
Beyond these extensions, the \matchete package provides versatile tools for manipulating Lagrangians and a user-friendly interface for defining arbitrary BSM theories and EFTs.
From a design perspective, the internal machinery already deploys functional field derivatives, which are crucial for determining Feynman rules.

In its present form, our implementation enables the automatic derivation of all tree-level Feynman rules for BSM theories and EFTs using functional methods and their export to the UFO format.
While we currently support only tree-level calculations, the extension to the one-loop level, i.e., the inclusion of the ultraviolet and $R_2$~counterterms (rational terms of the second kind)~\cite{Ossola:2006us,Ossola:2008xq,Degrande:2014vpa,Degrande:2020evl}, is under development.
Moreover, we provide supporting functionality for implementing spontaneous symmetry breaking~(SSB), in particular electroweak symmetry breaking~(EWSB), and generating gauge-fixing terms, significantly simplifying the extraction of UFOs in the broken phase of theories.
Utilizing these tools, we also provide the first complete $R_\xi$~gauge implementation of the SMEFT in the UFO format in the zero-width limit.\footnote{While \textsc{SmeftFR}~\cite{Dedes:2023zws} supports the derivation of Feynman rules in $R_\xi$~gauge and its UFO files contain the gauge parameters in the vertices, the gauge boson propagators are only in unitary and Feynman gauges in its implementation.}
With a straightforward user interface, simple model definitions, and integration into an EFT toolbox, our setup meaningfully simplifies the study of EFTs and concrete BSM models at colliders.
In particular, analyses of EFT validity are streamlined, as one only has to implement the model once in \matchete before automatically obtaining the model UFO, the matching onto the EFT, and the corresponding UFO of the EFT. 
We have also employed the code to generate UFO files for the SMEFT, including all dimension-six and -eight terms, with implementation of different electroweak input schemes and the option to incorporate arbitrary continuous flavor symmetries.
These predefined SMEFT UFOs can be found in the \matchete model database on \textsc{GitLab}~\href{https://gitlab.com/matchete/model-database/-/tree/master/UFO-models}{\faicon{gitlab}}~\cite{MatcheteDatabase} which also contains example notebooks that can serve as a guide for generating ones own UFOs.
All the new features are now available in the \matchete \textsc{v0.6.0} update, available at \url{https://matchete.gitlab.io/}.

We begin this article in Sec.~\ref{sec:UFO} by discussing the export of Feynman rules in the UFO format using \matchete. This section also touches on the derivation of said rules and outlines the new SSB functionality.
Subsequently, we discuss our SMEFT implementations in Sec.~\ref{sec:SMEFT}, focusing particularly on the inclusion of flavor symmetries and the extension to dimension eight.
In Sec.~\ref{sec:EFT-convergence}, we then apply this framework to study the validity of the EFT approach, taking a leptoquark search at the LHC with different production channels as a concrete example.
We conclude in Sec.~\ref{sec:conclusion}.
Various appendices provide additional information:
In Appendix~\ref{app:FR}, we review the derivation of tree-level Feynman rules from the quantum effective action.
Subsequently, we discuss in Appendix~\ref{app:SSB} how SSB and, in particular, the electroweak symmetry-breaking pattern can be implemented in \matchete to simplify model definitions while ensuring consistency.
Appendices~\ref{app:manual} and~\ref{app:examples} provide a brief manual of the core functionalities and several instructive code examples from the application in Sec.~\ref{sec:EFT-convergence}, respectively.
Finally, Appendix~\ref{app:operator-defs} provides a list of all EFT operator definitions used in this manuscript.

For readers solely interested in a guide on how to use the code, we recommend reading Sec.~\ref{sec:UFO} and Appendices~\ref{app:SSB}--\ref{app:examples}.
The tutorial notebooks provided in the built-in \textsc{Mathematica} documentation center and on the website~\cite{MatcheteWebsite} offer an easy entry point.

\section{Universal Feynman Output in Matchete}
\label{sec:UFO}
In this section, we provide an overview of the setup used to determine the tree-level Feynman rules and the corresponding UFO files in the \matchete framework.
Our intention is to highlight the general ideas and concepts, rather than providing a thorough documentation.
The latter can be found in Appendices~\ref{app:SSB} and~\ref{app:manual}, accompanied by usage examples in Appendix~\ref{app:examples}, alongside supplementary documentation available within \matchete through the built-in \textsc{Mathematica} documentation center or on the website~\cite{MatcheteWebsite}.
Here and in the following, we assume that readers are familiar with basic usage and model implementation in the \matchete package and refer to~\cite{Fuentes-Martin:2022jrf} and the documentation center for further details.

\subsection{Generating UFO Files}
\label{sec:UFO-generation}
MC event generators, such as \textsc{MadGraph}~\cite{Alwall:2011uj}, can access Feynman rules only in very specific file formats. 
Hence, the symbolic \textsc{Mathematica} output produced by \matchete must be adapted accordingly.
To that end, we adopt the common \textsc{Universal Feynman Output}~(UFO) format~\cite{Darme:2023jdn,Degrande:2011ua}. 

The UFO format represents a particle physics model as a collection of \textsc{Python} files, each encoding information about a specific component of the theory.
More specifically, a theory exported to the UFO format at tree level consists of six model-dependent files:
\begin{itemize}[-]
    \item 
    \texttt{particles.py} defines the particle content and their attributes;
    \item 
    \texttt{parameters.py} specifies the model parameters (masses, decay widths, couplings, etc.);
    \item 
    \texttt{couplings.py} expresses the coupling strengths of all vertices through the parameters;
    \item 
    \texttt{lorentz.py} encodes the Lorentz structures of all interaction vertices;
    \item 
    \texttt{coupling\_orders.py} defines the available coupling orders, such as \texttt{QED}, \texttt{QCD}, and~\texttt{NP};
    \item 
    \texttt{vertices.py} combines information from the other files to construct all interaction vertices.
\end{itemize}
In principle, the UFO format also allows for optional files such as \texttt{decays.py}, which contains analytic expressions for the decay widths of all particles. 
This is not currently included in our setup, but it can be obtained directly in simulation tools such as \textsc{MadSpin}~\cite{Artoisenet:2012st} within the \textsc{MadGraph} framework, and a future implementation in \matchete is foreseen as well. 
Another optional (and potentially model-dependent) file is \texttt{propagators.py}, which enables specifying custom propagators, necessary, for example, when working in the $R_\xi$~gauge.
If not provided, the default propagators of the event generators are used.\footnote{
\matchete always includes the file \texttt{propagators.py} with standard propagators for completeness. However, these propagators are currently not used in the \texttt{particles.py} file, i.e., they are not explicitly assigned to particles, in which case the event generator's default propagators are employed.
The only exception is the case of $R_\xi$~gauge, for which one custom vector-boson and Goldstone-boson propagator is included per gauge parameter (the Goldstone boson mass is set equal to the vector boson mass and the gauge parameter dependence of the mass is included through its propagator). 
The current implementation of the $R_\xi$~gauge yields gauge-invariant results only in the limit of vanishing decay widths for the vector bosons and the corresponding Goldstone bosons. 
Non-zero widths break the gauge invariance and a consistent implementation would require adopting the complex-mass scheme~\cite{Denner:1999gp,Denner:2005fg,Denner:2006ic}, which is left for future work. 
We refer the reader to Ref.~\cite{Schwinn:2003fp} and Sec.~5 of Ref.~\cite{Frederix:2018nkq} (and references therein) for further details.}
In \textsc{MadGraph} these defaults are available for unitary and Feynman gauges only.

In addition to these model-specific files, the UFO format requires several generic and model-independent files, namely \texttt{function\_library.py}, \texttt{object\_library.py}, \texttt{write\_param\_card.py}, and the initialization file \texttt{\_\_init\_\_.py}. 
As indicated by their names, these files provide standard libraries, helper functions, and default implementations required by simulation tools.
There is, in principle, one remaining piece of model dependence in the \texttt{\_\_init\_\_.py} file, in the variable \texttt{gauge=[0,1]}, which lists the gauges supported by the UFO, where \texttt{0} (\texttt{1}) denotes unitary (Feynman) gauge.\footnote{
If a UFO in Feynman gauge is provided to event generators such as \textsc{MadGraph} with the setting \texttt{gauge=[0,1]}, the \textsc{MadGraph} default is to load this model still in unitary gauge by dropping all Goldstone bosons and switching to unitary gauge propagators internally. 
The actual Feynman gauge model is selected when running \textsc{MadGraph} with the command \texttt{set~gauge~Feynman}. 
If the UFO is exported using Feynman gauge in \matchete, the variable \texttt{gauge=[0,1]} will be used, allowing for the use of the UFO in both Feynman and unitary gauges. 
If, instead, the UFO is exported using unitary gauge, the parameter is set to \texttt{gauge=[0]} instead. 
Exporting in $R_\xi$~gauge sets \texttt{gauge=[1]}, since the automatic switch to unitary gauge propagators is not possible due to the custom propagators being used in the $R_\xi$~case.}

The UFO files described above are sufficient to enable simulations at tree level in perturbation theory, which is the level currently supported by our \matchete implementation. 
In principle, the UFO framework also supports calculations at the one-loop level, which require the inclusion of UV and $ R_2 $ counterterms and several corresponding files~\cite{Darme:2023jdn}.
While going beyond leading order is, of course, crucial for performing precision phenomenology studies, a tree-level analysis is sufficient for a wide range of New-Physics effects.
Thus, we leave a one-loop UFO implementation in the \matchete framework for future work and cover the leading-order setup here.

To export a theory implemented with Lagrangian~\mmaInlineCell[]{Input}{L} in \matchete to the UFO format, we provide the routine 
\begin{mmaCell}{Input}
   ExportUFO[L]
\end{mmaCell}
generating all files listed at the beginning of this section.
Most of the relevant information for this purpose is directly extracted from the model definition and the Lagrangian. 
This relies internally on the~\mmaInlineCell[]{Input}{FeynmanRules} routine.
Compared to the one-loop matching of EFTs, for which \matchete was originally designed, it is necessary to include additional information in the model definitions relevant for the UFO generation.
To that end, we introduce the option~\mmaInlineCell[]{Input}{UFO} for the functions \mmaInlineCell[]{Input}{DefineField}, \mmaInlineCell[]{Input}{DefineGaugeGroup}, and \mmaInlineCell[]{Input}{DefineCoupling}.
It can be used to define PDG codes for particles and the variable names used in \textsc{Python} for particles and couplings.\cprotect\footnote{Note that specifying a PDG code for every particle in the export is \emph{mandatory}. For example, the gluon PDG number can be specified by giving the option \mmaInlineCell[]{Input}{UFO\,->\,<|"pdg"\,->\,21|>} to \mmaInlineCell[]{Input}{DefineGaugeGroup}.}

\begin{figure}[tb]
    \centering
    \begin{minipage}{0.85\textwidth}
    \begin{lstlisting}[language=json,firstnumber=1]
    {
      ...,
      "parameters" : {
        "aEM"       : {
          "name"        : "aEM",
          "nature"      : "external",
          "type"        : "real",
          "value"       : 7.816774798718049e-3,
          "texname"     : "\\alpha_{\\text{EM}}",
          "lhablock"    : "SMINPUTS",
          "lhacode"     : [1],
          "description" : "Electromagnetic coupling at MZ"
        },
        ...
        "ge"        : {
          "name"        : "ge",
          "nature"      : "internal",
          "type"        : "real",
          "value"       : "cmath.sqrt(4*cmath.pi*aEM)",
          "texname"     : "ge"
        },
        ...
      }
    }
    \end{lstlisting}
    \end{minipage}
    \cprotect\caption{Exemplary entries in the parameter card JSON file generated by \mmaInlineCell[]{Input}{DefaultParamCard} and given to \mmaInlineCell[]{Input}{ExportUFO}.
    Shown are the fine-structure constant as external input and the internal electromagnetic gauge coupling defined in terms of the former.
    The snapshot shown here uses the option \mmaInlineCell[]{Input}{DefaultValues\,->\,True} which includes default values for the SM inputs. By default all parameters are initialized with zero as value instead.
    }
    \label{fig:json}
\end{figure}

A~parametrization of all couplings and numerical default values for all free parameters must be specified as well.
By default, this is done through a graphical user interface that opens when calling~\mmaInlineCell[]{Input}{ExportUFO}.
Alternatively, one can also generate a parameter card using
\begin{mmaCell}{Input}
   DefaultParamCard[L]
\end{mmaCell}
creating a JSON file in the current working directory.
This file contains a minimal parametrization of all parameters in the given Lagrangian~\mmaInlineCell[]{Input}{L}, respecting flavor symmetries and Hermiticity of all couplings where applicable.
In particular, it relates all internal parameters of the theory to a minimal set of external inputs.
The default parameter card obtained this way can (and should) then be modified according to the needs of the user.
By default all parameters are initialized with \lstinline[columns=fixed,language=json]{"value":"0"}, except for parameters appearing in denominators of couplings which are set to~\lstinline[columns=fixed,language=json]{"value":"1"} since these can lead to divergences if set to vanish. 
In addition, one can use the option \mmaInlineCell[]{Input}{DefaultValues\,->\,True} to include default input values for the SM parameters, which can serve as a helpful starting point for models that do not modify the EW~symmetry breaking pattern of the SM.

Some example entries for the parameter card are shown in Fig.~\ref{fig:json}.
The JSON file must include a \lstinline[columns=fixed,language=json]{"parameters"} section.
Inside this section, each parameter entry follows the standard UFO format and includes fields such as \lstinline[columns=fixed,language=json]{"name"}, \lstinline[columns=fixed,language=json]{"type"}, \lstinline[columns=fixed,language=json]{"nature"}, \lstinline[columns=fixed,language=json]{"value"}, and \lstinline[columns=fixed,language=json]{"texname"}. 
It is crucial to distinguish between external parameters, which are assigned real numerical values and linked to \textit{Les Houches} blocks via \lstinline[columns=fixed,language=json]{"lhablock"} and \lstinline[columns=fixed,language=json]{"lhacode"}, and internal parameters, whose values are symbolic expressions depending on other parameters.\footnote{A parameter value can also be set to \lstinline[columns=fixed,language=json]{"ZERO"}. This specific choice will eliminate the parameter entirely from the UFO output generated by \matchete, which can be convenient, for example, to ensure massless particles, such as gauge bosons or light quarks, or to significantly reduce the size of the UFO.}
For detailed information, we refer to the original UFO publications~\cite{Darme:2023jdn,Degrande:2011ua}.

Before generating a UFO, it can also be convenient to define coupling orders to facilitate isolation of specific interactions during a simulation. This is achieved with the command
\begin{mmaCell}{Input}
   DefineCouplingOrder["QCD", gs, 1];
\end{mmaCell}
where, as an example, we defined the coupling order~\mmaInlineCell[]{Input}{"QCD"} for the strong coupling constant~\mmaInlineCell[]{Input}{gs}. 
The third argument denotes a relative hierarchy compared to other coupling orders. 
For instance, if we also assign \mmaInlineCell[]{Input}{DefineCouplingOrder["QED", ge, 2]}, two insertions of~\mmaInlineCell[]{Input}{ge} would be considered of the same order as four insertions of~\mmaInlineCell[]{Input}{gs}.

\subsection{Deriving Feynman Rules}
\label{sec:FR}
A~basic requirement for studying any NP model (whether a renormalizable or an effective theory) using MC simulation is access to an organized set of all Feynman rules in the model under consideration. 
The Feynman rules can be derived for any (effective) quantum field theory using standard methods, e.g., the generating functional.
For tree-level MC simulations, it is sufficient to derive only the Feynman rules based on the classical action of the model.
The master formula for a tree-level $n$-point Feynman rule takes the form:
\begin{align} \label{eq:Feynman-rules}
    \big\langle \eta_{I_1} \, \eta_{I_2} \cdots \eta_{I_n} \big\rangle_\text{1PI,\,tree} 
    &= 
    \left. i \frac{\delta^n S[\eta]}{\delta\eta^{I_1} \, \delta\eta^{I_2} \cdots \delta\eta^{I_n}} \right|_{\eta=0}
    \,,
\end{align}
where $S[\eta]$ is the classical action of the theory under consideration.
The fields~$\eta$ are conventionally taken in momentum space, so we replace partial derivatives~($\partial$) by the momenta~($p$) of the fields they are acting on, substituting $\partial_\mu \to \eminus i \,p_\mu$ following the convention that all momenta flow into the vertex.\footnote{This matches the convention of popular textbooks~\cite{Peskin:1995ev,Schwartz:2014sze}.}
The collective index $I_k$~denotes all regular indices of the field and its momentum. 
For identical fields the functional derivatives automatically provide the appropriate symmetry factors.
A~detailed derivation of the master formula in Eq.~\eqref{eq:Feynman-rules} is presented in Appendix~\ref{app:FR}.

While the derivation of Feynman rules poses a straightforward algebraic problem, computations are prone to errors due to combinatorial complexity.
The determination of Feynman rules is therefore best implemented in computer algebra codes, as done in \textsc{FeynRules}~\cite{Christensen:2008py,Alloul:2013bka} and the \matchete~\cite{Fuentes-Martin:2022jrf} implementation introduced here.
For any (non-)renormalizable theory implemented in \matchete with the Lagrangian~\mmaInlineCell[]{Input}{L}, the complete set of tree-level Feynman rules is derived with the function call
\begin{mmaCell}{Input}
   FeynmanRules[L]
\end{mmaCell}
This routine applies Eq.~\eqref{eq:Feynman-rules} with all relevant combinations of fields. 
The function call returns a \textsc{Mathematica} association keyed to the external fields of each vertex. 

As a concrete example, consider the SM Lagrangian in the unbroken phase, which can be loaded with the standard call 
\begin{mmaCell}{Input}
   LSM = LoadModel["SM"];
\end{mmaCell}
The Feynman rules are then calculated automatically by running the \mmaInlineCell[]{Input}{FeynmanRules} routine: 
\begin{mmaCell}{Input}
   FullSimplify/@ FeynmanRules[\mmaUnd{LSM}] // NiceForm
\end{mmaCell}
\begin{mmaCell}{Output}
   <|\{G,G,G\} -> \mmaSub{g}{s}\,\mmaSup{f}{ABC}\,\big(\mmaSub{g}{\(\mu\nu\)}(\mmaSup{\mmaSub{p}{\(\rho\)}}{1}-\mmaSup{\mmaSub{p}{\(\rho\)}}{2})\,+\,\mmaSub{g}{\(\mu\rho\)}(-\mmaSup{\mmaSub{p}{\(\nu\)}}{1}+\mmaSup{\mmaSub{p}{\(\nu\)}}{3})\,+\,\mmaSub{g}{\(\nu\rho\)}(\mmaSup{\mmaSub{p}{\(\mu\)}}{2}-\mmaSup{\mmaSub{p}{\(\mu\)}}{3})\big)\,\mmaStr{\mmaSub{\(\langle\)\mmaSup{G}{A\(\mu\)}\(\rangle\)}{1}}\,\mmaStr{\mmaSub{\(\langle\)\mmaSup{G}{B\(\nu\)}\(\rangle\)}{2}}\,\mmaStr{\mmaSub{\(\langle\)\mmaSup{G}{C\(\rho\)}\(\rangle\)}{3}}, 
   ..., \{B,e,\mmaSup{e}{\(\ast\)}\} -> II\,\mmaSub{g}{Y}\,\mmaSub{\(\delta\)}{pr}\,\mmaStr{\mmaSub{\(\langle\)\mmaSup{B}{\(\rho\)}\(\rangle\)}{3}}\,\big(\mmaStr{\mmaSub{\(\langle\)\mmaSup{\(\bar{e}\)}{r}\(\rangle\)}{2}}\(\cdot\)\,\mmaSub{\(\gamma\)}{\(\rho\)}\mmaSub{P}{R}\,\(\cdot\)\mmaStr{\mmaSub{\(\langle\)\mmaSup{e}{p}\(\rangle\)}{1}}\big), ...|>
   
\end{mmaCell}
where \mmaInlineCell[]{Output}{\mmaSup{e}{\(\ast\)}} denotes the charge conjugated field (\mmaInlineCell[]{Output}{\mmaDef{Conj}[\mmaUnd{e}]}).
We show only the three-gluon vertex and the hypercharge gauge interaction for the right-handed leptons (the majority of rules, indicated with the ellipsis, have been omitted for demonstration purposes).
The external fields of each vertex appear as the keys in the association, with the corresponding Feynman rule as a value.
The expressions in gray brackets symbolize the external legs, e.g., spinors and polarization vectors (cf.~Tab.~\ref{tab:externallegs}).

\subsection{Symmetry Breaking} \label{sec:matchete_symmetry-breaking}
The initial- and final-state fields used in MC simulation tools are the mass-basis states of the broken phase of electroweak symmetry, i.e., the physical states. 
As a new feature in \matchete \textsc{v0.6.0}, we have therefore included methods to implement symmetry breaking of the symmetric-phase Lagrangians, which define the SM, the SMEFT, and various BSM extensions. 
The symmetry-breaking information can be embedded within a model file for convenience; for instance, the SM Lagrangian can be loaded with the call
\begin{mmaCell}{Input}
  LSM = LoadModel["SM+breaking"];
\end{mmaCell}
This returns the symmetric-phase SM Lagrangian (similarly to \mmaInlineCell[]{Input}{LoadModel["SM"]}), but it also loads all the definitions needed to take the Lagrangian to the broken phase. 
Setting up the extra information needed to go to the broken phase requires some care, and we refer the reader to Appendix~\ref{app:SSB} for further details on the implementation of the SM EWSB. 
The package also provides model files implementing EWSB in the SMEFT. 
We leave it for future work to discuss how this extends to various BSM symmetry-breaking patterns. 

With all the symmetry-breaking definitions already loaded, the broken phase of the model (expanded about the vacuum) is determined with the call 
\begin{mmaCell}{Input}
  Lbrk = ImplementVacuumConditions[ToBrokenPhase[\mmaUnd{LSM}], 
    SubstituteMasses-> False] // NiceForm
\end{mmaCell}
\begin{mmaCell}[]{Output}
  -\mmaFrac{1}{4}\mmaFrac{1}{\mmaSub{g}{s}\mmaSup{}{2}}\mmaSup{G}{\(\mu\nu\)A2}-\mmaFrac{1}{4}\mmaFrac{1}{\mmaSup{SCe}{2}}\mmaSup{\mmaOver{SCA}{}}{\(\mu\nu\)2}-\mmaFrac{1}{2}\mmaSup{SCWbar}{\(\mu\nu\)}\mmaSup{SCW}{\(\mu\nu\)}+\mmaFrac{1}{4}\mmaFrac{\mmaSup{v}{2}\mmaSup{SCe}{2}}{\mmaSup{sw}{2}}\mmaSup{SCWbar}{\(\mu\)}\mmaSup{SCW}{\(\mu\)}-\mmaFrac{1}{4}\mmaSup{\mmaOver{SCZ}{}}{\(\mu\nu\)2}+\mmaFrac{1}{8}\mmaFrac{\mmaSup{v}{2}\mmaSup{SCe}{2}}{\mmaSup{sw}{2}\mmaSup{cw}{2}}\mmaSup{SCZ}{\(\mu\)2}+ ... 
  ... +II\mmaSqrt{2}\mmaFrac{1}{v}\mmaSubSup{V}{CKM}{rp}\mmaSup{Mu}{r}\(\chi\)(\mmaSubSup{\mmaOver{gou}{\_}}{a}{r}\(\,\cdot\,\)\mmaSub{P}{L}\(\,\cdot\,\)\mmaSup{god}{ap})+II\mmaSqrt{2}\mmaFrac{1}{v}\mmaSubSup{V}{CKM}{rp}\mmaSup{Md}{p}\(\chi\)(\mmaSubSup{\mmaOver{gou}{\_}}{a}{r}\(\,\cdot\,\)\mmaSub{P}{R}\(\,\cdot\,\)\mmaSup{god}{ap})+II\mmaSqrt{2}\mmaFrac{1}{v}\mmaSup{Me}{p}\(\chi\)(\mmaSup{\mmaOver{\(\nu\)}{\_}}{p}\(\,\cdot\,\)\mmaSub{P}{R}\(\,\cdot\,\)\mmaSup{goe}{p})
\end{mmaCell}
omitting the majority of the terms in the rather large expression for the sake of brevity. We observe, for instance, the appearance of mass terms for the $ W $ and $ Z $ bosons. 
The function \mmaInlineCell[]{Input}{ToBrokenPhase} simply substitutes broken-phase expansions of all fields and Clebsch--Gordan coefficients into the symmetric-phase Lagrangian with only minimal automated simplification. 
It is then useful to call the \mmaInlineCell[]{Input}{ImplementVacuumConditions} routine, which simplifies the coefficients of the resulting Lagrangian under the assumption that the theory parameters correspond to a vacuum of the theory (absence of tadpoles, canonical kinetic terms, absence of mass mixing between fields etc.).
The conditions identified and implemented by \mmaInlineCell[]{Input}{ImplementVacuumConditions} can be inspected by calling \mmaInlineCell[]{Input}{GetVacuumConditions[ToBrokenPhase[\mmaUnd{LSM}]]} to verify that everything proceeded as expected.
The broken-phase Lagrangian can then be passed to the methods for deriving Feynman rules and exporting to UFO format as desired. 

Another feature of our implementation is that we can derive gauge-fixing and ghost terms automatically, even for broken-phase Lagrangians, where gauge-fixing is usually technically involved. 
This is achieved with the method \mmaInlineCell[]{Input}{GaugeFixLagrangian}, which supports both the unitary and the $R_\xi $~gauge (the gauge fields of the unbroken subgroup are always fixed to the $R_\xi $~gauge). 
Choosing the Feynman gauge, i.e., setting $\xi=1$, can be done only later during the export to the UFO format using the option \mmaInlineCell[]{Input}{\mmaDef{R}\(\xi\)->1} for the \mmaInlineCell[]{Input}{ExportUFO} and \mmaInlineCell[]{Input}{DefaultParamCard} functions.
This can be useful in realistic models, where constructing these terms can be non-trivial. 
Isolating the gauge-fixing and ghost terms for the broken phase of the SM for the $R_\xi $~gauge looks like (where we subtract the original Lagrangian to retain only the new gauge-fixing and ghost terms)
\begin{mmaCell}{Input}
  GaugeFixLagrangian[\mmaUnd{Lbrk}, \mmaDef{Gauge}-> \mmaDef{R}\(\xi\)] - \mmaUnd{Lbrk} // NiceForm
\end{mmaCell}

\begin{mmaCell}[]{Output}
  \mmaFrac{1}{2}\mmaFrac{v\(\,\)SCe}{sw}\mmaOver{\(\chi\)}{\_}\mmaSub{\(\partial\)}{\(\mu\)}\mmaSup{SCW}{\(\mu\)}+\mmaFrac{1}{2}\mmaFrac{v\(\,\)SCe}{sw}\(\chi\)\mmaSub{\(\partial\)}{\(\mu\)}\mmaSup{SCWbar}{\(\mu\)}+\mmaFrac{1}{2}\mmaFrac{v\(\,\)SCe}{sw\(\,\)cw}\mmaSup{\(\chi\)}{0}\mmaSub{D}{\(\mu\)}\mmaSup{SCZ}{\(\mu\)}-II(aghostWbar\mmaSub{\(\partial\)}{\(\mu\)}\mmaSub{D}{\(\mu\)}ghostWbar)-II(\mmaSubSup{\(\hat{\omega}\)}{G}{A}\mmaSub{\(\partial\)}{\(\mu\)}\mmaSub{D}{\(\mu\)}\mmaSubSup{\(\omega\)}{G}{A})+ ... 
  ... -\mmaFrac{SCe\(\,\)cw}{sw}\mmaSup{SCW}{\(\mu\)}(\mmaSub{\(\partial\)}{\(\mu\)}\mmaSub{\(\hat{\omega}\)}{SCW}\mmaSub{\(\omega\)}{SCZ})+\mmaFrac{SCe\(\,\)cw}{sw}\mmaSup{SCW}{\(\mu\)}(\mmaSub{D}{\(\mu\)}\mmaSub{\(\hat{\omega}\)}{SCZ}ghostWbar)-\mmaFrac{SCe\(\,\)cw}{sw}\mmaSup{SCWbar}{\(\mu\)}(\mmaSub{D}{\(\mu\)}\mmaSub{\(\hat{\omega}\)}{SCZ}\mmaSub{\(\omega\)}{SCW})
\end{mmaCell}
Here the (anti-)ghost field of the vector field $ V $ is denoted by $ \omega_V $~($\hat{\omega}_V$). 
Note the somewhat unconventional factor of~$ i $ on the ghost kinetic terms. One can generally rescale anti-ghost (or ghost) fields freely, and with this choice we have ensured that the anti-ghost fields are Hermitian, i.e., $\hat{\omega}_V^\dagger = \hat{\omega}_V$~\cite{Kugo:1977yx}. 
With this choice the Feynman rules of our ghost propagators are multiplied by $ -i$ and the ghost vertices with $ +i $ (every ghost loop is nevertheless unchanged). 
Currently \matchete allows for exporting only tree-level UFOs where all ghost contributions can be neglected.
Therefore, all ghost terms are dropped during the export in this initial version of the interface.

It is often convenient to write vector masses, along with those of the associated ghost and Goldstone bosons, in terms of a single mass parameter, rather than retaining their full functional dependence on the fundamental parameters of the theory. 
Therefore, the functions \mmaInlineCell[]{Input}{ImplementVacuumConditions} and \mmaInlineCell[]{Input}{GaugeFixLagrangian} take the option \mmaInlineCell[]{Input}{SubstituteMasses}. 
With the choice \mmaInlineCell[]{Input}{SubstituteMasses->True}, the masses of the electroweak gauge bosons, their Goldstone bosons, and the radial Higgs mode are conveniently replaced by their canonical mass parameters (defined during their initialization with \mmaInlineCell[]{Input}{DefineField}).

\subsection{Conventions}
\label{sec:conventions}
Before continuing with the applications of our setup, we find it convenient to state the conventions employed in the code.
Since its \textsc{v0.3.0} release, \matchete has absorbed the gauge couplings into the gauge fields such that the kinetic term of a gauge field assumes the form \smash{$-\frac{1}{4g^2} F_{\mu\nu}^I F^{\mu\nu I}$} and the covariant derivative of a field~$\phi$ reads $D_\mu \phi = \left(\partial_\mu - i T^I A^I_\mu\right)\phi$ (without any couplings), where $T^I$ and $A^I_\mu$ are the generators and gauge fields summed over the appropriate representations and gauge groups.
For the output when deriving Feynman rules and exporting the UFO files, we, however, change to the more common convention by rescaling the gauge fields $A_\mu^I \to g A_\mu^I$.
Thus, the Feynman rules are given in the convention where the covariant derivative reads $D_\mu \phi = \left(\partial_\mu - i g T^I A^I_\mu\right)\phi$ and the field strength tensors take the form
\begin{align}
F_{\mu\nu}^I = 
\begin{cases}
D_\mu A^I_\nu - D_\nu A_\mu^I 
& \text{for massive vectors} \\[1.5ex]
\partial_\mu A_\nu^I - \partial_\nu A_\mu^I 
+g f^{IJK} A_\mu^J A_\nu^K 
& \text{for gauge fields}
\end{cases},
\label{eq:fieldstrength}
\end{align}
\noindent
where $g$ is the gauge coupling and $f^{IJK}$ are the structure constants of the corresponding non-Abelian gauge group and vanish in the Abelian case.
In principle, the UFO files could also be exported using the normal \matchete convention. 
However, the gauge boson propagators would all become proportional to the gauge coupling squared, which would both render them model dependent and complicate the assignment of consistent coupling orders. 
Hence, we refrain from using this convention for the UFO interface and exclusively support the more common \textit{textbook} convention.

The Levi-Civita tensor presents another difference between the \matchete ($\epsilon^{0123}=1$) and UFO ($\epsilon_{0123}=1$) conventions.
The relative minus sign for the conversion is included in the UFO files generated by \mmaInlineCell[]{Input}{ExportUFO}, but \emph{not} in \mmaInlineCell[]{Input}{FeynmanRules}, which being a native \matchete function uses the \matchete convention.

\paragraph{Subtleties of the UFO and \textsc{MadGraph} interface.}
\textsc{MadGraph} and ALOHA~\cite{deAquino:2011ub} use a few special conventions for Feynman rules that must be respected when interfacing via UFO. 
The main subtlety is that \textsc{MadGraph} applies the concept of fermion flow to fermionic vertices. 
The algorithm \textsc{MadGraph} employs for constructing an amplitude from the given Feynman rules for fermionic interactions was first worked out by Denner, Eck, Hahn, and K\"ublbeck~(DEHK) in~\cite{Denner:1992me,Denner:1992vza}.
Unlike the standard \textit{textbook} treatment, this approach neglects all Grassmann minus signs and charge conjugation matrices. 
The relative sign between two interfering diagrams is then automatically reconstructed from the different contractions of the external legs.
If a fermion-flow clash is detected, the fermion flow is reversed for the appropriate particles, effectively selecting the charge conjugate vertex (for details see Sec.~2.2 of~\cite{Alwall:2011uj}).
The reason for taking the charge conjugation matrix~$C$ to unity when ignoring the Grassmann signs can be found by looking at the transposition of a Majorana mass term:
\begin{equation}
    \left( \psi^\intercal C \psi + \overline \psi C \overline \psi^\intercal \right)^\intercal = - \left( \psi^\intercal C^\intercal \psi + \overline \psi C^\intercal \overline \psi^\intercal \right) = (-1) (-1) \left( \psi^\intercal C \psi + \overline \psi C \overline \psi^\intercal \right),
\end{equation}
where the first minus sign comes from the anti-commutation of fermionic fields and the second minus sign comes from the transposition of~$C$. 
If the Grassmann sign is left out, the charge conjugation matrix has to be set to unity because otherwise the Majorana mass term would vanish. 
To accomplish this, we implement a special mode for the Feynman rules and UFO export routines, which can be activated with the option \mmaInlineCell[]{Input}{"FermionFlow"\,->\,"DEHK"}.\cprotect\footnote{This \mmaInlineCell[]{Input}{"DEHK"} mode is the default for the function \mmaInlineCell[]{Input}{ExportUFO}, whereas \mmaInlineCell[]{Input}{FeynmanRules} uses the common \textit{textbook} convention including Grassmann minus signs and the $C$~matrix (dubbed \mmaInlineCell[]{Input}{"Standard"} mode) by default.}
In this mode, the Grassmann property of functional derivatives is deactivated, and charge conjugation matrices are set to unity. 
Contrary to the default behavior in \matchete, where charge conjugation matrices are always moved to the left of a Dirac chain, charge conjugation matrices are positioned directly next to the conjugated field using the relation $C\gamma_\mu C^{-1} = -\gamma_\mu^\intercal$ before removal. This ensures that other objects in the Dirac chain are consistently transposed. 

In general, we support UFO export of four-fermion vertices with charge conjugates or fermion flow violation with the convention of omitted charge conjugation matrices~\cite{Denner:1992me,Denner:1992vza}; however, this is not supported in \textsc{MadGraph} and may not produce correct results in other event generators either. 
It is, therefore, not possible for \textsc{MadGraph} to use the UFO export for any $B$-violating SMEFT operators.
For convenience, we have introduced a new option for all the SMEFT model files distributed with \matchete, which can be called through
\mmaInlineCell[]{Input}{LoadModel[..., ModelParameters\,->\,\{BLViolation\,->\,False\}]}
to directly drop all $B$- or $L$-violating EFT operators already at the point of loading the model.

Two other conventions can be changed using options:
Contrary to the \matchete convention $\sigma^{\mu\nu} = \frac{i}{2}[\gamma^\mu,\gamma^\nu]$, \textsc{MadGraph} seems to use the rescaled $\sigma^{\mu\nu}_{\sscript{MG}} = \frac{i}{4}[\gamma^\mu,\gamma^\nu]$ through its dependence on the ALOHA library.\footnote{The UFO format~\cite{Degrande:2011ua,Darme:2023jdn} does not specify any convention for the $ \sigma^{\mu\nu} $ matrix.} 
We conform to this standard for the UFO output by using the default option \mmaInlineCell[]{Input}{\mmaStr{"\(\sigma\)Normalization"}\,->\,}\mmaInlineCell[]{Output}{II/4} for \mmaInlineCell[]{Input}{ExportUFO}. 
Furthermore, we explicitly change the sign of all external momenta in the Feynman rules, effectively changing to outgoing momenta. This directly contradicts the prescriptions specified in the UFO format and instead conforms with the internal ALOHA conventions~\cite{deAquino:2011ub}.\footnote{We have verified that the UFO export in \textsc{FeynRules}~\cite{Christensen:2008py,Alloul:2013bka} always flips the sign of the external momenta, conforming to the same standard as \textsc{MadGraph}.} 
With these changes, we reproduce \textsc{MadGraph} output of other UFO implementations as discussed in Sec.~\ref{sec:validation}.
The change of direction for external momenta is controlled through the option \mmaInlineCell[]{Input}{"ExternalMomentum"\,->\,"outgoing"}, which is the default for \mmaInlineCell[]{Input}{ExportUFO}, whereas \mmaInlineCell[]{Input}{FeynmanRules} uses \mmaInlineCell[]{Input}{"incoming"} as default instead.
In addition, \matchete can remove Lorentz structures vanishing in four spacetime dimensions, in particular four-fermion operators of the form $\sigma^{\mu\nu} P_L \otimes \sigma_{\mu\nu} P_R$, since these structures are not supported by \textsc{MadGraph} and vanish at tree level in any event.
This is controlled through the option \mmaInlineCell[formatline=\mmanobreak]{Input}{"IncludeD4Vanishing"\,->\,False} (default) of \mmaInlineCell[]{Input}{ExportUFO}.

\section{SMEFT Implementation up to \texorpdfstring{$d \leq 8$}{d<=8}}
\label{sec:SMEFT}

Following the introduction of the UFO generation, Feynman rule derivation, and the symmetry breaking in the previous section, we now turn to a discussion of our SMEFT implementations. 
At mass dimension six, two operator bases are currently available in \matchete: the \textit{Warsaw basis}~\cite{Grzadkowski:2010es} and the \textit{Mainz basis}~\cite{MainzBasis} (see also Ref.~\cite{Born:2026xkr} for the complete operator list). 
At dimension eight, we also provide a UFO implementation of a modified version of the \textit{Murphy basis}~\cite{Murphy:2020rsh}; see also~\cite{Li:2020gnx}.
These SMEFT implementations are available in the \matchete model database on \textsc{GitLab}~\href{https://gitlab.com/matchete/model-database/-/tree/master/UFO-models}{\faicon{gitlab}}~\cite{MatcheteDatabase}.

In the following, we discuss how flavor symmetries can be incorporated in our setup, we review the EWSB patterns in the $d \leq 8$ SMEFT, and present the extension of the $M_W$ electroweak input scheme to $d=8$.
Afterwards, we discuss how one can extract the $d=8$ SMEFT UFO files from our implementation, and validate our setup against existing SMEFT implementations~\cite{Brivio:2020onw,Dedes:2023zws,Hays:2018zze} based on \textsc{FeynRules}~\cite{Alloul:2013bka}.

\subsection{Flavor Symmetries}
\label{sec:flavor-symmetries}
One of the common challenges for SMEFT studies is the plethora of free parameters in the EFT description.
For example, all baryon- and lepton-number-conserving SMEFT operators at dimension six give rise to 2\,499 free parameters~\cite{Alonso:2013hga} while this number grows to 36\,971 at dimension eight~\cite{Henning:2015alf,Fonseca:2019yya}.
These parameters predominantly originate from the flavor sector via the two- and in particular four-fermion operators.
Hence, most of these parameters, if sizable, would lead to overly large flavor-violating contributions incompatible with experimental observations.
One way to address both issues in a consistent and systematic manner is to impose flavor-symmetry assumptions in the SMEFT~\cite{Faroughy:2020ina,Greljo:2022cah}.
This allows for drastically reducing the number of free parameters while also suppressing large flavor-violating contributions.

The largest flavor symmetry compatible with the SM gauge group is a $\mathrm{U}(3)^5 = \mathrm{U}(3)_q \times \mathrm{U}(3)_u \times \mathrm{U}(3)_d \times \mathrm{U}(3)_\ell \times \mathrm{U}(3)_e$ symmetry, where every chiral fermion type is assumed to transform as a flavor triplet under its own $\mathrm{U}(3)$~factor.
Taken as an exact symmetry, this assumption forbids all flavor-violating effects, including the SM Yukawas. 
Treating the SM Yukawas as spurions breaking the $\mathrm{U}(3)^5$~symmetry leads to the Minimal Flavor Violation~(MFV) framework~\cite{Chivukula:1987py,DAmbrosio:2002vsn}, which aligns all flavor violation to the SM Yukawas.
A~less restrictive alternative is given by the $\mathrm{U}(2)^5 = \mathrm{U}(2)_q \times \mathrm{U}(2)_u \times \mathrm{U}(2)_d \times \mathrm{U}(2)_\ell \times \mathrm{U}(2)_e$ flavor symmetry~\cite{Barbieri:2011ci,Barbieri:2012uh,Blankenburg:2012nx} where the first two generations of every chiral fermion type are taken to be flavor-doublets of the corresponding $\mathrm{U}(2)$~factor, while the third generation fermions are all assumed to be singlets.

Compared to the $\mathrm{U}(3)^5$~assumptions, the $\mathrm{U}(2)^5$ case allows for richer third-generation dynamics while still offering a sufficient protection from large flavor-violating effects.
The number of free parameters at dimension six in the baryon- and lepton-number-conserving sector of the SMEFT shrinks to 47 and 147 assuming an exact $\mathrm{U}(3)^5$ and $\mathrm{U}(2)^5$ symmetry, respectively~\cite{Faroughy:2020ina}, highlighting the drastic reduction of parameters when considering additional global symmetries.
Besides these examples, a plethora of other flavor assumptions can be imposed on the SMEFT.
For an overview of various options see, e.g., Refs.~\cite{Faroughy:2020ina,Greljo:2022cah,Greljo:2025mwj}.
Flavor symmetries are also employed in several existing SMEFT UFO implementations. 
For example, \textsc{SMEFTsim}~\cite{Brivio:2020onw} provides their UFO files in the general scenario, as well as for an exact $\mathrm{U}(3)^5$ symmetry, MFV at linear order in the spurions, and $\mathrm{U}(2)$~factors in the quark sector.

Our implementation in \matchete provides a general setup for including flavor assumptions in the UFO files.
This functionality allows users to define any custom continuous flavor symmetry, which is then automatically incorporated into the UFO file.
As an example, consider the following symmetry group~\cite{Feldmann:2008ja,Faroughy:2020ina,Ethier:2021bye}
\begin{align}
    \mathrm{U}(2)_{q,u}^2 \times \mathrm{U}(3)_{d,\ell,e}^3
    =
    \mathrm{U}(2)_q \times \mathrm{U}(2)_u \times \mathrm{U}(3)_d \times \mathrm{U}(3)_\ell \times \mathrm{U}(3)_e
    \,.
\end{align}
This specific choice is phenomenologically motivated because it allows the top-quark Yukawa coupling without any spurion insertions, thereby accommodating a large top Yukawa.
However, as shown in~\cite{Faroughy:2020ina}, when breaking this symmetry in order to explain all down-quark masses, one needs to impose a specific alignment among the spurions to recover a sufficient suppression of flavor-changing amplitudes, making this group more fine-tuned and less compelling.

The $ \mathrm{U}(2)_{q,u}^2 \times \mathrm{U}(3)_{d,\ell,e}^3$ flavor symmetry can be implemented in \matchete by first defining global groups for every simple and Abelian group factor:
\begin{mmaCell}{Input}
  DefineGlobalGroup[SU2q, SU[2]];     DefineGlobalGroup[U1q, U1];
  DefineGlobalGroup[SU2u, SU[2]];     DefineGlobalGroup[U1u, U1];
  DefineGlobalGroup[SU3d, SU[3]];     DefineGlobalGroup[U1d, U1];
  DefineGlobalGroup[SU3l, SU[3]];     DefineGlobalGroup[U1l, U1];
  DefineGlobalGroup[SU3e, SU[3]];     DefineGlobalGroup[U1e, U1];
\end{mmaCell}
It is only possible to define individual simple and $ \U(1) $ groups in \matchete, which is why $\mathrm{U}(N)$ has to be implemented by defining separate $\mathrm{SU}(N) $ and $ \mathrm{U}(1)$ group factors.
The flavor symmetry can be assigned to the fields of the model through an association such as
\begin{mmaCell}{Input}
  flavorSym = <|
   q -> \{\{1,2\}   -> \{SU2q[fund], U1q[1]\}\},
   u -> \{\{1,2\}   -> \{SU2u[fund], U1u[1]\}\},
   d -> \{\{1,2,3\} -> \{SU3d[fund], U1d[1]\}\},
   \mmaUnd{l} -> \{\{1,2,3\} -> \{SU3l[fund], U1l[1]\}\},
   e -> \{\{1,2,3\} -> \{SU3e[fund], U1e[1]\}\}
  |>;
\end{mmaCell}
where the third generation for the fields \mmaInlineCell[]{Input}{q} and~\mmaInlineCell[]{Input}{u} (which is not explicitly specified in the rule) is automatically taken to be a singlet.

The flavor symmetry can be imposed on, e.g., the Mainz basis SMEFT Lagrangian, as follows:
\begin{mmaCell}{Input}
  LMainz = LoadModel["SMEFT_Mainz+breaking"]
  flavorInvariants\,=\,ImposeFlavorSymmetry[LMainz,\,flavorSym,\,Except\,->\,\{Yu,Yd,Ye\}]
\end{mmaCell}
where it is crucial that \mmaInlineCell[]{Input}{LMainz} is the SMEFT Lagrangian before EWSB, since the flavor symmetry is defined for the chiral fields. 
The last argument is used to exclude the Yukawa couplings \mmaInlineCell[]{Input}{Yu}, \mmaInlineCell[]{Input}{Yd}, and~\mmaInlineCell[]{Input}{Ye} from the symmetry such that they retain their general form.\footnote{If the Yukawas have been replaced (as usual) by the diagonal fermion mass matrices before, this has no effect and the masses remain in the most general form in any case, since they are not present in the unbroken phase Lagrangian.}
This routine derives the flavor transformation properties of all flavored couplings from the field content of the corresponding operators. 
It then determines the invariants under the given flavor representation product, combines them with any additional permutation symmetries of the coefficient, and returns the resulting invariant tensor structures in the form of \mmaInlineCell[]{Input}{SparseArray} objects.
These can be inspected using 
\begin{mmaCell}{Input}
  ArrayRules[flavorInvariants[\mmaUnd{cHd}][\!\![1,1]\!\!]]
\end{mmaCell}
\begin{mmaCell}{Output}
  \{\{1,1\}->1, \{2,2\}->1, \{3,3\}->1, \{_,_\}->0\}
\end{mmaCell}
where we have chosen the Wilson coefficient~$C_{Hd}^{pr}$ of the operator $Q_{Hd}^{pr}= i (H^\dagger \overset{\leftrightarrow}{D}_\mu H) (\bar{d}^p \gamma^\mu d^r)$ as an illustrative example. 
Displayed as a matrix, this object assumes the form
\begin{align}
    \begin{pmatrix}
        1 & 0 & 0 \\
        0 & 1 & 0 \\
        0 & 0 & 1
    \end{pmatrix},
\end{align}
and is easily recognized as the trivial invariant. 
When generating a parameter card afterwards, either using \mmaInlineCell[]{Input}{DefaultParamCard} or directly within the function \mmaInlineCell[]{Input}{ExportUFO}, as discussed in Sec.~\ref{sec:UFO-generation}, these flavor-invariant structures can then be used in order to parametrize the given coupling.
For the example of~$C_{Hd}$ at hand, this means only a single free real parameter is introduced and all diagonal entries of~$C_{Hd}$ are set equal to this parameter, while all off-diagonal entries are entirely removed from the UFO file.
The flavor symmetry implementation has been validated against the results of Ref.~\cite{Faroughy:2020ina}.

We remark that finding a minimal flavor-symmetric parametrization relies on all permutation symmetries of the coefficients being properly defined. 
This is the case for the vast majority of the SMEFT operators.
However, operators with three or more repeated fields, such as the baryon-number-violating $d=6$ operator $Q_{qqq}^{prst} = \varepsilon^{abc} \varepsilon^{il} \varepsilon^{jk} ({q}^\intercal_{aip} C q_{bjr})({q}^\intercal_{cks} C \ell_{lt})$ and some of the $d=8$ SMEFT operators in the class {\bf $\psi^4 H D$}, exhibit multi-term symmetries that are not currently supported in \matchete. 
We refer to~\cite{Li:2020gnx} for an alternative $d=8$ basis with explicit flavor permutation symmetries; however, not all of these symmetries can currently be implemented in \matchete.

\subsection{Electroweak Symmetry Breaking}
\label{sec:SMEFT-SSB}

Section~\ref{sec:matchete_symmetry-breaking} introduced the (semi-)automated SSB functionality in \matchete while details about the SM implementation are collected in Appendix~\ref{app:SSB}.
Here, we briefly review the symmetry-breaking relations for the SMEFT that have been discussed for $d=6$ in~\cite{Alonso:2013hga} and for $d=8$ in~\cite{Hamoudou:2022tdn}. 
The SMEFT operators modify the scalar potential terms, which in our conventions read
\begin{equation}
    \mathcal{L}_H = -\frac{\lambda}{2} \left(H^\dagger H - \frac{v^2}{2}\right)^2 + \frac{C_H}{\Lambda^2}\left(H^\dagger H\right)^3 + \frac{C_{H^8}}{\Lambda^4} \left(H^\dagger H\right)^4
    \label{eq:HiggsPotentiald8}
\end{equation}
up to $d=8$ following the Warsaw~\cite{Grzadkowski:2010es} and Murphy~\cite{Murphy:2020rsh} bases and naming schemes for the SMEFT operators. 
To make the power counting manifest, we have introduced the NP scale~$\Lambda$ here and in the remainder of Sec.~\ref{sec:SMEFT}. 
\matchete, on the other hand, employs dimensionful coefficients. 
Minimization of the effective potential at the semi-classical level yields the modified Higgs vacuum expectation value~(VEV)
\begin{align}
    v_T &= \sqrt{2\langle H^\dagger H\rangle}
        = v \left[ 
        1 + 
        \frac{3v^2}{4 \lambda \Lambda^2} C_H 
        + \frac{v^4}{2 \lambda \Lambda^4} \left( 
            \frac{63}{16 \lambda} C_H^2 
            + C_{H^8}
        \right)
    \right].
\end{align}
We then choose the following decomposition of the scalar doublet
\begin{equation}
    \label{eq:Higgs_decomposition}
    H 
    =  
    \begin{pmatrix} 
        -i \left(1+c_{\chi^+,\text{kin}}\right) \chi^+ 
        \\ 
        \frac{1}{\sqrt{2}} \left[ v_T +\left(1+c_{H,\text{kin}}\right)h + i \left(1 + c_{\chi^0,\text{kin}}\right) \chi^0 \right]
    \end{pmatrix},
\end{equation}
where we include normalization factors $c_{H,\text{kin}}$, $c_{\chi^0,\text{kin}}$, and $c_{\chi^+,\text{kin}}$ to compensate SMEFT corrections to the kinetic terms of the physical scalar~$h$ and the would-be Goldstone bosons~$\chi^{0,\pm}$. 
With this choice all broken-phase scalars are canonically normalized.
Up to $d=8$, these normalization constants read
\begin{align}
\begin{split} 
    c_{H,\text{kin}} 
    &= 
    \frac{v_T^2}{\Lambda^2} \left(
        C_{H\Box} 
        - \frac{C_{HD}}{4} 
    \right)
    + \frac{v_T^4}{2\Lambda^4} \left[ 
        3 \left( C_{H\Box} - \frac{C_{HD}}{4} \right)^2 
        - \frac{1}{4} C_{H^6}^{(1+2)} 
    \right] 
    \, ,
    \\
    c_{\chi^0,\text{kin}} 
    &= 
    -\frac{v_T^2}{4\Lambda^2} C_{HD}
    + \frac{v_T^4}{8\Lambda^4} \left(
        \frac{3}{4} C_{HD}^2
        - C_{H^6}^{(1+2)}
    \right)
    \, ,
    \qquad\qquad
    c_{\chi^+,\text{kin}} 
    = 
    -\frac{v_T^4}{8\Lambda^4} C_{H^6}^{(1-2)}
    \, ,
\end{split}
\end{align}
where we employ the notation $C^{(n \pm m)}_x \equiv C^{(n)}_x \pm  C^{(m)}_x$.

As in the SM, the generated mass matrix for the gauge bosons in the broken phase is non-diagonal. 
Furthermore, SMEFT operators modify the normalization of the gauge bosons and generate mixed kinetic terms. 
In this section, we use the normalization of vector bosons that is used for the calculation of the Feynman rules in \matchete, namely the convention where the covariant derivative reads $D_\mu \phi= (\partial_\mu - igT^IA_\mu^I) \phi$. 

In the broken phase and setting $W_\mu^\pm = \frac{1}{\sqrt{2}} \left( W_\mu^1 \mp iW_\mu^2 \right)$, the kinetic Lagrangian of the weak gauge bosons is parametrized as
\begin{equation}
    \begin{aligned}
        \mathcal{L}_{\text{EW},\text{kin}} = &-\frac{\alpha}{2} W_{\mu\nu}^+ W^{-\mu\nu} - \frac{\beta}{4} W_{\mu\nu}^3 W^{3\mu\nu} - \frac{\gamma}{4} B_{\mu\nu}B^{\mu\nu} - \frac{\delta}{2} W_{\mu\nu}^3 B^{\mu\nu} \\ 
        &+ \epsilon \frac{g_2^2v_T^2}{4} W_\mu^+ W^{-\mu} + \frac{\zeta}{2} \frac{v_T^2}{4}\left(g_2 W_\mu^3 - g_1 B_\mu\right)^2
    \end{aligned}
\end{equation}
with
\begin{align}
    \alpha &= 1 - 2\frac{v_T^2}{\Lambda^2} C_{HW} - \frac{v_T^4}{\Lambda^4} C_{W^2 H^4}^{(1)}
    \,, 
    &
    \delta &= \frac{v_T^2}{\Lambda^2} C_{HWB} + \frac{v_T^4}{2\Lambda^4} C_{WBH^4}^{(1)}
    \,,
    \nonumber\\
    \beta &= 1 - 2\frac{v_T^2}{\Lambda^2} C_{HW} - \frac{v_T^4}{\Lambda^4} C_{W^2 H^4}^{(1+3)}
    \,, 
    &
    \epsilon &= 1 + \frac{v_T^4}{4 \Lambda^4} C_{H^6}^{(1-2)}
    \,,
    \\
    \gamma &= 1 - 2\frac{v_T^2}{\Lambda^2} C_{HB} - \frac{v_T^4}{\Lambda^4} C_{B^2H^4}^{(1)}
    \,, 
    &
    \zeta &= 1 + \frac{v_T^2}{2\Lambda^2} C_{HD} + \frac{v_T^4}{4\Lambda^4} C_{H^6}^{(1+2)}
    \,.
    \nonumber
\end{align}
The relation $\alpha = \beta$ breaks down at $ d= 8 $. 
In order to transform to the physical mass basis and canonical normalization, we first introduce rescaled gauge couplings~$\bar{g}_i$
\begin{equation}
    \begin{aligned}
        g_1 &= \sqrt{\gamma} \, \overline g_1, 
        &
        \quad g_2 &= \sqrt{\alpha} \, \overline g_2, 
        &
        \quad g_3 &= \left[ 
            1 
            - \frac{v_T^2}{\Lambda^2} C_{HG} 
            - \frac{v_T^4}{2\Lambda^4} \left(
                C_{HG}^2
                + C_{G^2H^4}^{(1)}
            \right)
        \right] \overline g_3. 
    \end{aligned}
\end{equation}
We then introduce the mass-eigenstate gauge fields $\mathcal{G}_\mu^A$, $\mathcal{W}_\mu^\pm$, $\mathcal{Z}_\mu$ and~$\mathcal{A}_\mu$ in the broken phase, which are obtained through simultaneous canonization of kinetic terms and diagonalization of mass terms~\cite{Grinstein:1991cd}:
\begin{align}
        \mathcal{G}_\mu^A &= \left[ 
            1 
            - \frac{v_T^2}{\Lambda^2} C_{HG} 
            -
            \frac{v_T^4}{2\Lambda^4} \left( 
                C_{HG}^2
                + C_{G^2H^4}^{(1)}
            \right)
        \right] G_\mu^A,
        \qquad
        \mathcal{W}_\mu^\pm = \sqrt{\alpha} \,W_\mu^\pm,
        \qquad
        \begin{pmatrix} W^3_\mu \\ B_\mu \end{pmatrix} = X \begin{pmatrix} \mathcal{Z}_\mu \\ \mathcal{A}_\mu \end{pmatrix}, 
        \nonumber \\
        X &= \begin{pmatrix} \frac{1}{\sqrt{\beta}} & 0 \\ 0 & \frac{1}{\sqrt{\gamma}} \end{pmatrix} R \begin{pmatrix} \left( 1 + \frac{\delta}{\sqrt{\beta\gamma}}\right)^{-\frac{1}{2}} & 0 \\ 0 & \left( 1 - \frac{\delta}{\sqrt{\beta\gamma}}\right)^{-\frac{1}{2}}\end{pmatrix} R^T \begin{pmatrix} \cos\bar\theta & \sin\bar\theta \\ -\sin\bar\theta & \cos\bar\theta \end{pmatrix}, 
        \\
        R &= \frac{1}{\sqrt{2}} \begin{pmatrix} 1 & -1 \\ 1 & 1 
        \end{pmatrix}.
        \nonumber
\end{align}
The transformation matrix~$X$ successively rescales the fields, diagonalizes the kinetic terms, and then diagonalizes the mass terms with the modified mixing angle~$\bar\theta$. 
We reproduce the relations for the mixing angle and for the modified electromagnetic coupling from~\cite{Hamoudou:2022tdn}:
\begin{subequations}
\begin{align}
    \begin{split}
        \cos\overline \theta 
        &= 
        \frac{1}{\sqrt{\overline g_1^2 + \overline g_2^2}} \bigg[ \overline g_2 + \frac{v_T^4}{2\Lambda^4} \frac{\overline g_2 \overline g_1^2}{\overline g_1^2 + \overline g_2^2} C_{W^2H^4}^{(3)} + \frac{v_T^4 \overline g_2}{8\Lambda^4} \frac{6\overline g_1^2 \overline g_2^2 - \overline g_2^4 - 5\overline g_1^4}{(\overline g_1^2 + \overline g_2^2)^2} (C_{HWB})^2 
        \\
        &\quad+ \frac{v_T^2\overline g_1}{2\Lambda^2} \frac{\overline g_1^2 - \overline g_2^2}{\overline g_1^2 + \overline g_2^2} \left( C_{HWB} + \frac{v_T^2}{2\Lambda^2} C_{WBH^4}^{(1)} + \frac{v_T^2}{\Lambda^2} C_{HWB} \left( C_{HW} + C_{HB} \right) \right) \bigg] 
        \,,
    \end{split}
\\[0.2cm]
    \begin{split}
        \overline e 
        &= 
        \frac{\overline g_1 \overline g_2}{\sqrt{\overline g_1^2 + \overline g_2^2}} \bigg[ 1 - \frac{v_T^2}{\Lambda^2} \frac{\overline g_1 \overline g_2}{\overline g_1^2 + \overline g_2^2} C_{HWB} + \frac{v_T^4}{2\Lambda^4} \frac{\overline g_1^2}{\overline g_1^2 + \overline g_2^2} C_{W^2H^4}^{(3)} - \frac{v_T^4}{2\Lambda^4} \frac{\overline g_1 \overline g_2}{\overline g_1^2 + \overline g_2^2} C_{WBH^4}^{(1)} 
        \\
        &\quad- \frac{v_T^4}{\Lambda^4} \frac{\overline g_1 \overline g_2}{\overline g_1^2 + \overline g_2^2} C_{HWB} (C_{HW} + C_{HB}) + \frac{3}{2} \frac{v_T^4}{\Lambda^4} \frac{\overline g_1^2 \overline g_2^2}{(\overline g_1^2 + \overline g_2^2)^2} (C_{HWB})^2 \bigg]
        \,.
    \end{split}
\end{align}
\end{subequations}

Lastly, the Yukawa terms also receive SMEFT corrections in the broken phase. 
The relevant terms up to $d=8$ read
\begin{equation}
    \begin{aligned}
        \mathcal{L}_\text{Yukawa} = \bigg[ &\left( -[Y_d]^{pr} + \frac{[C_{dH}]^{pr}}{\Lambda^2} (H^\dagger H) + \frac{[C_{qdH^5}]^{pr}}{\Lambda^4} (H^\dagger H)^2 \right) H \bar q^p d^r
        \\ + &\left(-[Y_u]^{pr} + \frac{[C_{uH}]^{pr}}{\Lambda^2} (H^\dagger H) + \frac{[C_{quH^5}]^{pr}}{\Lambda^4} (H^\dagger H)^2 \right) \widetilde H \bar q^p u^r 
        \\ + &\left(-[Y_e]^{pr} + \frac{[C_{eH}]^{pr}}{\Lambda^2} (H^\dagger H) + \frac{[C_{\ell eH^5}]^{pr}}{\Lambda^4} (H^\dagger H)^2 \right) H \bar \ell^p e^r \bigg]  + \text{H.c.}
        \,.
    \end{aligned}
\end{equation}
After inserting the Higgs field decomposition~\eqref{eq:Higgs_decomposition}, these terms yield the fermion mass matrices
\begin{equation}
    M_\psi = \frac{v_T}{\sqrt{2}} \left( 
        Y_\psi 
        - \frac{v_T^2}{2\Lambda^2} C_{\psi H}
        - \frac{v_T^4}{4\Lambda^4} C_{\Psi \psi H^5}
    \right)
\end{equation}
with the appropriate $\psi \in \{u,d,e\}$ and $\Psi \in \{q,\ell\}$. 
The Yukawa interactions of the fermions with the physical Higgs boson $ h $ have coupling matrices
\begin{equation}
    \mathcal{Y}_\psi = \frac{1+c_{H,\text{kin}}}{\sqrt{2}} \left(
        Y_\psi 
        - \frac{3 v_T^2}{2 \Lambda^2} C_{\psi H} 
        - \frac{5 v_T^4}{4 \Lambda^4} C_{\Psi \psi H^5} 
    \right).
\end{equation}
Unlike the case of the ordinary SM, $M_\psi$~and~$\cY_\psi$ need not be simultaneously diagonalizable, allowing for flavor-violating couplings of the physical Higgs boson. 

We can rotate to the fermion mass basis with chiral unitary transformations $\psi_{L/R} \rightarrow U_{\psi_{L/R}} \psi_{L/R}$, the only physical combination of which is
\begin{equation}
    V_{\sscript{CKM}} = U_{u_L}^\dagger U_{d_L}\,,
\end{equation}
which defines the Cabibbo–Kobayashi–Maskawa~(CKM) matrix.
In our \matchete implementation, the Pontecorvo–Maki–Nakagawa–Sakata~(PMNS) matrix for the leptons is currently set to unity but can be implemented 
analogously to the CKM matrix at a later stage (e.g., when considering non-zero values for the Weinberg operator).
Four alignment choices for the CKM matrix are available in the SMEFT model files distributed with \matchete: 
the down-alignment with $U_{d_L}=\unit$, the up-alignment with $U_{u_L}=\unit$, a general choice where both matrices are kept non-trivial, and the simplified case where the CKM is set to the unit matrix $V_{\sscript{CKM}}=\unit$. 
For example, the down-aligned Warsaw basis Lagrangian (prepared for SSB) can be loaded using
\begin{mmaCell}{Input}
  LWarsaw = LoadModel["SMEFT_Warsaw+breaking", 
      ModelParameters -> \{\mmaUnd{QuarkAlignment} -> "DownAlignment"\}];
\end{mmaCell}
The other choices are obtained by replacing~\mmaInlineCell[]{Input}{"DownAlignment"} by one of the following strings \mmaInlineCell[]{Input}{"UpAlignment"}, \mmaInlineCell[]{Input}{"NoAlignment"}, or \mmaInlineCell[]{Input}{"UnitCKM"}.

\subsection{Electroweak Input Scheme}
\label{sec:input_scheme}

In SMEFT studies, the electroweak SM parameters are usually fixed by a choice of four independent input observables. 
For the $d=6$ Warsaw basis, we provide implementations of two different input schemes:\footnote{We do not provide an implementation of input schemes for the Mainz basis and, instead, refer to~\cite{MainzBasisInput} for details.} the $M_W$~scheme using the input observables $\{M_W,M_Z,G_F,M_h\}$ (the $W$- and $Z$-boson masses, the Fermi constant, and the Higgs-boson mass) and the $\alpha$~scheme with inputs $\{\alpha_\sscript{EM}(M_Z),M_Z,G_F,M_h\}$ ($\alpha_\sscript{EM}$ is the electromagnetic fine-structure constant).
Other schemes can be implemented by the user in a straightforward manner by modifying the parameter card.\footnote{%
The \matchete implementation introduces generic shifts, such as~$ \delta g_2^{(6)}$, for all SM parameters, and their values are later set in the parameter card. This allows for using a single model file for several different input schemes.}
The $W$~mass receives corrections from other SMEFT operators if it is not used as an input. 
For a consistent EFT truncation, the $W$~propagators should thus be expanded to the desired order in the power counting.
While such a feature is implemented in \textsc{SMEFTsim}~\cite{Brivio:2020onw}, it is not currently supported in \matchete, which keeps the EFT contributions to the mass in the denominator of the propagator. 
For a discussion of the differences between the two approaches, we refer to~\cite{Brivio:2020onw}.

At $d=8$ for the modified Murphy basis, we adopt only the $M_W$~scheme, where the gauge boson propagators do not get EFT corrections.
The $M_W$~input-scheme relations at $d=8$ were previously provided for another basis similar to Murphy's basis by \textsc{SmeftFR} in Appendix~A of~\cite{Dedes:2023zws} (see also~\cite{Hays:2020scx} for a partial geometric determination); however, we find several discrepancies as discussed at the end of this section.
Other input schemes can again be implemented by the user once the appropriate shifts are derived.
We briefly review the derivation of the input shifts in general and proceed to apply this formalism to the $M_W$~scheme up to $d=8$.

The relations between Lagrangian parameters and experimental inputs are determined in two steps: 
First, the predictions for the input observables~$\cO_n$ are calculated within the SMEFT. 
We parametrize these results in the general form
\begin{equation}
    \mathcal{O}_n = \mathcal{O}_n^{(4)}(g) \left[1 + \delta \mathcal{O}_n^{(6)}(g,C) + \delta \mathcal{O}_n^{(8)}(g,C)\right],
    \label{eq:input-observable-expansion}
\end{equation}
where $\mathcal{O}_n^{(d)} = \mathcal{O}(\Lambda^{4-d})$. 
The SMEFT Wilson coefficients are denoted by~$C$, and $g$~represents the set of electroweak Lagrangian parameters. 
Second, the relations~\eqref{eq:input-observable-expansion} are solved for the EW~parameters perturbatively up to the required order in~$1/\Lambda$, which yields relations of the form
\begin{equation}
    g_i = g_i^{(4)}(\mathcal{O}) \left[1 + \delta g_i^{(6)}(\mathcal{O},C) + \delta g_i^{(8)}(\mathcal{O},C)\right]
    \label{eq:input-coupling-expansion}
\end{equation}
in a similar notation. 
This procedure requires a number of input observables identical to the number of EW~parameters.
A general treatment of SMEFT input schemes, also including loop corrections, can be found in~\cite{Biekotter:2023xle} while a systematic procedure for the derivation of input relations can be found in~\cite{Brivio:2020onw}.

The Higgs VEV is determined by the Fermi constant, which in turn is inferred from the muon decay width. 
To determine the SMEFT prediction for this observable, one should first match onto the relevant terms of the Low-Energy Effective Field Theory (LEFT)~\cite{Jenkins:2017jig} 
\begin{equation}
    \label{eq:LEFT}
    \mathcal{L}_\text{LEFT} \supset [\mathcal{C}_{LL}]^{pr} \, (\overline \nu^p \gamma_\mu P_L \nu^r)(\overline{e} \gamma^\mu P_L \mu) + [\mathcal{C}_{LR}]^{pr} \, (\overline{\nu}^p \gamma_\mu P_L \nu^r)(\overline{e} \gamma^\mu P_R \mu) + \text{H.c.},
\end{equation}
by integrating out the $\mathcal{Z}$ and $\mathcal{W}^\pm$ bosons.
The tree-level result for the matching coefficients up to $d=8$ (obtained with \matchete) reads 
\begin{subequations}
\begin{align}
    [\mathcal{C}_{LL}]^{pr} 
    &= 
    -\frac{2}{v_T^2} \delta^{1r} \delta^{2p}
    + \frac{1}{\Lambda^2} \bigg\{ 
        \delta^{pr} [C_{H\ell}^{(1+3)}]^{12}  
        - 2 \delta^{1r} [C_{H\ell}^{(3)}]^{p2}  
        - 2 \delta^{2p} [C_{H\ell}^{(3)}]^{1r}  
        + [C_{\ell\ell}]^{12pr} + [C_{\ell\ell}]^{pr12} 
    \bigg\}
    \nonumber\\
    &\quad+ \frac{v_T^2}{2\Lambda^4} \bigg\{ 
    \delta^{pr} \Big( 
        [C_{\ell^2H^4D}^{(1)}]^{12} 
        + 2 [C_{\ell^2H^4D}^{(2)}]^{12} 
        - C_{HD} [C_{H\ell}^{(1+3)}]^{12}  
    \Big) 
    \nonumber\\
    &\qquad+ \delta^{1r} \delta^{2p} C_{H^6}^{(1-2)}  
    - 2 \delta^{2p} [C_{\ell^2H^4D}^{(2+i3+4)}]^{1r} 
    - 2 \delta^{1r} [C_{\ell^2H^4D}^{(2-i3+4)}]^{p2} 
    \\
    &\qquad- 2 [C_{H\ell}^{(1+3)}]^{12} [C_{H\ell}^{(1-3)}]^{pr} 
    - 4 [C_{H\ell}^{(3)}]^{1r} [C_{H\ell}^{(3)}]^{p2} 
    + [C_{\ell^4H^2}^{(1-2)}]^{12pr} 
    + [C_{\ell^4H^2}^{(1+2)}]^{pr12} 
    \bigg\}
    \,,
    \nonumber\\[0.2cm]
\begin{split}
    [\mathcal{C}_{LR}]^{pr} 
    &= 
    \frac{1}{\Lambda^2} \bigg\{ 
        [C_{\ell e}]^{pr12} 
        + \delta^{pr} [C_{He}]^{12} 
    \bigg\} 
    \\
    &\quad+ \frac{v_T^2}{2\Lambda^4} \bigg\{ 
        [C_{\ell^2e^2H^2}^{(1-2)}]^{pr12} 
        + \delta^{pr} [C_{e^2H^4D}]^{12} 
        - [C_{He}]^{12} \left( \delta^{pr} C_{HD} + 2[C_{H\ell}^{(1-3)}]^{pr} \right) 
    \bigg\} 
    \,,
\end{split}
\end{align}%
\end{subequations}%
adopting the notation $C_x^{(n\pm im)} \equiv C_x^{(n)}\pm i C_x^{(m)}$.
These relations have also been calculated up to $d=6$ in~\cite{Jenkins:2017jig} and to $d=8$ in~\cite{Hamoudou:2022tdn}; however, our relation for $\mathcal{C}_{LL}$ differs by the terms proportional to $C_{\ell^2H^4D}^{(4)}$, the operator of which gives a contribution to the coupling of the $\mathcal{W}$~boson to left-handed leptons:
\begin{equation}
    \begin{aligned}
    [Q_{\ell^2H^4D}^{(4)}]^{pr} &= \varepsilon^{IJK} (\overline \ell^p \gamma^\mu \tau^I \ell^r) (H^\dagger \tau^J H) D_\mu (H^\dagger \tau^K H) \\
    & \xrightarrow[]{\;\;\mathrm{SSB}\;\;} \ \frac{v_T^4}{2} \frac{\overline g_2}{\sqrt{2}} \left[ (\overline{\nu}^p \gamma^\mu P_L e^r) \mathcal{W}^+_\mu + (\overline e^p \gamma^\mu P_L \nu^r) \mathcal{W}^-_\mu \right] + \dots 
    \,.
    \end{aligned}
\end{equation}
Here, $\tau^I$~denotes the Pauli matrices and $\varepsilon^{IJK}$ the fully anti-symmetric tensor of the adjoint $\mathrm{SU}(2)_L$-indices.

The Fermi constant is treated as a pseudo-observable, a stand-in for the decay width of the muon, and is defined by
\begin{equation}
    \label{eq:Fermiconstant}
    G_F = \frac{1}{2\sqrt{2}} \, \sqrt{ 
    \big| [\mathcal{C}_{LL}]^{pr} \big|^2 + \big| [\mathcal{C}_{LR}]^{pr} \big|^2 
    }
    \,,
\end{equation}
where the interference term is negligible and has been omitted. 
Einstein summation over all neutrino flavors~($p,r$) is understood, since the neutrino flavors cannot be discriminated experimentally.
Following the parametrization~\eqref{eq:input-observable-expansion}, the expansion of the Fermi constant reads
\begin{align}
    G_F^{(4)} = &\ \frac{1}{\sqrt{2} \, v_T^2}
    \,, 
    \qquad\qquad\qquad
    \delta G_F^{(6)} 
    = 
    \frac{v_T^2}{\Lambda^2}\left[ [C_{H\ell }^{(3)}]^{11} - \frac{1}{2} [C_{\ell\ell}]^{1221} + (1 \leftrightarrow 2) \right]
    ,
    \nonumber
\\[0.2cm]
    \delta G_F^{(8)}
    =
    &- \frac{1}{2}\left|\delta G_F^{(6)}\right|^2
    + \frac{v_{T}^4}{16\Lambda^4} \Bigg\{ 3 \Big|[C_{He}]^{12}\Big|^2 + 2 [C_{He}]^{21} [C_{\ell e}]^{pp12} + \frac{1}{2} \Big| [C_{\ell e}]^{pr12} \Big|^2
    \nonumber\\
    &\quad+
    7 \Big|[C_{H\ell}^{(1)}]^{12}\Big|^2 
    - 9 \Big|[C_{H\ell}^{(3)}]^{12}\Big|^2 
    - 2 [C_{H\ell}^{(1)}]^{21} [C_{H\ell}^{(3)}]^{12}
    + 8 \Big| [C_{H\ell}^{(3)}]^{1p} \Big|^2 
    \nonumber\\
    &\quad+ 16 [C_{H\ell}^{(3)}]^{11} [C_{H\ell}^{(3)}]^{22}
    + 2 \, [C_{H\ell}^{(1+3)}]^{21} \Big( [C_{\ell\ell}]^{12pp} + [C_{\ell\ell}]^{pp12} \Big)
    \label{eq:dGF8}\\
    &\quad- 4 [C_{H\ell}^{(3)}]^{1p} \Big( [C_{\ell\ell}]^{21p2} + [C_{\ell\ell}]^{p221} \Big)
    - 4 [C_{H\ell}^{(3)}]^{p1} \Big( [C_{\ell\ell}]^{122p} + [C_{\ell\ell}]^{2p12} \Big)
    \nonumber\\
    &\quad- 2 C_{H^6}^{(1-2)} 
    + 8 [C_{\ell^2H^4D}^{(2+4)}]^{11} 
    - 4 [C_{\ell^4H^2}^{(1)}]^{1221} 
    \nonumber\\
    &\quad+ \left[ 
        \Big|[C_{\ell\ell}]^{12pr}\Big|^2
        + \Big|[C_{\ell\ell}]^{pr12}\Big|^2
        + [C_{\ell\ell}]^{21rp} [C_{\ell\ell}]^{pr12}
        + [C_{\ell\ell}]^{12pr} [C_{\ell\ell}]^{rp21}
    \right]
    + (1 \leftrightarrow 2)
    \Bigg\} .
    \nonumber
\end{align}
The final $ (1 \leftrightarrow 2) $ indicates the symmetric sum covering the permutation of the flavor indices~`1' and~`2'.\footnote{This applies only to the explicit numeric indices and not to the indices labeled $p$ and~$r$ when they assume these values.}
In the input scheme implementation for our tree-level SMEFT studies, we neglect the running of the Wilson coefficients from the scale set by the muon mass to the experimental scales at the LHC.

The symmetry-breaking relations from Sec.~\ref{sec:SMEFT-SSB} yield the following explicit results for the input masses within the SMEFT:
Starting with the $\mathcal{W}$-boson mass, we find
\begin{align}
    M_W^{(4)} 
    &= 
    \frac{1}{2} \overline g_2 v_T 
    \,, 
    &
    \delta M_W^{(6)} 
    &= 
    0
    \,, 
    &
    \delta M_W^{(8)} 
    &= 
    \frac{v_T^4}{8\Lambda^4} C_{H^6}^{(1-2)}
\end{align}
while we obtain the $\mathcal{Z}$-boson mass corrections
\begin{align}
    M_Z^{(4)} 
    &= 
    \frac{v_T}{2} \sqrt{\overline g_1^2 + \overline g_2^2}
    \,, 
    \qquad\qquad\qquad
    \delta M_Z^{(6)} 
    = 
    \frac{v_T^2}{4\Lambda^2}\left[ 
        C_{HD} 
        + 4\frac{\overline g_1 \overline g_2}{\overline g_1^2 + \overline g_2^2} C_{HWB} 
    \right], 
    \nonumber\\
    \delta M_Z^{(8)} 
    &= 
    \frac{v_T^4}{\Lambda^4} 
    \left\{ 
        \frac{\overline g_1 \overline g_2}{4(\overline g_1^2 + \overline g_2^2)} 
        \left[ 
            C_{HWB} 
            \big( 
                C_{HD} 
                + 4C_{HW} 
                + 4C_{HB}
            \big) 
            + 2C_{WBH^4}^{(1)} 
        \right]
    \right.
    \\
    &\quad\left. 
    + \frac{\overline g_2^2}{2(\overline g_1^2 + \overline g_2^2)} C_{W^2H^4}^{(3)} 
    + \frac{1}{32} 
    \left( 
        4 C_{H^6}^{(1+2)} 
        - C_{HD}^2 
        + 16 C_{HWB}^2 
    \right) 
    - \frac{\overline g_1^2 \overline g_2^2}{2(\overline g_1^2 + \overline g_2^2)^2} C_{HWB}^2 
    \right\}
    \nonumber
\end{align}
and the mass of the physical Higgs boson 
\begin{align}
    M_h^{(4)} 
    &= 
    v_T \sqrt{\lambda}
    \,, 
    \qquad\qquad\qquad
    \delta M_h^{(6)} 
    = 
    -\frac{v_T^2}{4\Lambda^2} \left[ \frac{6}{\lambda} C_H -4 C_{H\Box} + C_{HD} \right], 
    \\
    \delta M_h^{(8)} 
    &= 
    -\frac{v_T^4}{8\Lambda^{4}} 
    \left\{ 
        \frac{3}{\lambda} 
        \left[ C_H 
            \left(
                \frac{3}{\lambda} C_H
                + 4 C_{H\Box} 
                - C_{HD}
            \right)
            + 4 C_{H^8}
        \right]          
    - \frac{3}{4}\big( C_{HD} - 4 C_{H\Box} \big)^2 
    + C_{H^6}^{(1+2)}
    \right\}.
    \nonumber
\end{align}

Having calculated the four $ M_W $-scheme observables within the $ d=8 $ SMEFT, we next determine the EW~parameters, using the algorithm described in Ref.~\cite{Brivio:2020onw}.
For the Higgs vacuum expectation value we find
\begin{align} \label{eq:vT_shift}
    v_T^{(4)} 
    &= 
    \frac{1}{2^{1/4} \sqrt{G_F}}
    \,, 
    &
    \delta v_T^{(6)} 
    &= 
    \frac{1}{2} \delta G_F^{(6)}
    \,, 
    &
    \delta v_T^{(8)} 
    &= 
    \frac{1}{2}\delta G_F^{(8)} + \frac{3}{8} \big(\delta G_F^{(6)}\big)^2
\end{align}
while the corrections to the Higgs quartic coupling are given by
\begin{align}
\begin{split}
    \lambda^{(4)} 
    &= 
    \sqrt{2} M_h^2 G_F
    \,, 
    \qquad\qquad\qquad
    \delta\lambda^{(6)} 
    = 
    -2\delta M_h^{(6)} - \delta G_F^{(6)}, 
    \\
    \delta\lambda^{(8)} 
    &= 
    -2\delta M_h^{(8)} - \delta G_F^{(8)} +3 \big(\delta M_h^{(6)}\big)^2 - 2 \lambda \frac{\partial \,\delta M_h^{(6)}}{\partial \lambda} \left( \delta G_F^{(6)} + 2 \delta M_h^{(6)} \right)
    \,.
\end{split}
\end{align}
The gauge coupling shifts read 
\begin{align}
    \overline g_2^{(4)} 
    &= 
    2^{5/4}\sqrt{G_F}M_W
    \,, 
    &
    \delta\overline g_2^{(6)} 
    &= 
    - \frac{1}{2}\delta G_F^{(6)}
    \,, 
    &
    \delta\overline g_2^{(8)} 
    &= 
    -\delta M_W^{(8)} 
    - \frac{1}{2}\delta G_F^{(8)} 
    - \frac{1}{8} \big( \delta G_F^{(6)} \big)^2
    \,,
\end{align}
and
\begin{align} \label{eq:g1_shift}
\begin{split}
    \overline g_1^{(4)} 
    &=
    2^{5/4}\sqrt{G_F}\sqrt{M_Z^2-M_W^2}
    \,,
    \qquad\qquad
    \delta\overline g_1^{(6)} = 
    - \frac{M_Z^2}{M_Z^2-M_W^2}\delta M_Z^{(6)} - \frac{1}{2}\delta G_F^{(6)}
    \,,
    \\[0.2cm]
    \delta \overline{g}_1^{(8)}
    &=
    - \frac{1}{2} \delta G_F^{(8)}
    - \frac{1}{8} \big(\delta G_F^{(6)}\big)^2 + \frac{M_W^2}{M_Z^2-M_W^2} \delta M_W^{(8)}
    - \frac{M_Z^2}{M_Z^2-M_W^2} \Bigg[ 
        \delta M_Z^{(8)} 
        + \frac{1}{2} \delta G_F^{(6)} \delta M_Z^{(6)}
    \\
    &\qquad 
        + \left(
            \frac{1}{2} \frac{M_W^2}{M_Z^2-M_W^2} 
            - 1
        \right) \big(\delta M_Z^{(6)}\big)^2
        - \frac{M_Z^2}{M_Z^2-M_W^2} \left(\overline{g}_1 \frac{\partial \, \delta M_Z^{(6)}}{\partial \overline{g}_1}\right) \delta M_Z^{(6)}
    \Bigg]
    \,,
\end{split}
\end{align}
where
\begin{subequations}
\begin{align}
    \lambda \frac{\partial \, \delta M_h^{(6)}}{\partial \lambda} 
    &=
    \frac{3}{4 G_F^2 M_h^2} \frac{C_H}{\Lambda^2}
    \,,
    \\[0.2cm]
\begin{split} 
    \overline{g}_1 \frac{\partial\,\delta M_Z^{(6)}}{\partial \overline g_1} 
    &=
    -\overline{g}_2 \frac{\partial\,\delta M_Z^{(6)}}{\partial \overline g_2}
    =
    \frac{1}{\sqrt{2}\,G_F} \frac{M_W}{M_Z} \sqrt{1-\frac{M_W^2}{M_Z^2}} \left( 2 \frac{M_W^2}{M_Z^2} - 1 \right) \frac{C_{HWB}}{\Lambda^2}
    \,.
\end{split}
\end{align}
\end{subequations}
A parameter card implementing this scheme is available in the model database on \textsc{GitLab}~\href{https://gitlab.com/matchete/model-database/-/tree/master/UFO-models}{\faicon{gitlab}}~\cite{MatcheteDatabase}.

We compared our results for the $M_W$~input scheme against the findings in Appendix~A of Ref.~\cite{Dedes:2023zws}, finding full agreement at dimension six.\footnote{This is up to a typo in the expression for ${\bar{g}_{D6}^\prime} \equiv \bar{g}_1^{(4)} \bar{g}_1^{(6)}$ in~\cite{Dedes:2023zws} where $\Delta M \equiv \sqrt{M_Z^2-M_W^2}$ in the denominator must be replaced by $\Delta M^2$.}
However, we observe various differences at dimension eight.
For example, Ref.~\cite{Dedes:2023zws} does not consider the sum over neutrino flavors for the dimension-six squared contributions to the Fermi constant~$G_F$.
This not only affects the terms with sums over flavor indices~($p,r$) in Eq.~\eqref{eq:dGF8} but also various numerical prefactors of the other terms.
Also the real-part operation, $\mathrm{Re}(\,\cdot\,)$, should be applied to interference terms in~\cite{Dedes:2023zws}.
Ref.~\cite{Dedes:2023zws} uses a different operator basis, trading the operators {$Q_{H^6}^{(1)}= \big(H^\dagger H\big)^2 \big( D_\mu H^\dagger D^\mu H \big)$} and {$Q_{H^6}^{(2)}= \big(H^\dagger H\big) \big(H^\dagger \tau^I H\big) \big( D_\mu H^\dagger \tau^I D^\mu H \big)$} in favor of the operators {$Q_{H^6 \Box}=\big(H^\dagger H\big)^2 \Box \big(H^\dagger H\big)$} and {$Q_{H^6 D^2}=\big(H^\dagger H\big) \big(H^\dagger D_\mu H\big)^\ast \big(H^\dagger D^\mu H\big)$}.
These two operator sets are related by a field redefinition that modifies the Higgs vacuum expectation value~$v_T$ by the amount~{$\Delta v_T=-\frac{v_T^{5}}{4\Lambda^4} C_{H^6\Box}$}.
Considering this, we find agreement for the parts of the expressions for~{$\delta v^{(8)}_{T}$} and~{$\delta\overline{g}_2^{(8)}$} that are not due to the differences in~{$\delta G_F^{(8)}$}.
By contrast, there are further discrepancies in {$\delta\lambda^{(8)}$} and {$\delta\overline{g}_1^{(8)}$} 
compared with~\cite{Dedes:2023zws} that are not explained by the differences in the Fermi constant.\footnote{For $\delta \lambda^{(8)}$ there are apparent sign differences for the terms proportional to $C_{H^6\Box}$ and $\mathcal{B}_6 C_H$ and \cite{Dedes:2023zws} is apparently missing some of the dimension-six squared contributions from the $d=8$ Higgs-mass correction~$\delta M_h^{(8)}$.}
Since Ref.~\cite{Dedes:2023zws} does not provide a description of the derivation of their input shifts, we refrain from providing a more detailed comparison.

\subsection{How to Generate the \texorpdfstring{$d=8$}{d=8} UFO} 
\label{sec:d=8_UFO}

Our implementation of the $d=8$ SMEFT is based on the Murphy basis~\cite{Murphy:2020rsh}, but exhibits several differences compared to the operator tables in~\cite{Murphy:2020rsh}:
\begin{itemize}
    \item A factor of $i$ has been added to all operators in class $\boldsymbol{XH^4D^2}$ to render their Wilson coefficients real.
    \item The operators of class $\boldsymbol{\psi^2H^2D^3}$ are not Hermitian (as noted in~\cite{Hays:2018zze}). Therefore, the Hermitian conjugate is added.
    \item We define the two-sided derivative as $H^\dagger \overset{\leftrightarrow}{D}_\mu H = i H^\dagger (D_\mu H) - i (D_\mu H^\dagger) H$ with an extra factor of~$i$, making it Hermitian. As a consequence, the operators of class $\boldsymbol{\psi^2XH^2D}$ 
    containing a Hermitian derivative obtain an extra factor of~$i$ as in~\cite{Hays:2018zze} so that the Wilson coefficients are Hermitian. Furthermore, in all operators of class $\boldsymbol{\psi^2X^2D}$ that do not contain an explicit factor of~$i$ in~\cite{Murphy:2020rsh}, an extra factor of~$i$ is added through our two-sided derivative. Lastly, all operators of class $\boldsymbol{\psi^4D^2}$ with a pair of two-sided derivatives have an additional factor of~$(-1)$ w.r.t.~\cite{Murphy:2020rsh}.
    Operators with two-sided derivatives that were already defined with a factor of~$i$ in~\cite{Murphy:2020rsh} are not modified.
    \item A charge conjugation matrix has been added to the first Dirac chain of the operators $Q_{lqd^2H^2}$ and $Q_{eq^2dH^2}$ in~\cite{Murphy:2020rsh} to ensure Lorentz invariance. The corrected operators read
    \begin{equation}
        \begin{aligned}
            Q_{\ell qd^2H^2} &= \varepsilon^{abc}\varepsilon^{jk}\varepsilon^{mn}(\ell^\intercal_{jp} C q_{amr})(d^\intercal_{bs} C d_{ct})H_k H_n, \\
            Q_{eq^2dH^2} &= \varepsilon^{abc}\varepsilon^{jk}\varepsilon^{mn}(e^\intercal_p C d_{ar})(q^\intercal_{bjs}C q_{cmt})H_k H_n.
        \end{aligned}
    \end{equation}
    \item The operators $Q_{lq^2uHD}^{(1,2,3)}$ in~\cite{Murphy:2020rsh} are not gauge invariant under~$\mathrm{SU}(2)_L$ because they contain the contraction $q_j \varepsilon_{jn} {H^{n}}^\ast$ and analogs. This is corrected in our implementation by removing the anti-symmetric tensor from the contraction with $H^\ast$. The corrected operators read
    \begin{equation}
        \begin{aligned}
            Q_{\ell q^2uHD}^{(1)} &= i\varepsilon^{abc}\varepsilon^{km}D_\mu {H^{n}}^\ast (q^\intercal_{amp}C\gamma^\mu u_{br})(q^\intercal_{cns}C \ell_{kt}), \\
            Q_{\ell q^2uHD}^{(2)} &= i\varepsilon^{abc}\varepsilon^{jm}D_\mu {H^{n}}^\ast(q^\intercal_{amp}C\gamma^\mu u_{br})(q^\intercal_{cjs}C \ell_{nt}), \\
            Q_{\ell q^2uHD}^{(3)} &= i\varepsilon^{abc} \varepsilon^{km}H^{n\ast}(q^\intercal_{amp}C\gamma^\mu u_{br})(D_\mu q^\intercal_{cns}C \ell_{kt}).
        \end{aligned}
    \end{equation}
    \item The operator $Q_{lqudD^2}^{(2)}$ in~\cite{Murphy:2020rsh} vanishes due to the contraction of spinors with different chiralities. We replace this operator with
    \begin{equation}
        Q_{\ell qudD^2}^{(2)(\text{new})} = \varepsilon^{abc} \varepsilon^{ij} (\ell^\intercal_{ip} C \sigma^{\mu\nu} q_{ajr}) (D_\mu d^\intercal_{bs} C D_\nu u_{ct})
    \end{equation}
    from the basis in~\cite{Li:2020gnx}.
\end{itemize}

As already mentioned, the input scheme is implemented at the level of the parameter card. 
We introduce coupling variables corresponding to the parameter shifts~\eqref{eq:input-coupling-expansion} and truncate the Lagrangian consistently at~$d=8$. 
The input scheme relations~(\ref{eq:vT_shift}--\ref{eq:g1_shift}) are then implemented in the parameter card file in \textsc{Python} syntax. 
A~full parameter card file for SMEFT at~$d=8$, including the implementation of the input scheme, is provided in the \matchete model database on \textsc{GitLab}~\href{https://gitlab.com/matchete/model-database/-/tree/master/UFO-models}{\faicon{gitlab}}~\cite{MatcheteDatabase}.
Furthermore, we provide a \textsc{Mathematica} notebook for the automatic determination of the input scheme relations with \matchete and the translation to \textsc{Python} syntax in the ancillary material. 

We define the coupling order \texttt{QCD} with hierarchy~$1$ and the coupling order \texttt{QED} with hierarchy~$2$, such that two QCD insertions count as one QED insertion, following the choice in the default Standard Model UFO shipped with \textsc{MadGraph5}~\cite{Alwall:2011uj}. 
Similarly to other SMEFT implementations~\cite{Brivio:2020onw,Dedes:2023zws,Degrande:2020evl}, we define the coupling order \texttt{NP} for insertions of effective vertices, where \texttt{NP=1} indicates one insertion of a $d=6$ vertex. We choose the coupling order hierarchy as~$99$, as it is done in~\cite{Brivio:2017btx}.
With this definition, new physics vertices will be neglected by default, and the user can actively select the truncation order by specifying the maximum order of \texttt{NP} during event generation. 
In order to avoid negative \texttt{QED} orders due to powers of~$v_T$, we also assign one \texttt{QED} order to~$1/\Lambda$, i.e., $d=6(8)$ Wilson coefficients are assigned the order \texttt{QED=2}(\texttt{4}).
Moreover, we give the option of defining one coupling order for each $d=8$ operator class with the naming \texttt{NPdim6} and \texttt{NPdim8<i>}, where \texttt{<i>} is the index of the operator class in~\cite{Murphy:2020rsh}. Further coupling orders for finer control can be defined by the user as desired; note, however, that too many coupling orders (e.g. for every single $d=8$ Wilson coefficient) can increase the export time significantly. 

We provide two model files for the $d=8$ Lagrangian in the broken phase. 
The model \mmaInlineCell[]{Input}{"SMEFT_D8+breaking"} implements the symmetry-breaking relations for the SMEFT up to $d \leq 8$ in our modified version of the Murphy basis~\cite{Murphy:2020rsh} and can be used to derive the broken-phase Lagrangian from scratch, similar to the case of the $d=6$ Warsaw (\mmaInlineCell[]{Input}{"SMEFT_Warsaw+breaking"}) and Mainz (\mmaInlineCell[]{Input}{"SMEFT_Mainz+breaking"}) basis model files, using:
\begin{mmaCell}{Input}
  L8   \,=\,LoadModel["SMEFT_D8+breaking"];
  L8brk\,=\,ImplementVacuumConditions[ToBrokenPhase[L8, \mmaDef{GoldstoneBoson}\,->\,False]];
  L8gf \,=\,GaugeFixLagrangian[L8brk, Full\,->\,False];
\end{mmaCell}
The option \mmaInlineCell[]{Input}{\mmaDef{GoldstoneBoson}\,->\,False} ensures that Goldstone bosons from the Higgs doublet are omitted when going to the broken phase.
This is required in our $d=8$ implementation, as the large number of generated terms otherwise leads to excessive memory consumption and eventually a crash of the \textsc{Mathematica} kernel.
Thus, our full $d=8$ SMEFT implementation is restricted to the unitary gauge.\footnote{When generating UFO files including only small subsets of the $d=8$ operators in the SSB, other gauges can be used too.}
One cannot simply omit the definitions of the Goldstone bosons in the model file in the first place, since the information about the scalar degrees of freedom in the broken and unbroken phase provided through \mmaInlineCell[]{Input}{FieldDecomposition} is used by \mmaInlineCell[]{Input}{GaugeFixLagrangian}.
The latter function is given the option \mmaInlineCell[]{Input}{Full\,->\,False} since the input Lagrangian is already partially gauge fixed to unitary gauge, and only the gluon and photon have to be fixed to $R_\xi$ gauge in this step. 

The method \mmaInlineCell[]{Input}{ImplementVacuumConditions} uses various built-in \matchete routines for manipulating and simplifying Lagrangians, such as integration-by-parts identities.
Therefore, the broken-phase EFT operator structures are rearranged with respect to their unbroken-phase counterparts. 
Using the (semi-)automatic SSB implementation, one has no control over the exact EFT operator forms used.\footnote{The operator structures can even differ between different \matchete versions due to updated operator scoring in the simplification routines.}
Exact operator forms can be maintained only by implementing the complete broken-phase Lagrangian manually, which is an arduous task.
Despite omitting the Goldstone bosons from the broken phase, the expressions encountered in intermediate steps are enormous, and their manipulations require a substantial amount of compute and memory. 
On an Apple~M3~Pro chip with 18\,GB of memory, the code above runs for approximately 180~minutes and uses 25\,GB of memory at the peak (using swap).
For comparison, the same steps for the SSB in the $d=6$ SMEFT take only around 25\,seconds and 280\,MB of memory.

To allow users to bypass this computationally demanding step, we provide a second model file \mmaInlineCell[]{Input}{"SMEFT_D8_brokenPhase"} containing the broken-phase $d=8$ Lagrangian in the unitary gauge (including all required definitions) such that it can be loaded directly without computation.
A~subsequent call of \mmaInlineCell[]{Input}{ExportUFO[L8gf,\,\mmaDef{Legs}\,->\,6]} (the maximum number of external legs is restricted to be less than or equal to six in order to ignore the complicated seven or eight gauge boson vertices) takes another 170~minutes and uses around 40\,GB of memory at the peak.
This changes to 45~minutes and 7\,GB of memory when limiting to four or less external legs with \mmaInlineCell[]{Input}{\mmaDef{Legs}\,->\,4} and 40~seconds and 550\,MB for the $d=6$ SMEFT without restricting the number of external legs.
The resulting directory of UFO files is 165\,MB in size and hence too large to be used with common event generators on reasonable computing resources. 
Therefore, we provide the SMEFT $d=8$ UFO files with all operator classes but considering only vertices with up to four external legs, which can be used with \textsc{MadGraph} but still experience severe performance problems.
To remedy this shortcoming, we provide functionality to select smaller parts of interest in order to quickly generate smaller UFO files, which can be used for more efficient simulations.
The complexity of the model can be reduced by selecting single operators or classes of Wilson coefficients, by imposing flavor symmetries on the Lagrangian,\footnote{For a discussion of the limitations of flavor symmetries applied to the Murphy basis~\cite{Murphy:2020rsh} see Sec.~\ref{sec:flavor-symmetries}.} or by reducing the maximum number of external legs in the computation of the Feynman rules.
To simplify the export of the model, we provide as a starting point a documentation notebook in the model database on \textsc{GitLab}~\href{https://gitlab.com/matchete/model-database/-/tree/master/UFO-models}{\faicon{gitlab}}~\cite{MatcheteDatabase} and in the built-in documentation center, which explains the whole process in detail. 
The notebook contains lists of Wilson coefficients sorted by operator classes and mass dimension and examples of selecting or removing terms from the Lagrangian. Furthermore, it implements the input scheme shifts and defines the coupling orders. There are options for substituting different parameters for the fermion masses entering the Yukawa interactions and for the propagator masses. This makes it possible to define, for instance, a massless $b$-quark that still couples to the physical Higgs boson. Finally, the user can adjust the parameter card and choose the maximum number of vertex legs.

\subsection{Validation}
\label{sec:validation}

The generated SMEFT UFO models have been numerically validated against existing models following the proposal in~\cite{Durieux:2019lnv} with an existing dedicated \textsc{MadGraph5} plugin.\footnote{Available at \url{https://code.launchpad.net/~rwgtdim6/mg5amcnlo/plugin_eft_contrib} for the \textsc{MadGraph5} version available at \url{https://code.launchpad.net/~maddevelopers/mg5amcnlo/dim6_eft}.} 
The plugin evaluates squared matrix elements in both models for a variety of processes with different parameter choices. 
In each calculation, all Wilson coefficients but one are set to zero while the remaining one is set to the value $C^{(6)}/\Lambda^2 = 10^{\eminus 6}\,\mathrm{GeV}^{\eminus 2}$ or $C^{(8)}/\Lambda^4 = 10^{\eminus 12}\,\mathrm{GeV}^{\eminus 4}$. 
The comparison is performed for the squared SM contribution, the squared EFT contribution and the interference with the SM. 
The cross-check is considered successful if all relative differences are below $10^{\eminus 3}$. We apply the following choices:
\begin{itemize}
    \item the $\{G_F, M_W, M_Z, M_h\}$ input scheme is used to avoid SMEFT corrections to propagators;
    \item particle widths are set to zero to eliminate the influence of gauge choices;
    \item the EFT expansion in $1/\Lambda$ is truncated consistently at the amplitude level, i.e., at quadratic~($\Lambda^{\eminus 2}$) order for $d=6$ models and at quartic~($\Lambda^{\eminus 4}$) order for $d=8$ models;
    \item numerical Standard Model inputs are chosen as $\{G_F = 1.16637 \times 10^{\eminus 5}~\mathrm{GeV}^{\eminus 2},\ M_W = 80.3692~\mathrm{GeV},\ M_Z = 91.1880~\mathrm{GeV},\ M_h = 125.2~\mathrm{GeV},\ M_t = 172.56~\mathrm{GeV},\ \alpha_s = 0.118 \}$;
    \item the CKM matrix is set to unity for simplicity.
\end{itemize}
As various $d=6$ Feynman rules implementations have already been cross-checked in~\cite{Durieux:2019lnv}, it is sufficient to compare with only one of them. 
To that end, the $d=6$ \matchete SMEFT UFO has been thoroughly validated against \textsc{SMEFTsim~3.0}~\cite{Brivio:2020onw} covering a wide range of $2\rightarrow 2$ and $2 \rightarrow 3$ processes. 
The $d=8$ \matchete SMEFT UFO has been validated against a number of models that implement a subset of $d=8$ operators: 
All bosonic $d=8$ operators have been cross-checked with \textsc{SmeftFR~v3}~\cite{Dedes:2023zws}, which implements the Warsaw basis and all bosonic operators at $d=8$. 
Operators of the classes $\boldsymbol{\psi^2XH^2D}$, $\boldsymbol{\psi^2H^4D}$, and $\boldsymbol{\psi^2H^2D^3}$ have been cross-checked with the model from~\cite{Hays:2018zze}, which implements all operators relevant for $pp \rightarrow hW$. 
Furthermore, we have checked four-fermion interactions at $d=8$ against internal implementations in \textsc{FeynRules}~\cite{Alloul:2013bka}. 

We find that almost all relative deviations in the squared matrix elements between \matchete and the reference models remain well below~$10^{\eminus 3}$. There are a few exceptions with large relative differences, which are due to numerical artifacts. They occur when an exact cancellation of contributions produce a vanishing amplitude in one model, which might only be realized numerically in the other. We provide PDF files for the numerical validation in the ancillary material. The files contain lists of squared matrix elements for different parameters and different models. The first model is the reference model, while the second model is the one generated with \matchete.
As all reference models have been created with \textsc{FeynRules}, we further performed partial analytical checks of the output files. 
Lastly, the scalar leptoquark UFO model used in Sec.~\ref{sec:EFT-convergence} has been checked against the UFO provided in~\cite{Dorsner:2018ynv}.

\section{Application: EFT Validity for Leptoquark Searches}
\label{sec:EFT-convergence}

\newcommand{\slq}{\ensuremath{\tilde{S}_1} }
\definecolor{RWTHdark}{RGB}{0, 84, 159}
\definecolor{RWTHlight}{RGB}{144,186,223}
\definecolor{RWTHbordeaux}{RGB}{161,16,53}
\newcommand{\dark}[1]{\textcolor{RWTHdark}{#1}}
\newcommand{\light}[1]{\textcolor{RWTHlight}{#1}}
\newcommand{\madgraph}{MG5}
\newcommand{\Madgraph}{\textsc{MadGraph5\_aMC@NLO}}

This section provides an example of applying \matchete's new UFO interface in a realistic BSM analysis. 
We consider the scalar $\tilde{S}_1 \sim (\overline{\textbf{3}},\textbf{1})_{4/3}$ leptoquark\footnote{That is, $\tilde{S}_1$ is a color anti-triplet, an isospin singlet, and has hypercharge $ 4/3$.} and study the validity of the EFT description of its interactions at the LHC for Drell--Yan and related processes, assuming a coupling exclusively to third-generation fermions.

\begin{figure}[t]
    \centering
    \begin{minipage}{0.3\textwidth}
        \centering
        \includegraphics[]{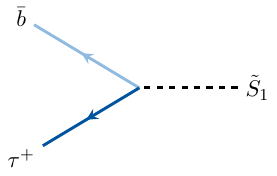}
        \subcaption{}
    \end{minipage}
    \hfill
    \begin{minipage}{0.3\textwidth}
        \centering
        \includegraphics[]{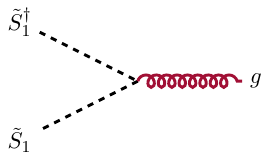}
        \subcaption{}
    \end{minipage}
    \hfill
    \begin{minipage}{0.3\textwidth}
        \centering
        \includegraphics[]{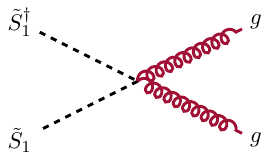}
        \subcaption{}
    \end{minipage}
    \caption{%
    Interaction vertices involving the \( \tilde{S}_1 \) leptoquark: 
    \textbf{(a)} Yukawa interaction $\tilde{S}_1 \bar{\tau} \bar{b}$; 
    \textbf{(b)} QCD three-point vertex \( \tilde{S}_1 \tilde{S}_1^\ast g \); 
    \textbf{(c)} QCD four-point vertex \( \tilde{S}_1 \tilde{S}_1^\ast gg \).}
    \label{fig:interactions}
\end{figure}

Leptoquarks~\cite{Buchmuller:1986zs,Dorsner:2016wpm}, coupling to both the SM leptons and quarks, are predicted in many BSM theories, such as Grand Unified Theories~\cite{Pati:1973uk,Pati:1974yy,Georgi:1974sy}, technicolor~\cite{Dimopoulos:1979es,Dimopoulos:1979sp}, and composite models~\cite{Schrempp:1984nj} and have recently received increased interest as a possible explanation~\cite{Buttazzo:2017ixm,Crivellin:2017zlb,Becirevic:2018afm,Angelescu:2018tyl,Angelescu:2021lln} of the $B$-anomalies~\cite{BaBar:2012obs,BaBar:2013mob,Belle:2015qfa,LHCb:2015gmp}.
In the following analysis, however, we focus exclusively on the LHC signatures of the leptoquark and how well they can be described by a low-energy EFT. 
To that end, we consider the production channels relevant to the ATLAS search~\cite{ATLAS:2023vxj} for the $\tilde{S}_1$ leptoquark decaying into the $b\tau$ final state at the LHC. 
The restriction to third-generation fermion couplings is motivated by the $B$-anomalies and flavor-physics constraints. Previous studies of the EFT validity for leptoquark searches have been undertaken, e.g., in~\cite{Allwicher:2022gkm,Allwicher:2024mzw,Boughezal:2022nof,Corbett:2024evt,Chang:2025ohh}.
The BSM Lagrangian of the $\tilde{S}_1$ leptoquark reads
\begin{equation}
    \mathcal{L}_\mathrm{BSM} 
    =
    \mathcal{L}_\mathrm{SM} 
    + (D_\mu \tilde{S}_1 )^\dagger(D^\mu \tilde{S}_{1}) 
    - M_{\tilde{S}_{1}}^2 \tilde{S}_{1}^\dagger \tilde{S}_{1}
    + \left[\kappa \, \tilde{S}_1^a (\overline{b}^{c}_{Ra} \tau_{R})+\text{H.c.}\right],
\label{eq:L}
\end{equation}
where $\mathcal{L}_\mathrm{SM}$ is the SM Lagrangian, $M_{\tilde{S}_1}$ the leptoquark mass, and $\kappa$ its flavor-specific coupling to third-generation quarks and leptons.\footnote{For simplicity, we consider here only baryon-number--conserving interactions, where we assign $L=-1$ and $B=-\frac{1}{3}$ to the $\tilde{S}_1$. In principle, the SM gauge symmetry allows for another interaction term of the form $\tilde{S}_1^\ast (\bar{u}_R^{cp} u_R^r)$ which violates baryon number and, hence, induces proton decay.} The superscript~`${}^c$' denotes the charge conjugate of a fermion, i.e., $\bar{b}^c = b^\intercal C$ with $C=i\gamma^2\gamma^0$.
For this analysis we consider only the gluon and fermion couplings of~$\tilde{S}_1$.
The corresponding interactions are displayed in Fig.~\ref{fig:interactions}, showing \textbf{(a)} a Yukawa fermion-flow-violating vertex; \textbf{(b)} a single gluon interaction; and \textbf{(c)} a double gluon interaction.

In the following, we consider all channels for the production of the \smash{$\tilde{S}_1$}~leptoquark in association with $b$-tagged jets studied by the ATLAS collaboration in~\cite{ATLAS:2023vxj} and investigate how well the EFT description works for the various processes. The $b$~tagging is employed to isolate the third-generation couplings and to suppress the background~\cite{Haisch:2022lkt}.
While the ATLAS search excludes scalar leptoquark masses below \(1.28\)\,TeV for \(\kappa=1.0\) and below \(1.53\)\,TeV for \(\kappa=2.5\), we nevertheless include lower masses as illustrative benchmark points to study the breakdown of the EFT expansion.

The processes underlying the considered channels are
\begin{align*}
	1. & \ pp \to \tau^+ \tau^- \,,
    & 
    2. & \ pp \to \tau^+ \tau^- \, b \ (+\,\mathrm{c.c.}) \,,
    & 
    3. & \ pp \to \tau^+ \tau^- \, b \, \bar{b} \,,
\end{align*}
which will be discussed in turn in the following sections. Throughout our analyses, we truncate at the amplitude level; that is, we retain 
only terms up to a fixed order in $1/M_{\tilde{S}_1}$ in the 
amplitude and do not consider truncations at the level of the 
cross section or individual interference terms as motivated for Drell--Yan studies in Ref.~\cite{Allwicher:2024mzw}. We adopt the five-flavor scheme for the PDFs and treat the bottom quark as massless.
To isolate the convergence of the EFT expansion, we consider only the pure leptoquark contributions and omit the SM amplitudes and their interference with the leptoquark amplitudes throughout this section. Accordingly, the cross sections denoted by \(\sigma_{\mathrm{BSM}}\) and \(\sigma_{\mathrm{EFT}}\) below are signal-only quantities and should not be interpreted as predictions for the complete physical final states. 
The SM interference can be relevant for non-resonant leptoquark production, in particular outside the high-\(p_{T}^{b}\) signal region used in the model-dependent ATLAS interpretation~\cite{ATLAS:2023vxj}. As a cross-check, we performed the simulations for the full $\tilde{S}_1$ model also with the UFO model provided by Ref.~\cite{Dorsner:2018ynv} (see also~\cite{Crivellin:2021ejk}), finding full agreement with our results.

In the following, we focus on the details of the simulations and on the phenomenological discussion.
The implementation of the $\tilde{S}_1$ model and its EFT in \matchete is presented in Appendix~\ref{app:examples} together with the determination of the Feynman rules and the extraction of the UFO files for both scenarios.
Further details are also given in an interactive tutorial notebook available both in the built-in \textsc{Mathematica} documentation center, included with every \matchete installation, and on the website~\cite{MatcheteWebsite}.

\subsection{Drell--Yan: \texorpdfstring{$p p \to \tau ^+\tau ^-$}{pp -> tau+ tau-}}

\begin{figure}[t]
    \centering
    \includegraphics[width=0.8\textwidth]{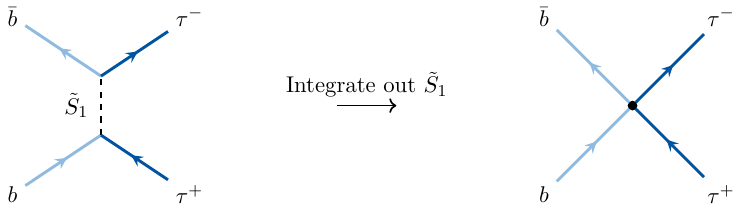}
    \caption{Tree-level exchange of $\slq$ in $b\bar b \to \tau^+\tau^-$ (left) and the corresponding EFT contact interaction (right).}
    \label{fig:4fermionMatching}
\end{figure}

Ignoring SM electroweak processes, only a single Feynman diagram---the $t$-channel leptoquark exchange---contributes to the Drell--Yan process ($p p \to \tau^+ \tau^-$) in the $\tilde{S}_1$~model at leading order.
The diagram and the corresponding four-fermion vertex in the EFT are shown in Fig.~\ref{fig:4fermionMatching}.
As already mentioned, the $\tilde{S}_1$-fermion vertex violates the usual fermion-number symmetry due to the presence of charge-conjugated fermions; however, the entire process conserves it.\footnote{Alternatively, one can also assign fermion number $F=-2$ to the leptoquark~\cite{Dorsner:2016wpm} in order to have fermion number conserved at every vertex. Nevertheless, the vertex still violates the fermion flow.}

\subsubsection{EFT Matching and Convergence}
\label{sec:Drell--Yan-matching}
Assuming a real coupling~$\kappa$ for simplicity, the amplitude for the $t$-channel leptoquark exchange, shown on the left-hand side of Fig.~\ref{fig:4fermionMatching}, reads 
\begin{align}
    i\mathcal{M}_{\sscript{BSM}}
    &= -i\kappa^2  \big[\bar{u}_\tau(p_4) P_L C \bar{v}_b^{c_2}(p_2)^\intercal\big]
    \frac{\delta_{c_{2}c_{1}}}{t - M_{\tilde S_1}^2}
    \big[u_b^{c_1}(p_1)^\intercal C P_R v_\tau(p_3)\big]
    \,.
    \label{eq:MBSMC}
\end{align}
Here, $C$~denotes the charge-conjugation matrix and $P_{R/L} = \frac{1}{2}(1\pm\gamma_5)$ are the chiral projection operators. 
The Mandelstam variable $t = (p_1 - p_3)^2 = (p_2 - p_4)^2$ corresponds to the momentum exchanged by the scalar leptoquark~$\slq$. 
Summing (averaging) over final (initial) state spins and colors, we find the averaged squared amplitude of the BSM theory
\begin{align}
    \begingroup
    \setlength{\fboxsep}{8pt}
    \fcolorbox{black}{white}{$\displaystyle
    \overline{|\mathcal{M}_{\sscript{BSM}}|^2}
    = \frac{\kappa^4}{12}\,
    \frac{t^2}{(t - M_{\tilde S_1}^2)^2}
    $}
    \endgroup
    \,.
    \label{eq:BSM-4}
\end{align}
The relevant operators in the EFT are a $\boldsymbol{\psi^4}$ operator at dimension six and a $\boldsymbol{\psi^4 D^2}$ operator at dimension eight, sharing the same four-fermion current structure. 
Integrating out the~$\slq$ and applying a Fierz transformation (see Appendix~\ref{app:examples-LQEFT}) generates 
\begin{equation}
    \mathcal{L}_{\sscript{EFT},\, d {=} 6} = 
    \frac{\kappa^2}{2\,M_{\slq}^2}\,
    (\bar{b}\,\gamma_\mu P_R\, b)(\bar{\tau}\,\gamma^\mu P_R\, \tau)\,,
\end{equation}
at dimension six. The dimension-eight contribution is given by 
\begin{equation}
    \mathcal{L}_{\sscript{EFT},\, d {=} 8} = -
\frac{\kappa^2}{M_{\slq}^2}\,
(\bar{\tau}_R\, b^c_R)  \frac{D^2}{M_{\slq}^2}(\bar{b}^c_R\, \tau_R)\,,
\end{equation}
showing, for compactness, the expression before Fierzing.
The dimension-six and -eight operators contribute to the same Feynman diagram, shown on the right-hand side of Fig.~\ref{fig:4fermionMatching}, and their Feynman rules are provided in Appendix~\ref{app:examples-LQEFT}. 
Their kinematic structure differs due to the additional derivatives at~$d=8$.
The EFT amplitude up to dimension eight is then given by
\begin{align}
    i\mathcal{M}_{\sscript{EFT},\, d\leq8}
    &= -\frac{i\kappa^{2}}{2  M_{\slq}^{2}}
    \delta_{c_{2}c_{1}}
    \left( 1 + \frac{t}{M_{\slq}^2}\right)
    \big[\bar{u}_\tau(p_4) \gamma _\mu P_R v_\tau(p_3) \big]
    \big[\bar{v}_b^{c_2}(p_2) \gamma ^\mu P_R u_b^{c_1}(p_1) \big] \,,
\end{align}
where the term proportional to $ t $ captures the entire dimension-eight contribution.

The averaged squared EFT amplitude is then given by
\begin{align}
    \begingroup
    \setlength{\fboxsep}{8pt}
    \fcolorbox{RWTHbordeaux}{white}{$\displaystyle
    \overline{|\mathcal{M}_{\sscript{EFT},\,d\leq 8}|^2}
    = \frac{\kappa^4}{12}\,
    \frac{(M_{\slq}^2 + t)^2}{M_{\slq}^8}\,
    t^2
    $}
    \endgroup
    \ =\
    \underbrace{%
    \begingroup
    \setlength{\fboxsep}{8pt}
    \fcolorbox{RWTHdark}{white}{$\displaystyle
    \frac{\kappa^4}{12}\,
    \frac{t^2}{M_{\slq}^4}
    $}
    \endgroup}_{=\overline{|\mathcal{M}_{\sscript{EFT},\,d=6}|^2}}
    \left[
        1
        + 2 \frac{t}{M_{\slq}^2}
        + \frac{t^2}{M_{\slq}^4}
    \right] 
    \,.
    \label{eq:EFT8-4}
\end{align}
On the right-hand side, the first term is the pure dimension-six squared amplitude, the second term is the interference of dimension six and eight, and the third term is the pure $d=8$ squared contribution.
Taylor expanding the full-model squared amplitude in Eq.~\eqref{eq:BSM-4} in $\big|t\big/M_{\slq}^2\big| \ll 1$, we find full agreement with the EFT amplitude in Eq.~\eqref{eq:EFT8-4} up to terms of order~$\cO(M_{\slq}^{\eminus8})$, which correspond to the missing interference of the $d=6$ and $d=10$ EFT amplitudes. 

In the following phenomenological analysis, we quantify the degree to which the EFT can reproduce the full BSM theory results by studying the relative deviation of the EFT cross sections from the full model result:
\begin{equation}
    R_i \equiv
    \frac{\sigma_{\sscript{BSM}} - \sigma_i}{\sigma_{\sscript{BSM}}}
    \qquad \text{for} \qquad
    i \in \big\{ \mathrm{EFT}_{d=6},\ \mathrm{EFT}_{d\leq8} \big\}.
    \label{eq:4fermiondeviation}
\end{equation}
From the amplitude-level comparison above, it is apparent that both the EFT and full-model squared amplitudes (and, hence, also the cross sections) scale as~$\kappa^4$. This renders $R_i$ independent of~$\kappa$, and the ratios will depend exclusively on~$M_{\slq}$.

\subsubsection{Total Cross Section}
We start by investigating the di-tau Drell--Yan tails without associated $b$-tagged jets. We employ a simplified parton-level setup inspired by the ATLAS search in Ref.~\cite{ATLAS:2023vxj} without attempting a recast of the experimental analysis.
The relevant \textsc{MadGraph} settings are summarized in Tab.~\ref{tab:settings-madgraph}.
\begin{table}[t]
    \centering
    \begin{tabular}{@{}ll@{}}
        \multicolumn{2}{@{}l}{\textbf{General (all processes)}} \\
        \addlinespace[2pt]
        Center-of-mass energy                   &  $\sqrt{s} = 13\,\mathrm{TeV}$ \\
        PDF set                  & NNPDF3.0 NLO (\texttt{lhaid 261000}) \\
        Renormalization/factorization\ scale        & dynamical, partonic center-of-mass energy $\sqrt{\hat{s}}$ \\
        Random seed              & $30$ (fixed) \\
        Five-flavor scheme       & \texttt{maxjetflavor} $=5$ \\
        Integration strategy     & \texttt{sde\_strategy} $=1$ \\
        \midrule
        \multicolumn{2}{@{}l}{\textbf{$pp \to \tau^+\tau^-$}} \\
        \addlinespace[2pt]
        Lepton transverse momentum & $p_{T,\ell} > 10\,\mathrm{GeV}$ \\
        Lepton pseudorapidity            & $|\eta_\ell| < 2.5$ \\
        \midrule
        \multicolumn{2}{@{}l}{\textbf{$pp \to \tau^+\tau^-\,b$ \ and \ $pp \to \tau^+\tau^-\,b\bar{b}$}} \\
        \addlinespace[2pt]
        Lepton transverse momentum & $p_{T,\ell} > 10\,\mathrm{GeV}$ \\
        Lepton pseudorapidity            & $|\eta_\ell| < 2.5$ \\
        $b$-jet transverse momentum & $p_{T,b} > 20\,\mathrm{GeV}$ \\
        $b$-jet pseudorapidity            & $|\eta_b| < 5.0$ 
    \end{tabular}
    \caption{Summary of the \textsc{MadGraph} settings and cuts used for the simulations presented in this section. The general settings apply to all processes, while the cuts are listed separately for the pure $pp \to \tau^+\tau^-$ process and for the channels with one or two final-state $b$~jets. We use \textsc{MadGraph5\_aMC@NLO}~3.6.3~\cite{Alwall:2011uj,Alwall:2014hca} with parton distribution functions taken from \textsc{NNPDF3.0}~\cite{NNPDF:2014otw}, accessed via \textsc{LHAPDF6}~\cite{Buckley:2014ana}.}
    \label{tab:settings-madgraph}
\end{table}

\begin{figure}[tb]
    \centering
    \includegraphics[width=0.9\textwidth]{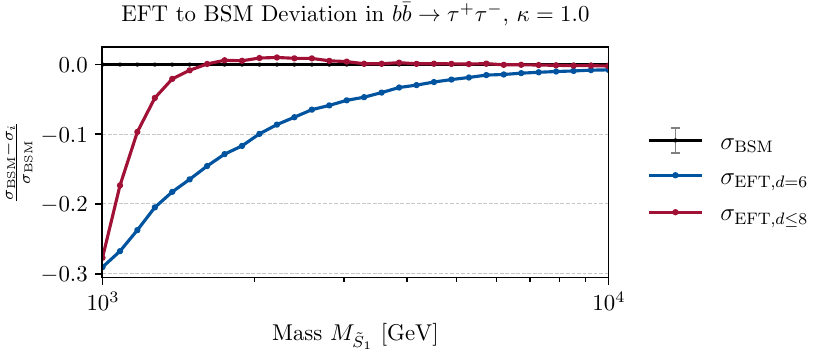}
    \caption{Relative deviation of the total $b\bar{b}\to\tau^+\tau^-$ cross section of the EFT predictions from the full BSM theory as a function of the leptoquark mass.
    The dimension-six contribution is shown in blue, whereas the full $d \leq 8$ result is given in red. 
    The ratios are independent of the leptoquark coupling $\kappa$.}
    \label{fig:eft-ratio-tt}
\end{figure}

The total cross section for the Drell--Yan process $b\bar{b}\to\tau^+\tau^-$ is shown in Fig.~\ref{fig:eft-ratio-tt}. 
Plotted are the relative deviations~$R_i$, as defined in Eq.~\eqref{eq:4fermiondeviation}, of the EFTs compared to the full BSM theory result as a function of the leptoquark mass~$M_{\slq}$.
The blue line corresponds to a truncation of the amplitude at $d=6$, whereas the red curve gives the result for a truncation at $d \leq 8$. All simulations use a fixed \textsc{MadGraph} integration seed such that the results are reproducible. As renormalization and factorization scale we have chosen (here and below) the partonic center-of-mass energy which is well defined in both the EFT and BSM theory.\footnote{The default scale-choice scheme in \textsc{MadGraph} depends on the topologies of the contributing diagrams and can choose different renormalization and factorization scales between EFT (contact vertex) and BSM theory (mediator), leading to an offset between EFT and BSM cross section.}
The simulation was performed setting $\kappa=1.0$, but as noted above, the ratio is independent of the value of $\kappa$.

At low leptoquark masses, the EFT (both $d=6$ and $d\leq 8$) deviates significantly from the full model: the expansion parameter $E^2/M_{\slq}^2$, with $E$ the typical energy scale of the process, becomes large and the EFT approach breaks down. For example, at masses of around 1\,TeV the deviation of the EFT results is about 30\,\%.
The $d=6$ EFT converges more slowly toward the full model, reaching a 10\,\% (5\,\%) deviation at $M_{\slq} \sim 2\,\mathrm{TeV}$ ($\sim 3\,\mathrm{TeV}$), whereas the $d\leq 8$ cross section converges rapidly, reaching 1\,\% deviation already at $M_{\slq} \approx 1.5\,\mathrm{TeV}$, in agreement with Ref.~\cite{Allwicher:2024mzw}. At $M_{\slq} = 1\,\mathrm{TeV}$ the $d\leq 8$ truncation still slightly outperforms $d=6$, but for lower masses the expansion parameter $E^2/M_{\slq}^2$ grows too large and $d\leq 8$ performs worse than $d=6$.

\subsubsection{Differential Cross Section for Invariant Mass $m_{\tau\tau}$}

\begin{figure}[tb]
    \centering
    \includegraphics[width=0.49\linewidth]{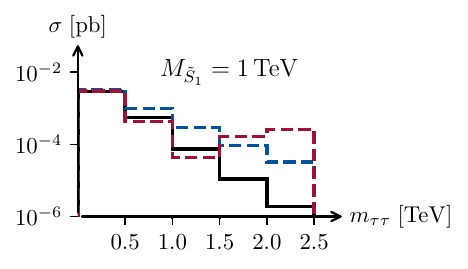}
    \includegraphics[width=0.49\linewidth]{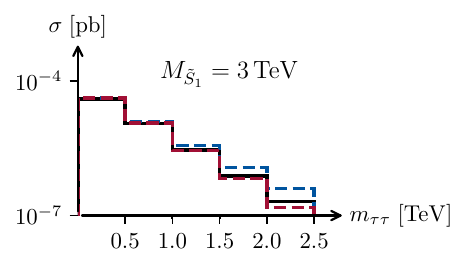}
    \\
    \includegraphics[width=0.49\linewidth]{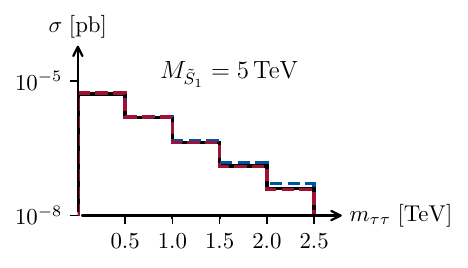}
    \includegraphics[width=0.49\linewidth]{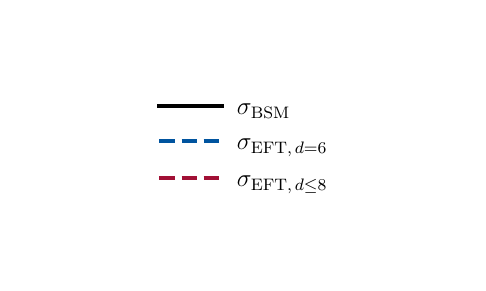}
    \caption{Simulated $\tau^+\tau^-$ cross sections as a function of the invariant mass~$m_{\tau\tau}$, shown for fixed leptoquark masses~$M_{\slq}$. For each mass, the cross section for BSM, EFT at order six, and EFT at order eight are plotted against the $m_{\tau\tau}$ interval.
    }
    \label{fig:diff_cross_section}
\end{figure}

When studying the EFT validity, it is imperative to investigate differential distributions, since the EFT contributions are often (as is the present case) energy enhanced and therefore provide larger contributions to the high-energy kinematic regimes.
Figure~\ref{fig:diff_cross_section} shows differential cross sections $d\sigma/dm_{\tau\tau}$ for $pp\to\tau^+\tau^-$.
The differential cross sections are provided for three benchmark leptoquark masses $M_{\slq}\in\{1,\,3,\,5\}\,\mathrm{TeV}$.
For each mass, the differential cross section is plotted in bins of the di-tau invariant mass~$m_{\tau\tau}$.
The full BSM results are shown in black, whereas the $d=6$ ($d \leq 8$) EFT cross section is displayed in blue (red).

As expected, the deviations become large for higher energies, i.e., larger values of $m_{\tau\tau}$, and for lower leptoquark masses.
The largest differences in the plots are found for $M_{\slq}=1\,\mathrm{TeV}$ and $m_{\tau\tau}\in[2.0,\,2.5]\,\mathrm{TeV}$.
In this bin (and the one below), the $d \leq 8$ results show an even larger difference with respect to the full model than the $d=6$ results. 
This is not surprising, since we have $m_{\tau\tau} > M_{\slq}$ in this bin leading to energy-enhanced $d=8$ contributions.
Conversely, the heavier the leptoquark and the lower $m_{\tau\tau}$, the better the agreement between the results. 
Moreover, we observe that the $d \leq 8$ description 
exhibits a broader range of validity for $m_{\tau\tau}<M_{\slq}$. 
For $M_{\slq}=3$ and $5\,\mathrm{TeV}$, the $d\leq 8$ prediction is essentially indistinguishable from the full model across the entire $m_{\tau\tau}$ range, whereas the $d=6$ result retains a slight excess in the highest bins.
These results highlight that the EFT may struggle to accurately describe high-energy kinematic regions. 
In fact, the EFT validity in these regions heavily depends on the masses assumed for the BSM states.

\subsection{Resonant Production: \texorpdfstring{$p p \to \tau ^+ \tau^- \ b$}{pp -> tau+ tau- b}}
\label{sec:sec_btt}

\begin{figure}[t]
    \centering
    \begin{subfigure}{\textwidth}
        \centering
        \includegraphics[]{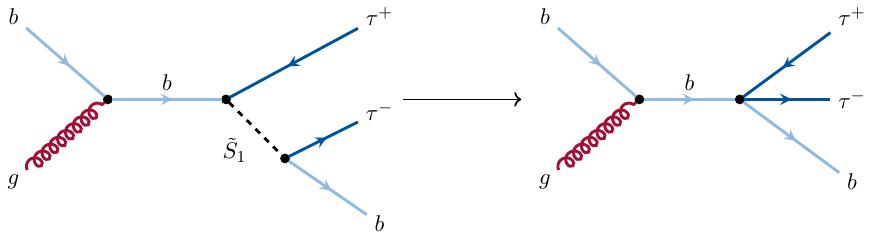}
        \caption{Diagram 1: Resonant leptoquark production.}
        \label{fig:dia1}
    \end{subfigure}
    \\[0.2cm]
    \begin{subfigure}{\textwidth}
        \centering
        \includegraphics[]{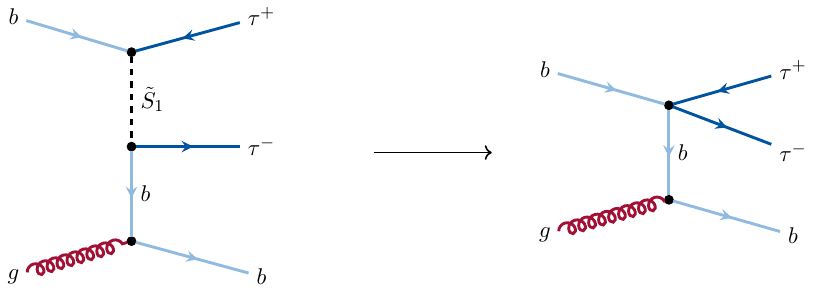}
        \caption{Diagram 2: Drell--Yan associated production.}
        \label{fig:dia2}
    \end{subfigure}
    \\[0.2cm]
    \begin{subfigure}{\textwidth}
        \centering
        \includegraphics[]{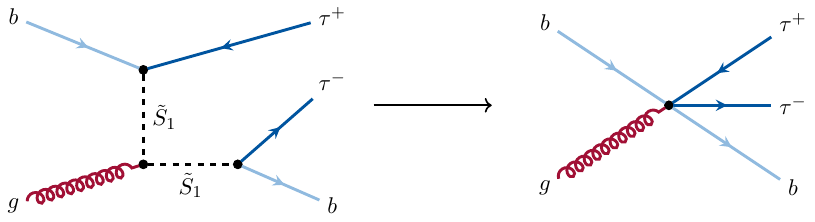}
        \caption{Diagram 3: Two-leptoquark exchange.}
        \label{fig:dia3}
    \end{subfigure}
    \caption{The three Feynman diagrams contributing to the process $b g \to \tau^+ \tau^- b$ in the full BSM theory (left) and the corresponding channels in the EFT (right).}
    \label{fig:gb-tautaub-diags}
\end{figure}

The second process we consider is the di-tau final state with an associated $b$~quark.
The underlying partonic scattering at the LHC is given by $bg \to \tau^+ \tau^-\,b$ and the three contributing topologies, shown in Fig.~\ref{fig:gb-tautaub-diags}, are
\textbf{(a)}~resonant leptoquark production;
\textbf{(b)}~$b$-associated Drell--Yan production;
\textbf{(c)}~two-leptoquark exchange.
On the left-hand side of Fig.~\ref{fig:gb-tautaub-diags}, the channels in the full BSM model are displayed, whereas the corresponding EFT vertices are given on the right.
Due to $CP$~invariance our results below also apply to the charge-conjugated process $\bar{b}g \to \tau^+\tau^-\,\bar{b}$.
Recall that our analysis is performed in the five-flavor scheme. While the EFT counterparts of topologies~\textbf{(a)} and~\textbf{(b)} begin at dimension-six, the topology~\textbf{(c)} first contributes at dimension eight. 
In the full theory, the leptoquark in~\textbf{(a)} can be produced resonantly, while in~\textbf{(b)} it is exchanged in the $t$-channel.
In~\textbf{(c)} one of the leptoquarks can be resonant, whereas the other is in the $t$-channel.

\subsubsection{The Total Cross Section of Each Subprocess}

The relative deviations of the EFT from the BSM cross sections [defined in Eq.~\eqref{eq:4fermiondeviation}] are shown in Figs.~\ref{fig:btt-convergence-1}--\ref{fig:btt-convergence-3} for each of the three channels described above, as a function of the leptoquark mass. 
The results are obtained with a benchmark value of the leptoquark coupling $\kappa=1$. As we shall see, the presence of a resonant leptoquark exchange in the full theory introduces a coupling dependence in the ratios.
The three channels exhibit very different characteristics: 

\begin{figure*}[tb]
    \centering
    \begin{subfigure}{0.47\textwidth}
        \centering
        \includegraphics[width=\textwidth]{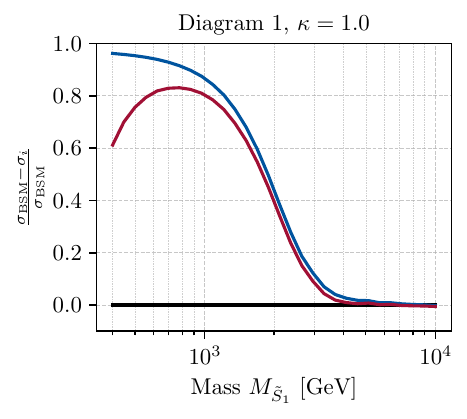}
        \caption{Diagram 1}
        \label{fig:btt-convergence-1}
    \end{subfigure}
    \begin{subfigure}{0.47\textwidth}
        \centering
        \includegraphics[width=\textwidth]{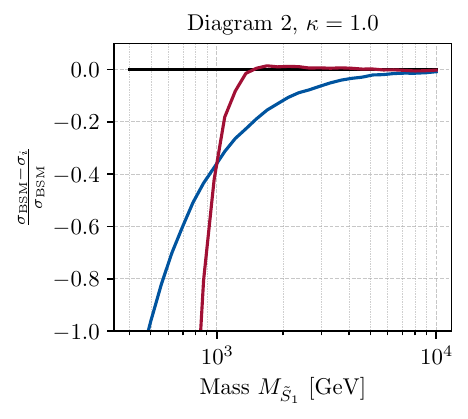}
        \caption{Diagram 2}
        \label{fig:btt-convergence-2}
    \end{subfigure}
    \\[-0.5cm]
    \begin{subfigure}{0.47\textwidth}
        \centering
        \includegraphics[width=\textwidth]{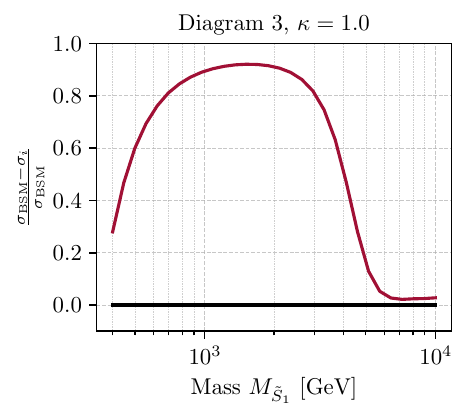}
        \caption{Diagram 3}
        \label{fig:btt-convergence-3}
    \end{subfigure}
    \begin{subfigure}{0.47\textwidth}
        \centering
        \includegraphics[width=\textwidth]{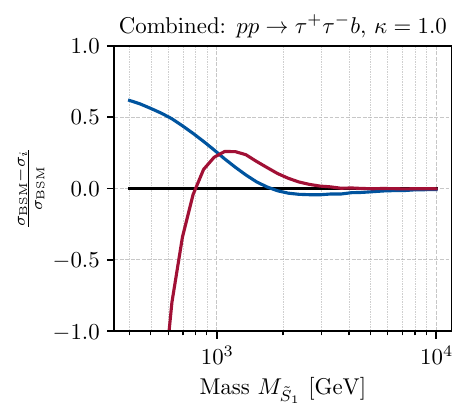}
        \caption{Combined (incl. interference)}
        \label{fig:btt-convergence-tot}
    \end{subfigure}
    \caption{%
    Subfigures \textbf{(a)}--\textbf{(c)} show the relative deviation of the EFT from the BSM cross section individually for the three subprocesses, shown in Fig.~\ref{fig:gb-tautaub-diags}, contributing to the partonic process $b g \to \tau^+ \tau^- b$.  \textbf{(d)}~displays the combined cross section, including interference between the different channels. The blue (red) lines show the results for the $d=6$ ($d \leq 8$) EFT cross section.}
\end{figure*}

\begin{itemize}
    \item[\textbf{(a)}]
    The resonant leptoquark production (Fig.~\ref{fig:btt-convergence-1}) exhibits a large deviation at low $M_{\slq}$. It is only for $M_{\slq} \gtrsim 3\,\mathrm{TeV}$ that the difference in cross section drops below 10\,\% and converges toward the full model at higher masses. Including dimension-eight contributions in the EFT does not improve convergence noticeably. This is because the leptoquark can be produced on-shell at the LHC at low masses, which causes the EFT series to break down: no finite truncation, whatever the order, can approximate the (close to) on-shell propagator.
    \item[\textbf{(b)}]
    The $b$-associated Drell--Yan production in Fig.~\ref{fig:btt-convergence-2} shows a pattern of convergence similar to that of the pure Drell--Yan scenario shown in Fig.~\ref{fig:eft-ratio-tt}.
    While both $d=6$ and $d \leq 8$ EFT descriptions converge well toward the full model with increasing~$M_{\slq}$, the inclusion of higher-dimensional contributions significantly improves the EFT validity for intermediate leptoquark masses.
    \item[\textbf{(c)}]
    The two-leptoquark-exchange contribution shown in Fig.~\ref{fig:btt-convergence-3} arises first at~$d= 8$ in the EFT.
    We find significant differences between the EFT and the full-model description for this channel, with the difference in cross sections falling below 10\,\% only for masses in excess of $ 5\,\mathrm{TeV}$.
    While one of the leptoquarks can be produced resonantly, the $g\slq\slq^\ast$~vertex is also energy enhanced due to a derivative in the corresponding Feynman rule, explaining the difficulty of approximating this topology within the EFT. 
    At the upper end of the mass range, the deviation reaches a minimum of $2\,\%$ at $M_{\slq}=7\,\mathrm{TeV}$ before increasing slightly again. Given the very small absolute cross sections in this region, this mild non-monotonic behavior is likely caused by residual numerical integration uncertainties. 
\end{itemize}
Overall, we find very different EFT-convergence patterns for the three topologies. The combined cross section, including interference, is shown in Fig.~\ref{fig:btt-convergence-tot}: it converges relatively fast, since the Drell--Yan topology~\textbf{(b)} dominates. Similar to the case of the $ pp\to \tau^+\tau^-$ process, the $d\leq 8$ description is worse than $d=6$ at very low~$M_{\slq}$, where the expansion parameter becomes large ($E/M_{\slq}>1$) and the EFT series no longer converges. 
As expected, when increasing~$M_{\slq}$ the $d \leq 8$ result converges faster to the full model.

\subsubsection{Dependence on \texorpdfstring{$b$}{b}-Jet Transverse Momentum~\texorpdfstring{$p_T^b$}{pTb}}
The convergence of the EFT toward the full model also depends on the $p_{T}$ of the $b$~quark in the final state. The $p_{T}^b$~cut controls the balance between resonant production \textbf{(a)}, which yields a hard $b$ from the on-shell $\slq$ decay, and associated Drell--Yan production~\textbf{(b)}, whose $b$ stems from initial-state gluon splitting and is predominantly soft and collinear to the beam. The two-leptoquark-exchange topology~\textbf{(c)} is suppressed at \smash{$\cO(M_{\slq}^{-8})$} and remains subleading. 
Similar to the analysis outlined by ATLAS~\cite{ATLAS:2023vxj}, we compare the cross sections with a low $p_{T}^{b}$~cut of $p_{T}^{b}>20$\,GeV for the associated $b$~jet to the high-$p_{T}^{b}$ region of the cross section defined by $p_{T}^{b}>200$\,GeV. 

\begin{figure}[tbp]
    \centering
    \includegraphics[width=0.95\textwidth]{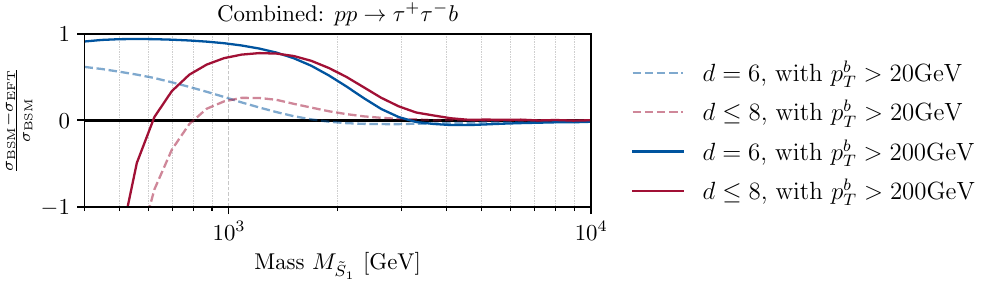}
    \caption{Relative deviation of the EFT from the BSM cross section for $pp\to\tau^+\tau^-b$, with a low cut on the $b$-jet~$p_T$ of $p_{T}^{b}>20\,\mathrm{GeV}$ (dotted lines) and for a high cut of $p_{T}^{b}>200\,\mathrm{GeV}$ (solid lines), at $d=6$ (blue) and $d\leq8$ (red).}
    \label{fig:pt-cut}
\end{figure}

In Fig.~\ref{fig:pt-cut} we see that the range of validity of the EFT description is very different for the two signal regions. While the low-$p_T^b$ region (dotted lines) is dominated by the Drell--Yan process~\textbf{(b)}, requiring $p_{T}^{b}>200$\,GeV (solid lines) leads to a dominant resonant contribution~\textbf{(a)}, worsening the EFT convergence. The deviation at $d=6$ increases from about $60\,\%$ to nearly $90\,\%$ at low~\smash{$M_{\slq}$}, and its rate of convergence is significantly reduced in the high-$p_T^b$ region, where the EFT provides an accurate approximation only at higher masses. 
A similar pattern arises at $d \leq 8$.
Notably, the inclusion of dimension-eight terms does not significantly enhance the overall rate of convergence and improves the full theory reconstruction only at very high~\smash{$M_{\slq}$}.
As expected, the contact interaction obtained by integrating out the~$\slq$ cannot reproduce its propagator shape close to the pole. This highlights the limited applicability of EFTs for studying resonant NP signals, where the EFT validity is limited to the very-high mass region; even the inclusion of higher-dimensional operators does not provide a sufficient improvement.
By contrast, it also illustrates that the low-$p_T^b$ region might indeed be used for reliable EFT analyses due to the better convergence of the EFT series there.

\subsubsection{Coupling Dependence of the Cross Section}
\label{sec:-tautaub-lambda-dependence}

\begin{figure}[tbp]
    \centering
    \vspace{-1.5cm}
    \begin{subfigure}{0.49\textwidth}
        \centering
        \includegraphics[width=\textwidth]{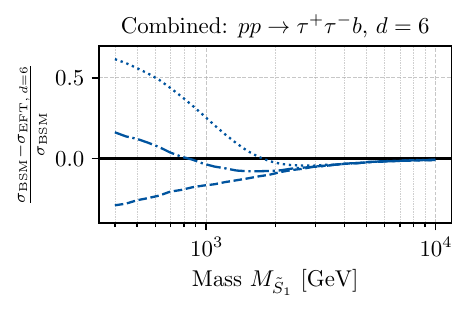}
        \caption{Total $pp \to \tau^+ \tau^- \ b$ cross section at $d=6$.}
        \label{fig:btt-lambda-6}
    \end{subfigure}
   \begin{subfigure}{0.49\textwidth}
        \centering
        \includegraphics[width=\textwidth]{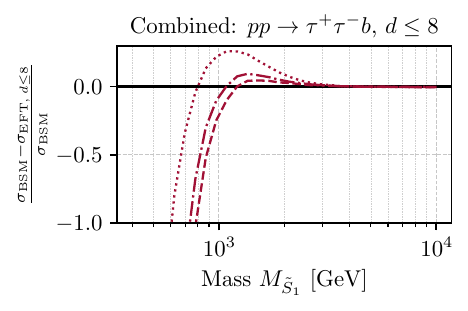}
        \caption{Total $pp \to \tau^+ \tau^- \ b$ cross section at $d \leq 8$.}
        \label{fig:btt-lambda-8}
    \end{subfigure}
    \\
    \begin{subfigure}{0.49\textwidth}
        \centering
        \includegraphics[width=\textwidth]{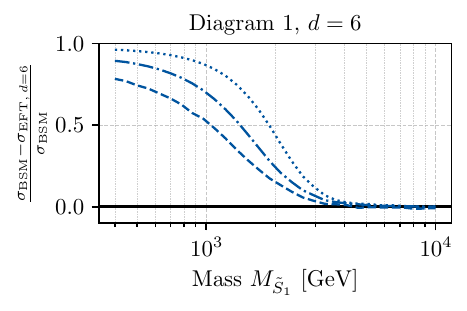}
        \caption{Resonant leptoquark channel (Fig.~\ref{fig:dia1}) at $d=6$.}
        \label{fig:btt-lambda-a-6}
    \end{subfigure}
   \begin{subfigure}{0.49\textwidth}
        \centering
        \includegraphics[width=\textwidth]{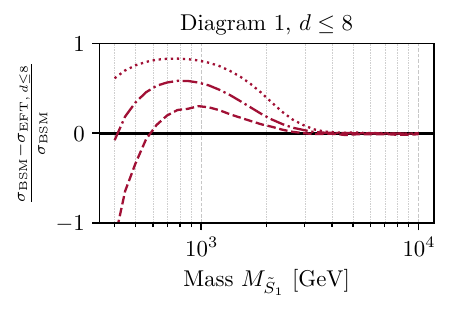}
        \caption{Resonant leptoquark channel (Fig.~\ref{fig:dia1}) at $d \leq 8$.}
        \label{fig:btt-lambda-a-8}
    \end{subfigure}
    \\
    \begin{subfigure}{0.49\textwidth}
        \centering
        \includegraphics[width=0.9\textwidth]{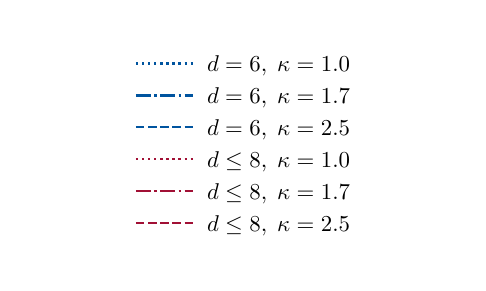}
        \\[1cm]
        \caption{Legend.}
        \label{fig:legend}
    \end{subfigure}
    \begin{subfigure}{0.49\textwidth}
        \centering
        \includegraphics[width=\textwidth]{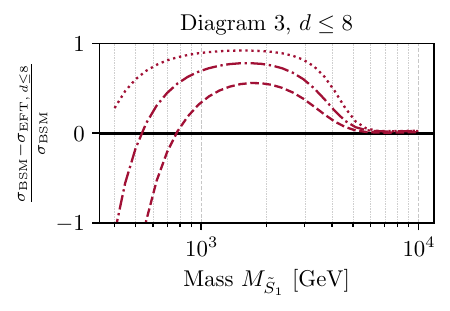}
        \caption{Two-leptoquark-exchange 
        (Fig.~\ref{fig:dia3}) at $d\leq8$.}
        \label{fig:btt-lambda-c-8}
    \end{subfigure}
    \caption{Relative deviation $\tfrac{\sigma_{\mathrm{BSM}}-\sigma_i}{\sigma_{\mathrm{BSM}}}$ for the process $pp \to \tau^+ \tau^- \ b$ for the EFT at dimension six (blue) and eight (red) from the BSM prediction for $\kappa = 1.0$, $1.7$, and~$2.5$. The relative deviation is shown for the total cross section followed by the individual resonant and double production channels.
    The Drell--Yan production with associated $b$~jet (Fig.~\ref{fig:dia2}) is not displayed here since it does not exhibit a coupling dependence (see main text).}
    \label{fig:btt-lambda}
\end{figure}

For the discussion above, we fixed the benchmark coupling~$\kappa=1$.
While we argued in Sec.~\ref{sec:Drell--Yan-matching} that the relative deviation~$R_i$~\eqref{eq:4fermiondeviation} of the EFT and BSM cross sections in the pure Drell--Yan process ($p p \to \tau^+ \tau^-$) is independent of the value of the coupling~$\kappa$, this is no longer the case for the process $p p \to \tau^+ \tau^- b$. To investigate the coupling dependence, we compare the three choices $\kappa \in \{1.0,\,1.7,\,2.5\}$ in Fig.~\ref{fig:btt-lambda}.
Figures~\ref{fig:btt-lambda-6} and~\ref{fig:btt-lambda-8} show the relative deviation for the total cross section of $p p \to \tau^+ \tau^- b$ for the $d=6$ (blue) and $d \leq 8$ (red) EFT description, respectively.
Figs.~\ref{fig:btt-lambda-a-6} and~\ref{fig:btt-lambda-a-8} show the corresponding deviations for the resonant leptoquark channel~\textbf{(a)} only, while Fig.~\ref{fig:btt-lambda-c-8} displays the $\kappa$~dependence of the double leptoquark channel~\textbf{(c)} which has no $d=6$ contribution.
In analogy with the pure Drell--Yan process the $b$-associated Drell--Yan topology~\textbf{(b)} does not exhibit a coupling dependence and is not shown here.

For all plots in Fig.~\ref{fig:btt-lambda}, we observe that the deviation between the EFT and BSM cross sections changes with the coupling, at any fixed leptoquark mass.
To explain this, we observe that the EFT cross section always scales as $\sigma_\sscript{EFT} \propto \kappa^4$.
This applies to both $d=6$ and $d=8$ contributions.
On the other hand, the coupling dependence of the resonant contributions in the full BSM model is more complicated and can be understood schematically from the squared propagator structure,
\begin{align}
|\mathcal{M}_{\sscript{BSM}}|^2 
    \propto
    \frac{\kappa^4}{\left(Q^2-M_{\slq}^2\right)^2 + M_{\slq}^2 \Gamma^2}
    \,,
    \label{eq:BSM-lambda-scaling}
\end{align}
where the decay width of the leptoquark scales as~$\Gamma \propto \kappa^2$ and $Q$ denotes the momentum transfer of the leptoquark. For the simulations, the leptoquark width is evaluated at tree level as \(\Gamma_{\tilde S_1}=\kappa^2 M_{\tilde S_1}/(16\pi)\), assuming \(\tilde S_1\to b\tau\) as the only decay channel and neglecting fermion masses.
One can see that the naive $\sigma_\mathrm{BSM} \propto \kappa^4$ scaling holds only far away from the pole of the propagator, where the width can be neglected. Close to the pole, the $ \kappa $-dependence of the width cancels against the $\kappa$-dependence of the numerator.

This is in agreement with our finding that for large $M_{\slq}$~values the coupling dependence decreases in all panels of Fig.~\ref{fig:btt-lambda}.
In addition, this also highlights again the coupling independence of the cross-section ratio for the Drell--Yan process, where the leptoquark is in the $t$-channel with a spacelike momentum transfer $Q^2<0$ and hence never on-shell.
However, for resonant production~\textbf{(a)} and two-leptoquark exchange~\textbf{(c)},\footnote{For two-leptoquark exchange~\textbf{(c)}, the schematic form in Eq.~\eqref{eq:BSM-lambda-scaling} does of course not apply due to the presence of a second leptoquark propagator. The arguments presented here apply nevertheless.} one of the leptoquarks can be resonantly produced, in which case the decay width in Eq.~\eqref{eq:BSM-lambda-scaling} reintroduces a $ \kappa $~dependence in the ratio~$R_i$.

\subsection{Pair Production: \texorpdfstring{$pp \to \tau ^+ \tau ^- b \bar{b}$}{pp -> tau+ tau- b b}}
\begin{figure}[tb]
    \centering
    \begin{subfigure}{.9\textwidth}
        \centering
        \includegraphics[width=\textwidth]{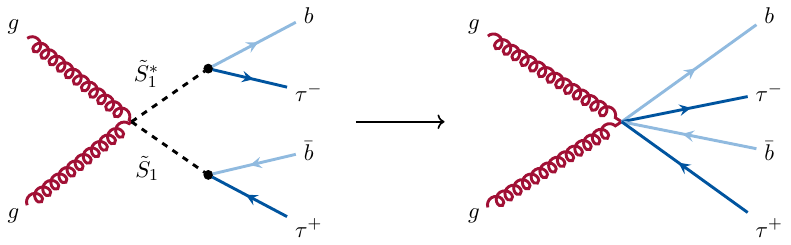}
        \caption{Dominant contribution at low $M_{\slq}$: Pair production.}
        \label{fig:g-g-b-b-tau-tau-pair}
    \end{subfigure}
    \\[0.2cm]
    \begin{subfigure}{.9\textwidth}
        \centering
        \includegraphics[width=\textwidth]{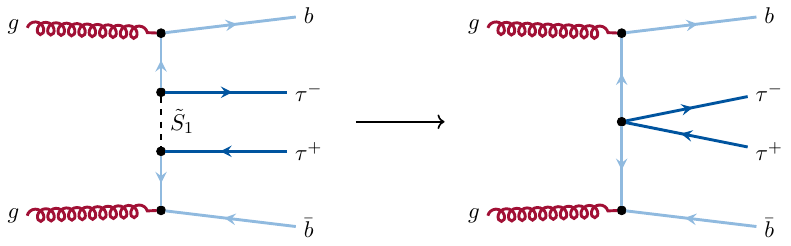}
        \caption{Dominant Contribution at high $M_{\slq}$: Drell--Yan.}
        \label{fig:g-g-b-b-tau-tau-DrellYan}
    \end{subfigure}
    \caption{Two sample Feynman diagrams dominating the process $p  p\to \tau^+ \tau^- b \bar{b}$ at low~\textbf{(a)} and high~\textbf{(b)} values of $M_{\slq}$ in the full BSM theory (left) and the corresponding channels in the EFT (right).
    }
    \label{fig:gb-tautaubb-diags}
\end{figure}

Lastly, we investigate the EFT validity for the third channel with two associated $b$-tagged jets in the final state, i.e., the scattering $pp \to \tau ^+ \tau ^- b \bar{b}$, considered by ATLAS~\cite{ATLAS:2023vxj}.
A~multitude of Feynman diagrams contribute to this channel, and we show only some sample topologies with the largest contributions to the total cross section in Fig.~\ref{fig:gb-tautaubb-diags}. 

\begin{figure}[tbp]
    \centering
    \begin{subfigure}{0.49\textwidth}
        \centering
        \includegraphics[width=\textwidth]{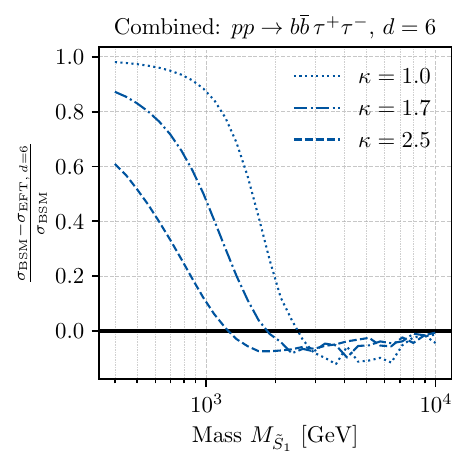}
        \caption{$pp \to \tau^+ \tau^- \ b \bar{b}$ at $d=6$.}
        \label{fig:bbtt-lambda-6}
    \end{subfigure}
   \begin{subfigure}{0.49\textwidth}
        \centering
        \includegraphics[width=\textwidth]{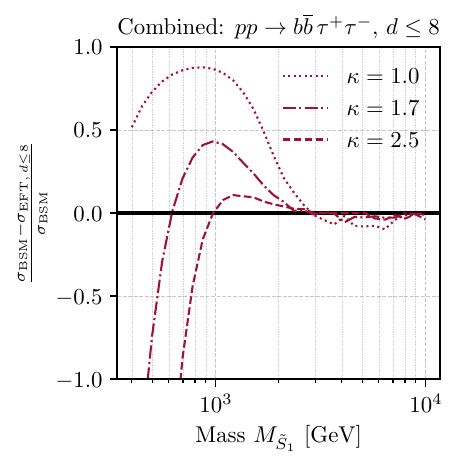}
        \caption{$pp \to \tau^+ \tau^- \ b \bar{b}$ at $d \leq 8$.}
        \label{fig:bbtt-lambda-8}
    \end{subfigure}
    \caption{Relative deviation of the total $pp\to b\overline{b}\,\tau^+\tau^-$ cross section of the EFT predictions from the full BSM theory as a function of the leptoquark mass. Cross sections are calculated for $\kappa = 1.0$, $1.7$, and~$2.5$.}
    \label{fig:eft-ratio-bbtt}
\end{figure}

The largest contribution at low masses comes from leptoquark pair production (Fig.~\ref{fig:g-g-b-b-tau-tau-pair}). In the EFT description, this process is a pure $d=8$ contribution, indicating that $d=6$ terms do not adequately approximate the full model in that parameter space. For higher leptoquark masses, the cross section is instead dominated again by Drell--Yan production with two associated $b$~jets as shown in Fig.~\ref{fig:g-g-b-b-tau-tau-DrellYan}.
The relative differences between the total EFT and leptoquark cross sections are shown in Fig.~\ref{fig:eft-ratio-bbtt} for three different coupling strengths $\kappa = 1.0, 1.7, 2.5$.

Figure~\ref{fig:bbtt-lambda-6} shows the relative deviation of the cross section for the EFT truncated at $d=6$, whereas Fig.~\ref{fig:bbtt-lambda-8} displays the corresponding truncation to $d \leq 8$.
Overall, we observe again the convergence of the EFT series for increasing leptoquark mass.
The $d \leq 8$ truncation converges faster toward the full model than the $d=6$ terms alone, though the difference is rather minimal.
For low masses, we find a significant dependence on the coupling, which reduces to a negligible level for $M_{\slq}\gtrsim 3\,\text{TeV}$.
This is explained by leptoquark pair production, which is a pure QCD process and does not depend on~$\kappa$, dominating the channel at low masses.
Hence, the smaller the $\kappa$~coupling is, the more dominant pair production becomes. This is not modeled well by the EFT, since it is a $d \geq 8$ effect with two (close to) on-shell BSM states.
For larger masses, the Drell--Yan contribution dominates the process, and the $\kappa$~dependence drops out in the ratio of cross sections.
In this large-mass regime, we also find more pronounced numerical fluctuations in our simulations of the full BSM theory with \textsc{MadGraph} due to the very small values of the cross section.\footnote{%
In this regime, the absolute cross sections drop to a level where the \textsc{MadGraph} integration becomes extremely sensitive to numerical fluctuations and instabilities, explaining the fluctuations seen in the high-mass region of Fig.~\ref{fig:eft-ratio-bbtt}. 
We have found that the choice of factorization and renormalization scale as well as the Monte Carlo seed can have a sizable impact on the value of the absolute cross section in that region.}

\section{Conclusions}
\label{sec:conclusion}

Monte Carlo simulations of BSM models and SMEFT scenarios at
the LHC require tools that can automatically derive the Feynman rules of generic theories and export them to the UFO format~\cite{Degrande:2011ua,Darme:2023jdn} for use by event generators. 
In this work, we present such a tool implemented within the \matchete~\cite{Fuentes-Martin:2022jrf} framework.
Key features of our setup include:
\begin{enumerate}[i)]
    \item the simple model definition interface provided by~\matchete;
    \item native integration with \matchete's matching functionality, streamlining EFT analyses, as showcased in Sec.~\ref{sec:EFT-convergence};
    \item functionality for (semi-)automatic spontaneous symmetry breaking and gauge fixing (including $R_\xi$~gauges), as discussed in Sec.~\ref{sec:matchete_symmetry-breaking};
    \item the automatic incorporation of arbitrary continuous flavor symmetries, as presented in Sec.~\ref{sec:flavor-symmetries};
    \item a complete implementation in \matchete of the SMEFT operator 
    basis through dimension eight, based on a modified version of the Murphy 
    basis~\cite{Murphy:2020rsh}, together with the $M_W$ electroweak input 
    scheme; see Secs.~\ref{sec:d=8_UFO} and~\ref{sec:input_scheme}. 
\end{enumerate}

The framework presented here is flexible and broadly applicable to a wide range of renormalizable BSM theories and EFTs.
The SMEFT implementation serves as a particularly clear illustration of these features.
Availability of the SMEFT up to dimension eight paves the way for thorough studies of the importance of higher-dimensional operators and the validity of the EFT approach in various processes. 
Our UFO implementations have been extensively validated by comparing squared matrix elements with reference implementations, following the approach of Ref.~\cite{Durieux:2019lnv} and using a dedicated plugin for \textsc{MadGraph}. 
At dimension six, we validated our implementation against \textsc{SMEFTsim}~\cite{Brivio:2020onw}, which in turn has previously 
been validated against several other SMEFT implementations.
No other complete implementation of the dimension-eight SMEFT exists at this point, so we validated the bosonic sector against \textsc{SmeftFR~v3}~\cite{Dedes:2023zws} and selected fermionic operator classes against the UFO accompanying Ref.~\cite{Hays:2018zze}; 
four-fermion interactions were additionally checked against independent internal \textsc{FeynRules} implementations. 
Results agreed with the corresponding reference implementations for all tested processes and operators classes within the numerical accuracy of the comparisons.

To illustrate some of the advantages offered by our tool, we have analyzed the validity of the EFT description of a leptoquark model, including $d=8$ contributions, in Sec.~\ref{sec:EFT-convergence}. We considered the production channels relevant to the ATLAS search~\cite{ATLAS:2023vxj} for a scalar leptoquark coupling only to third-generation fermions. For the three final states $pp \to \tau^+ \tau^-$, $pp \to \tau^+ \tau^- b$, and $pp \to \tau^+ \tau^- b \bar{b}$, we investigated how rapidly the EFT signal cross sections approach the full-model results as a function of the leptoquark mass. 

We found significant differences depending on the nature of the underlying hard-scattering topology and on whether the leptoquark can be produced on shell. For non-resonant $t$-channel exchange, including the dimension-eight contributions can substantially improve the agreement with the full model. By contrast, for resonant production, a local EFT cannot reproduce the propagator near its pole, and the dimension-eight terms provide little improvement. In processes receiving both resonant and non-resonant contributions, the apparent EFT validity therefore also depends on the kinematic cuts that determine the dominant topology. In all cases, both the $d=6$ and $d\leq8$ results approach the full model in the large leptoquark mass limit.

While the main part of this work focuses on the physics applications of the new functionality incorporated into the \matchete framework, further technical details and usage examples are provided in the appendices. The complete documentation is available through the built-in \textsc{Mathematica} documentation center included with \matchete and on the website~\cite{MatcheteWebsite}.

Our current implementation generates UFO files for tree-level simulations only. Extending it to the one-loop level requires the computation of ultraviolet and $R_2$~counterterms~\cite{Ossola:2006us,Ossola:2008xq} and is left for future work; see also Refs.~\cite{Degrande:2014vpa,Degrande:2020evl}. Furthermore, the symmetry-breaking functionality introduced here will serve as a basis for implementing automatic one-loop matching of heavy vector bosons in scenarios with spontaneously broken extended gauge groups in a future version of \matchete. For additional details, see Ref.~\cite{Thomsen:2024abg}. 
Finally, one could eventually extend the present work to further interfaces between \matchete and \textsc{FeynArts}~\cite{Hahn:2000kx} or \textsc{FeynCalc}~\cite{Shtabovenko:2016sxi}.

\phantomsection
\section*{Acknowledgments}

AET would like to thank Valentin Hirschi for several helpful discussions.
AET and FW would like to thank the Mainz Institute for Theoretical Physics~(MITP) of the Cluster of Excellence PRISMA+ (Project ID 390831469) for its hospitality and support during the completion of this manuscript. 
This research was supported by the Deutsche Forschungsgemeinschaft (DFG, German Research Foundation) under grant 396021762 -- TRR~257: \textit{Particle Physics Phenomenology after the Higgs Discovery}.
The work of AET was funded by the Swiss National Science Foundation~(SNSF) through the Ambizione grant: \textit{Matching and Running: Improved Precision in the Hunt for New Physics}, project number 209042.
The work of LH was supported by the SNSF through the SNSF Starting Grant: \textit{Automated Effective Field Theories} (225951).

\begin{appendix}

\section{Feynman Rules from the Effective Action} \label{app:FR}

For the derivation of the Feynman rules we follow a path-integral approach. 
Consider a quantum (effective) field theory described by the action~$S[\eta]$, where $\eta$ collectively denotes all fields.
The vacuum functional of this theory is
\begin{align}
    W[J] &= -i\hbar\log  \!\int \cD\eta \, \exp \left\{ \frac{i}{\hbar} \left(S[\eta] + J^I\eta_I \right) \right\} ,
    \label{eq:partition-function}
\end{align}
where a source term~$J$ is introduced for every field~$\eta$.\footnote{For better readability, we keep the perturbative expansion in powers of~$\hbar$ explicit.}
For compactness, we have grouped the spacetime coordinates~($x$) and internal indices~($a$) into a single index $I=(x,a)$, such that $J^I\eta_I \equiv \int\dd^Dx\, J^a(x)\eta_a(x)$ with implied sums over all fields and their discrete indices.
This is the generating functional of all connected correlation functions of this theory:
\begin{align}
    \big\langle \eta_{I_1} \, \eta_{I_2} \cdots \eta_{I_n} \big\rangle 
    &= \left. (\eminus i)^n\frac{\delta^n W[J]}{\delta J^{I_1} \, \delta J^{I_2} \cdots \delta J^{I_n}} \right|_{J=0} .
\end{align}

Performing a Legendre transformation with respect to the source terms, we obtain the quantum effective action
\begin{align} \label{eq:effective_action_def}
    \Gamma[\hat\eta] &= W[J] - J^I \hat\eta_I\, , \qquad 
    \hat\eta_I \equiv \left\langle \eta_I \right\rangle = \frac{\delta W}{\delta J^I} \,,
\end{align}
where $\hat\eta$ denotes the expectation value of the field~$\eta$ in the presence of sources, while we have $J^I = -\frac{\delta\Gamma[\hat\eta]}{\delta\hat\eta_I}$.
This is the generating functional of all one-particle-irreducible~(1PI) correlation functions:
\begin{align}
    \big\langle \hat\eta_{I_1} \, \hat\eta_{I_2} \cdots \hat\eta_{I_n} \big\rangle_{\! \sscript{1PI}} 
    &= i \frac{\delta^n \Gamma[\hat\eta]}{\delta\hat\eta^{I_1} \, \delta\hat\eta^{I_2} \cdots \delta\hat\eta^{I_n}} \bigg|_{\hat\eta = 0} .
\end{align}

The Feynman rules for all interaction vertices of a theory are determined by its tree-level 1PI correlation functions. 
Hence, it is sufficient to consider only the tree-level effective action~$\Gamma^{(0)}[\hat\eta]$, which can be determined following
\begin{align}
    \Gamma[\hat\eta]
    &= -i\hbar\log \! 
    \int\cD\eta \, \exp\! \left( \frac{i}{\hbar} \left\{ S[\eta] + J^I (\eta-\hat\eta)_I \right\} \right)
    \nonumber\\
    &= -i\hbar\log \! 
    \int \cD\eta \, \exp\!\left( \frac{i}{\hbar} \left\{ S[\bar\eta+\hbar^{\scriptscriptstyle 1/2}\eta] + J^I (\hbar^{\scriptscriptstyle 1/2}\eta + \bar\eta - \hat\eta)_I \right\} \right)
    \nonumber\\
    &= -i\hbar\log \! 
    \int \cD\eta \, \exp\! \left( \frac{i}{\hbar} \left\{ S[\bar\eta] + J^I (\bar\eta - \hat\eta)_I + \cO(\hbar) \right\} \right)
    ,
\end{align}
where we have shifted the integration variable by $\eta \to \bar\eta + {\hbar}^{\scriptscriptstyle 1/2}\eta$ in the second equality with the classical field configuration~$\bar\eta$ defined by the classical equation of motion $\left. \frac{\delta S[\eta]}{\delta \eta_I} \right|_{\scriptscriptstyle \eta=\bar\eta} = -J^I$.
In the last equality we have applied a saddle-point approximation $S[\bar\eta+\hbar^{\scriptscriptstyle 1/2}\eta] = S[\bar\eta] + \left.\frac{\delta S[\eta]}{\delta \eta_I}\right|_{\scriptscriptstyle \eta=\bar\eta} \hbar^{\scriptscriptstyle 1/2}\eta + \cO(\hbar)$.
Using the definition~\eqref{eq:effective_action_def}, one can show that $\bar\eta=\hat\eta + \cO(\hbar)$, and it follows that $\Gamma^{(0)}[\hat\eta] = S[\hat\eta]$.

Finally, we can write the master formula for deriving Feynman rules for all interaction vertices of a theory with action~$S[\eta]$ in the form
\begin{align} \label{eq:PI-Feynman-rules}
    \big\langle \eta_{I_1} \, \eta_{I_2} \cdots \eta_{I_n} \big\rangle_\text{1PI,\,tree} 
    &= i \frac{\delta^n S[\hat\eta]}{\delta\hat \eta^{I_1} \, \delta\hat \eta^{I_2} \cdots \delta\hat\eta^{I_n}} \bigg|_{\hat \eta = 0}
    .
\end{align}

Conventionally, Feynman rules refer to the momentum-space tree-level 1PI vertex function. 
This is conveniently interpreted in terms of the momentum-space fields~$\eta_a(k)$ which are related to their position-space counterparts~$\eta_a(x)$ by a Fourier transformation:
    \begin{equation}
    \eta_a(x) = \int \dfrac{\dd^D k }{(2\pi)^D} e^{\eminus i\,k\cdot x} \eta_a(k).
    \end{equation}
The \emph{Feynman rule} $ F_{a_1\ldots a_n}(k_1,\ldots, k_n) $ associated with the vertex between fields $ \eta_{a_1},\ldots,\eta_{a_n}$ is defined by 
    \begin{equation} \label{eq:def_FR}
    (2\pi)^D \delta\big( k_1 + \ldots + k_n\big) F_{a_1\ldots a_n}(k_1,\ldots, k_n) = i\, \dfrac{\delta }{\delta \eta_{a_1}(k_1)} \cdots \dfrac{\delta }{\delta \eta_{a_n}(k_n)} S[\eta] \bigg|_{\eta=0}.
    \end{equation}
Here $ k_i $ is interpreted as the \emph{incoming momentum} (to the vertex) of the field $ \eta_{a_i} $. The momentum-conserving delta function on the l.h.s. of~\eqref{eq:def_FR} is universal, following from translation invariance of the action.

To apply functional derivatives w.r.t. the momentum space fields in~\eqref{eq:def_FR}, it is useful to consider how an operator is represented when expanded in terms of these fields. Take a generic operator $ O $, which is a monomial in (derivatives of) fields $ \partial^{m_i} \eta_{a_i}(x) $ (with $ m_i\geq 0$).\footnote{Gauge-invariant operators, with covariant derivatives, can be written as a sum of such monomial operators for the purposes of extracting Feynman rules.} 
Its contribution to the action can be written in momentum space as 
    \begin{multline}
    S[\eta]\supset \int \dd^D x\,  O\Big(\partial^{m_1} \eta_{a_1}(x), \ldots , \partial^{m_n} \eta_{a_n}(x) \Big) \\ 
    = \int \dfrac{\dd^D k_1 }{(2\pi)^D} \cdots \dfrac{\dd^D k_n }{(2\pi)^D} (2\pi)^D \delta\big( k_1 + \ldots + k_n\big) O\Big( (\eminus i\, k_1)^{m_1} \eta_{a_1}(k_1), \ldots , (\eminus i\, k_n)^{m_n} \eta_{a_n}(k_n) \Big).
    \end{multline}
We may think of the expression as prescribing the replacement $ \partial_\mu \to \eminus i\, k_\mu $, where $ k_\mu $ is the momentum associated with the field. 
For identical fields the functional derivatives provide all the permutations between the various equivalent insertions. 
The momentum-conserving delta function reproduces the one that is factored out of the Feynman rule on the l.h.s. of~\eqref{eq:def_FR}. 
One simply repeats these steps for all operators in the theory. 

\section{Model Definition and Symmetry Breaking}
\label{app:SSB}

We summarize the semi-automated \matchete workflow needed to cover the case of EW symmetry breaking in the SM. The underlying methods are built with generality in mind rather than being hard-coded to this specific breaking pattern: further details and generalizations to other breaking patterns are left for future work. The SM Higgs mechanism breaks $ \SU(2)_L \times \U(1)_Y \to \U(1)_{\sscript{EM}}$ and all the physical particles are in well-defined representations---states of definite charge---under the remnant electromagnetic symmetry. Rather than having to manually write down the broken-phase Lagrangian, we find it less error-prone and tedious to specify the symmetric-phase Lagrangian and have included functionality to expand out the Lagrangian in the broken phase. The information needed to execute this decomposition in \matchete is 
\begin{enumerate}[i)]
    \item a specification of  the symmetry breaking pattern;
    \item details of how each irreducible representation of $ \SU(2)_L $ decomposes in terms of $ \U(1)_{\sscript{EM}} $ charges and how the $ \U(1)_Y $ charge contributes to the electromagnetic charge;
    \item a specification of how each $\SU(2)_L$ Clebsch--Gordan (CG) coefficient decomposes;
    \item specifications for how each symmetric-phase field decomposes in terms of broken-phase fields. 
\end{enumerate}
With this information \matchete can automatically substitute all relevant expressions into the symmetric-phase Lagrangian and carry out all tensor contractions. Various cross-checks of the input help users catch errors before implementing the symmetry breaking. The information is also sufficient to automatically derive gauge-fixing and ghost terms for the Lagrangian in the $ R_\xi $ or unitary gauge.

Before specifying the breaking pattern, the user must first define all gauge groups and fields following standard \matchete workflow, not only for the symmetric-phase but for the broken-phase fields and gauge groups, too. This provides context for all objects passed around to various functions.

\subsection{Group Algebra}
At the top level, we need to specify to the program what symmetry breaking pattern we are looking at. This is indicated with the command 
\begin{mmaCell}{Code}
  SetSymmetryBreakingPattern[{\mmaUnd{SU2L}, \mmaUnd{U1Y}}, \mmaUnd{U1em}]
\end{mmaCell}
and sets the stage for further details. In particular, this ensures that all subsequent functions can analyze the input for inconsistencies w.r.t. the declared breaking pattern, minimizing the risk of implementation errors.

All symmetric-phase contractions with indices from representations of the non-Abelian \mmaInlineCell[]{Input}{\mmaUnd{SU2L}} group should be decomposed in terms of their constituent parts when going to the broken phase. Practically, we instruct \matchete on how to write each of the non-trivial \mmaInlineCell[]{Input}{\mmaUnd{SU2L}} representations as vectors of representations of the stability group \mmaInlineCell[]{Input}{\mmaUnd{U1em}} (defining the branching rules of the representations). In this particular example, the stability group is Abelian and all its representations are one-dimensional. We call  
\begin{mmaCell}{Code}
  RepresentationDecomposition[\mmaUnd{U1Y}[1], \mmaUnd{U1em}[1]]
  RepresentationDecomposition[\mmaUnd{SU2L}[fund], {\mmaUnd{U1em}[1/2], \mmaUnd{U1em}[-1/2]}]
  RepresentationDecomposition[\mmaUnd{SU2L}[adj], {\mmaUnd{U1em}[1], \mmaUnd{U1em}[-1], Singlet}] 
\end{mmaCell}
The charges in the decomposition of each element essentially reproduce the familiar relation $ Q = Y + T_3 $ between the electric charge, hypercharge, and weak isospin; the first line of the input indicates that a unit of hypercharge becomes a unit of electric charge in the broken phase. The second line instructs \matchete that the doublet (fundamental) representation of $ \SU(2)_L $ decomposes into two one-dimensional representations in the broken phase, with electric charges~$ \pm \tfrac{1}{2} $.    

The last line in the input above indicates that the adjoint $\SU(2)_L$ index, \mmaInlineCell[]{Code}{\mmaUnd{SU2L}[adj]}, contains three components, two of charges $\pm 1$ and a singlet (neutral). The ordering of the components specifies a particular basis for the adjoint representation. While this ordering has no physical significance, \matchete makes certain assumptions about the input in order to construct the new invariants.\cprotect\footnote{For instance, \mmaInlineCell[]{Code}{{\mmaUnd{U1em}[1], Singlet, \mmaUnd{U1em}[-1]}} is more in line with a conventional basis choice for the isospin triplet but is \emph{invalid} in \matchete (it produces an error because the conjugate components are not arranged consecutively).} This is an important subtlety in the embedding of the representations of the stability group inside the representations of the original group.
One typically chooses bases for real representations, such as the adjoint, where invariant contractions are performed with a Kronecker delta. However, when rotating into a basis where each component is in a well-defined, possibly complex representation of the stability group, this may require a unitary but \emph{non-orthogonal} rotation matrix, $ U $. Consequently, contractions through the Kronecker delta change to 
\begin{equation} \label{eq:2ind_invariant_new_basis}
    \delta \longrightarrow U\transpose \delta U = \begin{pmatrix}
        0 & 1 & 0\\ 1 & 0 &0 \\ 0 & 0 & 1
    \end{pmatrix} 
\end{equation}
in the present example. In other words, the two-index invariant of the real representation need not be the identity matrix in the new basis. 
Branching rules for real representations of the original group ensure that complex representations of the stability group necessarily occur in pairs in the decomposition. \matchete requires the members of each such pair to be listed consecutively in the decomposition specified with \mmaInlineCell{Code}{RepresentationDecomposition}; it then constructs the two-index invariant in the new basis in the form of~\eqref{eq:2ind_invariant_new_basis}. 

The previous discussion may seem a little intangible, so take the case of the $ W $ bosons. In the original basis we label their components $ W^{1,2,3} $ and an invariant under global $ \SU(2) $ rotations is 
    \begin{equation}
    (W^I)^2 = (W^1)^2 + (W^2)^2 + (W^3)^2.
    \end{equation}
If we go to the basis with well-defined electromagnetic quantum numbers, the three states~$ W^{+,-,0} $ organize as
    \begin{equation}
    (W^I)^2 = 2 W^+W^- + (W^0)^2
    \end{equation}
instead. Observe that the contraction~\eqref{eq:2ind_invariant_new_basis} ensures that all terms in the original invariant are manifestly electrically neutral. The factor of two in the first term is consistent with, e.g., the expected normalization of masses and kinetic terms for complex fields.

Having specified how each representation of the original EW symmetry decomposes in the broken phase, we must also specify how the Clebsch--Gordan coefficients that appear in the Lagrangian decompose. For instance, each of the doublet indices of the two-index invariant $ \varepsilon^{ij} $ of $ \SU(2)_L $ takes values in the two distinct electric charges, $ \pm \tfrac{1}{2}$. We must specify the components of $ \varepsilon^{ij} $ in the chosen basis for all the $ 2\times 2 $ combinations of remnant representations that the original $ \SU(2)_L $ indices can take values in; effectively, we must specify a $ 2\times 2$ matrix:
\begin{mmaCell}{Input}
  CGDecomposition[eps[\mmaUnd{SU2L}][i, j], \{\{1, 2\}-> +1, \{2, 1\}-> -1\}]
\end{mmaCell}
Similarly, the adjoint $ \SU(2)_L $ index can take values in three separate representations of the remnant group, so the generators of the fundamental representation of $ \SU(2)_L $ are identified with $ 3\times 2\times 2$ tensors:
\begin{mmaCell}[]{Input}
  CGDecomposition[gen[\mmaUnd{SU2L}[fund]][J, i, j], \{
    \{2, 1, 2\} -> +1/Sqrt[2], (*T^+*)
    \{1, 2, 1\} -> +1/Sqrt[2], (*T^-*)
    \{3, 1, 1\} -> +1/2, \{3, 2, 2\} -> -1/2 (*T^3*)
  \} ]
\end{mmaCell}

The \mmaInlineCell[]{Code}{i}, \mmaInlineCell[]{Code}{j} indices are in the fundamental and anti-fundamental representations of $ \SU(2)_L $, respectively, and \mmaInlineCell[]{Code}{J} represents an adjoint index of this group. 

From a user point of view, it may seem daunting to get this decomposition correct. To alleviate these problems, we have implemented extensive checks to ensure that no components can be specified in entries that correspond to a non-trivial (charged) representation of the remnant groups. Such entries would produce non-invariant terms in the broken phase Lagrangian, and are clearly impossible. We also check that the decomposition of the CG has the same normalization as the original CG. Having specified the fundamental generators, \matchete automatically infers the decomposition of the $ \SU(2)_L $ structure constants, in the corresponding basis, by relying on commutation identities. Also the decomposition of all $ \delta $-invariants is automatically inferred based on the branching rules of the associated representations.   

\subsection{Field Decomposition}
Next, we should specify how each field of the unbroken phase is written in terms of the broken-phase ones. One may think of \mmaInlineCell{Input}{FieldDecomposition} as defining a set of replacement rules. For the original gauge fields, we set 
\begin{mmaCell}[]{Input}
  FieldDecomposition[W[\mmaUnd{\(\mu\)}, a], 
    \{\mmaUnd{\(e\)}[]/sw[] \mmaUnd{\(\cW\)}[\mmaUnd{\(\mu\)}], \mmaUnd{\(e\)}[]/ sw[] Bar[\mmaUnd{\(\cW\)}[\mmaUnd{\(\mu\)}]], (cw[]/sw[] \mmaUnd{\(e\)}[] \mmaUnd{\(\cZ\)}[\mmaUnd{\(\mu\)}] + \mmaUnd{\(\cA\)}[\mmaUnd{\(\mu\)}])\}]
  FieldDecomposition[B[\mmaUnd{\(\mu\)}], -sw[]/cw[] \mmaUnd{\(e\)}[] \mmaUnd{\(\cZ\)}[\mmaUnd{\(\mu\)}] + \mmaUnd{\(\cA\)}[\mmaUnd{\(\mu\)}]]
\end{mmaCell}
The guiding principle is that each index of the original fields for which the representation decomposes should respect the branching rules set up for that representation. 
The $ W^a_\mu$ field has one index \mmaInlineCell[]{Input}{a}---the adjoint of $ \SU(2)_L$---decomposing, while the Lorentz index is of course preserved into the broken phase. Accordingly, $ W^a_\mu $ decomposes into a rank-one tensor of dimension $ 3 $. By contrast $ B_\mu $ has no indices involved in the breaking, so it decomposes into a scalar. Indices unaffected by the breaking, such as the Lorentz index \mmaInlineCell[]{Input}{\mmaUnd{\(\mu\)}}, must also appear in all components on the right hand side of the replacements. 

In much the same way, we let the Higgs field (with its one fundamental index of $\SU(2)_L$) decompose as 
\begin{mmaCell}[]{Input}
  FieldDecomposition[\mmaUnd{H}[i], \{-I \mmaUnd{\(\chi\)}[], (\mmaUnd{v}[] + \mmaUnd{\(h\)}[] + I \mmaUnd{\(\chi\)0}[])/\(\sqrt{2}\)\}]
\end{mmaCell}
in terms of the Goldstone bosons, the neutral Higgs boson and the scalar VEV. 
The chiral fermions of the unbroken phase decompose in terms of the massive vector-like up~quarks, down~quarks, and charged leptons of definite electromagnetic charge (along with the chiral neutrinos). To discriminate the fields, we have chosen gothic letters for the broken-phase fields (users are free to use other choices):  
\begin{mmaCell}[]{Input}
  FieldDecomposition[q[a, i, p], \{Bar[CKM[r, p]] \mmaUnd{\(\mathfrak{u}\)}[a, r], \mmaUnd{\(\mathfrak{d}\)}[a, p]\}]
  FieldDecomposition[u[a, p], \mmaUnd{\(\mathfrak{u}\)}[a, p]]
  FieldDecomposition[d[a, p], \mmaUnd{\(\mathfrak{d}\)}[a, p]]
  FieldDecomposition[\mmaUnd{l}[i, p], \{\mmaUnd{\(\nu\)}[p], \mmaUnd{\(\mathfrak{e}\)}[p]\}]
  FieldDecomposition[e[p], \mmaUnd{\(\mathfrak{e}\)}[p]]
\end{mmaCell}
A short comment is in order for the left-handed quarks: We use mass eigenstates for the broken-phase quarks, but the up and down quarks are misaligned with the CKM matrix when embedded into $ q^{ai}_{\LL,p} $. The right-hand side of the replacement follows ordinary dummy-index summation convention, so we let the up quarks enter multiplied by the CKM matrix.\cprotect\footnote{\matchete can be instructed to treat the CKM matrix as unitary by defining it with \mmaInlineCell[]{Code}{DefineCoupling[CKM, Indices->{\mmaUnd{Flavor}, \mmaUnd{Flavor}}, Unitary->True]}.}  

As a final refinement, we will also want to substitute the symmetric-phase marginal couplings for the usual broken-phase parameters. In particular, replacing the Yukawa coupling matrices with the fermion mass matrices will let us specify that we are in a mass basis, by using the flavor-diagonal mass matrices of the fermion fields, \mmaInlineCell[]{Input}{\mmaUnd{M}\mmaUnd{\(\mathfrak{u}\)}}, \mmaInlineCell[]{Input}{\mmaUnd{M}\mmaUnd{\(\mathfrak{d}\)}}, and \mmaInlineCell[]{Input}{\mmaUnd{M}\mmaUnd{\(\mathfrak{e}\)}}. We use the \mmaInlineCell{Input}{SetSSBReplacements} method to set some replacement rules that are applied automatically during \mmaInlineCell{Input}{ToBrokenPhase} calls (with correct handling of dummy indices). For the SM, we let 
\begin{mmaCell}[]{Input}
  SetSSBReplacements[\{
    (* rotate Yukawa couplings to down-quark mass eigenbasis *)
    Yu[s_,t_] -> (Sqrt[2]/v[]) Bar[CKM[t, s]] M\mmaUnd{\(\mathfrak{u}\)}[t],
    Yd[s_,t_] -> (Sqrt[2]/v[]) Delta[\mmaUnd{Flavor}][s, t] M\mmaUnd{\(\mathfrak{d}\)}[t],
    Ye[s_,t_] -> (Sqrt[2]/v[]) Delta[\mmaUnd{Flavor}][s, t] M\mmaUnd{\(\mathfrak{e}\)}[t],
    (* replace electroweak gauge couplings *)
    gL[] -> \mmaUnd{\(e\)}[]\,/\,sw[],
    gY[] -> \mmaUnd{\(e\)}[]\,/\,cw[]
  \}];
\end{mmaCell}
The Kronecker delta functions in flavor space are needed to get matching indices between the diagonal mass matrices and the matrix Yukawa couplings. 

We have now fully defined the symmetry-breaking pattern and decompositions to \matchete. This is the information contained in the \mmaInlineCell{Input}{"SM+breaking"} model file distributed with the \matchete package and allows the \mmaInlineCell{Input}{ToBrokenPhase} method to transform the symmetric-phase Lagrangian to the broken phase (as in Sec.~\ref{sec:matchete_symmetry-breaking}). 
When supplementing new BSM fields on top of the SM, the changes to the symmetry breaking definitions can be as simple as defining a \mmaInlineCell{Input}{FieldDecomposition} for each new field (assuming that they do not participate in the breaking). Clearly, significant changes have to be made when the new fields participate in EWSB. Additional information on the subject is distributed with the \matchete package and can be found in the \textsc{Mathematica} documentation center after installation of the package (see, for instance, the tutorial page ``SM Electroweak Symmetry Breaking'').

\section{Basic Manual}
\label{app:manual}

We include here a compact overview and description for the central functions for deriving Feynman rules, generating of UFO files, and implementing SSB in \matchete.
Up-to-date, more detailed documentation is available within \matchete through the \textsc{Mathematica} documentation center.
We exclusively outline newly implemented functionality and refer to~\cite{Fuentes-Martin:2022jrf} for the general \matchete documentation and, in particular, details on the model implementation in the symmetric phase.
The documentation can also be accessed online~\cite{MatcheteWebsite}.

\begin{table}[tbp]
    \centering
    {
    \renewcommand{\arraystretch}{1.3}
    \begin{tabular}{p{0.15\textwidth}p{0.15\textwidth}p{0.44\textwidth}}
    \toprule
    \textbf{Field} & \textbf{\texttt{Ext} object} & \textbf{Physical object} \\
    \midrule
    Scalar {\color{gray} ${\langle\phi\rangle}_n$}   & \texttt{Ext}[$\phi$, n] & Identity (1) 
    \\
    Fermion {\color{gray} ${\langle\psi\rangle}_n$}  & \texttt{Ext}[$\psi$, n] & Spinor ($u$ for particles or $v$ for antiparticles)
    \\
    Vector {\color{gray} ${\langle A_\mu \rangle}_n$}  & \texttt{Ext}[$A_\mu$, n] & Polarization vector ($\varepsilon_\mu$)
    \\
    \bottomrule
    \end{tabular}
    \cprotect\caption{External legs are represented in \matchete by \mmaInlineCell[]{Input}{\mmaDef{Field}[\mmaDef{Ext}[lab, n], \ldots]}, where the usual label (\mmaInlineCell[]{Input}{lab}) of the field is replaced by \mmaInlineCell[]{Input}{\mmaDef{Ext}[lab, n]} and \mmaInlineCell[]{Input}{n} is a unique integer enumerating all external legs in one vertex. The \mmaInlineCell[]{Input}{\mmaDef{Ext}} objects represent the external spinors, polarization vectors, or simply unity, depending on the type of field, and are displayed (using \mmaInlineCell[]{Input}{NiceForm}) in gray with brackets as indicated in this table.}
    \label{tab:externallegs}
    }
  \end{table}

The new routines needed to extract the broken phase of a Lagrangian are 
\begin{itemize}
    \item 
    \mmaInlineCell[]{Input}{SetSymmetryBreakingPattern[originalGroup, remnantGroup]}
    \\
    instructs \matchete on which gauge group factors are replaced in the theory during SSB. Both the original group (\mmaInlineCell[]{Input}{originalGroup}) and the remnant group (\mmaInlineCell[]{Input}{remnantGroup}) can be lists of Abelian and simple Lie group factors. All group names should be previously defined gauge groups. 
    \item 
    \mmaInlineCell[]{Input}{RepresentationDecomposition[originalRep, decomposition]}
    \\
    instructs \matchete that the irreducible representation \mmaInlineCell[]{Input}{originalRep} of the original symmetry group decomposes into a direct sum of representations (\mmaInlineCell[]{Input}{decomposition}) of the remnant group. This effectively determines how an index of the original symmetry group decomposes in the broken phase.
    \item 
    \mmaInlineCell[]{Input}{CGDecomposition[cg, decomposition]}
    \\
    instructs \matchete that the CG coefficient \mmaInlineCell[]{Input}{cg} of the original symmetry decomposes in the broken phase. The \mmaInlineCell[]{Input}{decomposition} should be written as tensor blocks, where the indices of each block/element are dictated by the decomposition of the indices of \mmaInlineCell[]{Input}{cg}. The original \mmaInlineCell[]{Input}{cg} should be specified with open indices, which can also be used for any open indices in the blocks of the \mmaInlineCell[]{Input}{decomposition}.
    \item 
    \mmaInlineCell[]{Input}{FieldDecomposition[field, decomposition]}
    \\
    instructs \matchete that the \mmaInlineCell[]{Input}{field}, charged under the original symmetry group, decomposes into components, each of which is a linear combination of broken-phase fields with corresponding quantum numbers. The \mmaInlineCell[]{Input}{decomposition} is a block tensor, where the open indices of each element is dictated by the decomposition of the indices of \mmaInlineCell[]{Input}{field}.
    \item 
    \mmaInlineCell[]{Input}{SetSSBReplacements[couplingReplacements]}
    \\
    sets up replacement rules for symmetric-phase couplings, which are automatically applied in the symmetry-breaking routines. The replacements \mmaInlineCell[]{Input}{couplingReplacements} are a list of rules for the substitution of couplings. Open indices can be specified and will be used as patterns in replacements. 
    \item 
    \mmaInlineCell[]{Input}{ToBrokenPhase[expr]}
    \\
    transforms a symmetric-phase Lagrangian, operator, or similar to the broken phase using the decompositions of individual fields~(\mmaInlineCell[]{Input}{FieldDecomposition}) and Clebsch--Gordan coefficients~(\mmaInlineCell[]{Input}{CGDecomposition}). 
    All index sums are expanded out following the decomposition of the corresponding symmetric-phase representations. 
    The option \mmaInlineCell[]{Input}{\mmaDef{GoldstoneBoson}\,->\,False} can be used to drop all terms involving Goldstone bosons during the transformation to the broken phase (default value \mmaInlineCell[]{Input}{True}), effectively going directly to the unitary gauge. The EFT expansion can be truncated at any order with the \mmaInlineCell[]{Input}{EFTOrder} option.
    \item 
    \mmaInlineCell[]{Input}{ImplementVacuumConditions[lagrangian]}
    \\
    implements conditions on a broken-phase \mmaInlineCell[]{Input}{lagrangian} corresponding to the fields being in a kinetic and mass basis (no mixing between fields) and free of tadpoles. In short, it assumes appropriate mixing angles and VEV magnitudes. 
    By default \mmaInlineCell[]{Input}{ImplementVacuumConditions} will substitute the coefficients of the vector mass terms and vector-Goldstone boson kinetic mixing terms in favor of the canonical mass couplings associated with the vector fields.
    This behavior can be disabled with the option \mmaInlineCell[]{Input}{SubstituteMasses-> False}.  
    \item 
    \mmaInlineCell[]{Input}{GetVacuumConditions[lagrangian]}
    \\
    extracts the conditions on a broken-phase \mmaInlineCell[]{Input}{lagrangian} corresponding to the fields being in a kinetic and mass basis (no mixing between fields) and free of tadpoles and returns these as an association. It also determines the masses of the radial scalar modes and the massive vectors. 
    \item 
    \mmaInlineCell[]{Input}{GaugeFixLagrangian[lagrangian]}
    \\
    returns the gauge-fixed version of the \mmaInlineCell[]{Input}{lagrangian} including ghost fields. This routine works for both symmetric and broken-phase Lagrangians.\cprotect\footnote{It relies on all the information from the decomposition of fields and CGs to extract the relevant parts of the broken group algebra, so it will fail on manually specified broken-phase Lagrangians, that is, Lagrangians not obtained with \mmaInlineCell[]{Input}{ToBrokenPhase}.} The option \mmaInlineCell[]{Input}{Gauge-> \mmaDef{R\(\xi\)}} (default) selects the $ R_\xi $ gauge while \mmaInlineCell[]{Input}{Gauge-> Unitary} indicates the unitary gauge (with $ R_\xi $ for the remnant gauge fields). The option \mmaInlineCell[]{Input}{SubstituteMasses-> True} will return vector, ghost, and Goldstone boson masses in terms of the mass coupling associated with the vectors.
\end{itemize}

The core functionality for determining Feynman rules and generating UFO files is provided through the following five routines:
\begin{itemize}
    \item 
    \mmaInlineCell[]{Input}{FeynmanRules[lagrangian, fields (*optional*)]}
    \\
    derives the tree-level Feynman rules from the given Lagrangian~(\mmaInlineCell[]{Input}{lagrangian}).
    The second argument~(\mmaInlineCell[]{Input}{fields}) is optional. 
    It can take a list of field labels, in which case only the Feynman rule with the corresponding external legs is computed. 
    If it is not specified, the complete set of Feynman rules for all vertices is derived.
    This function returns an association that maps each external set of fields to the symbolic expressions for the Feynman rule of that vertex.
    External legs are represented by the symbol~\mmaInlineCell[]{Input}{\mmaDef{Ext}} in the output expression. 
    Its interpretation for the various field types is described in Tab.~\ref{tab:externallegs}.
    The option \mmaInlineCell[]{Input}{\mmaDef{Legs}} allows for limiting the number of external fields. 
    For example, \mmaInlineCell[]{Input}{\mmaDef{Legs}->6} restricts to vertices with up to six external legs (default \mmaInlineCell[]{Input}{\mmaDef{Legs}->All}).
    In addition, there are two options for controlling conventions: \mmaInlineCell[]{Input}{"ExternalMomentum"\,->\,"incoming"} (default) fixes the momentum direction and can be set to \mmaInlineCell[]{Input}{"outgoing"} otherwise.
    The option \mmaInlineCell[]{Input}{"FermionFlow"\,->\,"Standard"} (default) determines how fermionic Feynman rules are interpreted. 
    When set to \mmaInlineCell[]{Input}{"DEHK"} the algorithm from Refs.~\cite{Denner:1992me,Denner:1992vza} (which is also employed by \textsc{MadGraph}) is used as described in Sec.~\ref{sec:conventions}.
    \item 
    \mmaInlineCell[]{Input}{DefineCouplingOrder["name", couplings, hierarchy, maxOrder (*optional*)]}
    \\
    defines a new coupling order labeled by the given~\mmaInlineCell[]{Input}{"name"}.
    This order is assigned to the provided~\mmaInlineCell[]{Input}{couplings}, which can be either a single coupling label or a list of multiple labels.
    By default the couplings are all assigned order one. Other orders can be specified using powers of couplings or lists where the first entry is the coupling label and the second the order.
    The argument~\mmaInlineCell[]{Input}{hierarchy} allows for specifying a relative hierarchy when multiple coupling orders are defined.
    For example, if one coupling order is defined with \mmaInlineCell[]{Input}{hierarchy} set to~\mmaInlineCell[]{Input}{1} and another one where \mmaInlineCell[]{Input}{hierarchy} is set to~\mmaInlineCell[]{Input}{2}, two insertions of couplings of the former class are taken to be of the same order as one insertion of the latter class.
    As an optional fourth argument (\mmaInlineCell[]{Input}{maxOrder}) one can provide an integer that determines the maximum expansion order for that coupling order (defaults to~\mmaInlineCell[]{Input}{99}).
    \item 
    \mmaInlineCell[]{Input}{ImposeFlavorSymmetry[lagrangian, symmetry]}
    \\
    takes a flavor symmetry~(\mmaInlineCell[]{Input}{symmetry})---specified by an association that contains information about which flavor of which field transforms in which representation of the previously defined global flavor groups---and associates it to the couplings in the Lagrangian~(\mmaInlineCell[]{Input}{lagrangian}). 
    This is done by determining which fields the flavor indices of the couplings are contracted into in the Lagrangian. 
    The Lagrangian must contain the fields for which the symmetry is defined, usually the Lagrangian of the unbroken phase.
    The output of the method is an association with the coupling labels as keys and whose values are lists containing one \mmaInlineCell[]{Input}{SparseArray} per determined flavor-invariant structure.
    The option \mmaInlineCell[]{Input}{\mmaDef{Except}->\{\}} (default) allows for excluding certain couplings (whose labels should be added to the list) from the flavor assumptions, such that they retain their most general form. 
    This can be useful for, e.g., Yukawa couplings or masses.
    In addition, the option \mmaInlineCell[]{Input}{Simplify->True} (default) indicates that \matchete should attempt to simplify the basis of flavor invariants it uses by maximizing the number of entries in the invariants that are either 0~or~1. 
    When set to \mmaInlineCell[]{Input}{False} an unsimplified but orthogonal basis of invariants is used.
    For more details see Sec.~\ref{sec:flavor-symmetries}.
    \item 
    \mmaInlineCell[]{Input}{DefaultParamCard[lagrangian]}
    \\
    generates a default parameter card as a JSON file, containing all parameters present in the input Lagrangian~\mmaInlineCell[]{Input}{lagrangian} and the widths of all massive particles.
    The symmetries of all defined couplings are exploited to obtain a minimal parametrization of the theory that respects Hermiticity.
    The option \mmaInlineCell[]{Input}{\mmaDef{FlavorInvariants}\,->\,<||>} can be used to employ flavor symmetries for the minimal parametrization of all couplings. To that end, the empty association should be replaced by the output of \mmaInlineCell[]{Input}{ImposeFlavorSymmetry}.
    The output file can serve as a template and provide the basis for a custom parameter card.
    By default, all external input parameters are set to~\mmaInlineCell[]{Input}{0} (vanish). 
    To entirely remove a parameter from the UFO output, the special value \lstinline[columns=fixed,language=json]{"ZERO"} can be assigned in the parameter card. 
    The option \mmaInlineCell[]{Input}{DefaultValues->True} (default:  \mmaInlineCell[]{Input}{False}) instructs \matchete to include predefined numerical values for the SM inputs.
    This feature only works for simple BSM scenarios that do not modify the electroweak breaking pattern of the SM, i.e., it cannot be used for the SMEFT.
    The option \mmaInlineCell[]{Input}{Gauge->Automatic} (default) controls which gauge should be used. 
    By default, the function checks if \mmaInlineCell[]{Input}{lagrangian} is already gauge fixed and uses that gauge. 
    If it is not yet gauge fixed, the $R_\xi$ gauge is employed.
    Alternative values are \mmaInlineCell[]{Input}{\mmaDef{R\(\xi\)}} and \mmaInlineCell[]{Input}{Unitary} (see the documentation of \mmaInlineCell[]{Input}{GaugeFixLagrangian} for more details).
    The option \mmaInlineCell[]{Input}{\mmaDef{R\(\xi\)}->1} (default) indicates that the general $R_\xi$ gauge should be reduced to the Feynman gauge by setting all gauge-fixing parameters to $\xi=1$. 
    Instead, \mmaInlineCell[]{Input}{\mmaDef{R\(\xi\)}->General} will keep generic gauge parameters. 
    Finally, one can specify the location to which the parameter card should be saved using \mmaInlineCell[]{Input}{OutputDirectory->Automatic} (default) which uses the current working directory. Replacing \mmaInlineCell[]{Input}{Automatic} by a path allows for specifying any other location.
    \item 
    \mmaInlineCell[]{Input}{ExportUFO[lagrangian]}
    \\
    determines all tree-level Feynman rules from the Lagrangian~(\mmaInlineCell[]{Input}{lagrangian}) and translates them to the UFO format~\cite{Degrande:2011ua,Darme:2023jdn}. 
    As output, a directory containing the various resulting \textsc{Python} files is generated in the current working directory. Another directory can be provided with the optional argument \mmaInlineCell[]{Input}{OutputDirectory->"PATH"}, where \mmaInlineCell[]{Input}{"PATH"} should be the file path to the desired output location.
    Furthermore, the option \mmaInlineCell[]{Input}{InputFile->"PATH"} can be used to provide the \mmaInlineCell[]{Input}{"PATH"} to a custom parameter card in the JSON format as input.
    If the \mmaInlineCell[]{Input}{InputFile} option is not provided, the default parameter card will be used as template and a user interface dialogue opens that allows for modifications.
    The \mmaInlineCell[]{Input}{ExportUFO} method also takes the options \mmaInlineCell[]{Input}{DefaultValues}, \mmaInlineCell[]{Input}{\mmaDef{FlavorInvariants}}, \mmaInlineCell[]{Input}{Gauge}, and \mmaInlineCell[]{Input}{\mmaDef{R\(\xi\)}} with the same functionality as in \mmaInlineCell[]{Input}{DefaultParamCard}.
    Moreover, the options \mmaInlineCell[]{Input}{\mmaDef{Legs}}, \mmaInlineCell[formatline=\mbox]{Input}{"ExternalMomentum"}, and \mmaInlineCell[]{Input}{"FermionFlow"} are shared with \mmaInlineCell[]{Input}{FeynmanRules} except with default values of the latter two changed to \mmaInlineCell[]{Input}{"ExternalMomentum"\,->\,"outgoing"} and \mmaInlineCell[]{Input}{"FermionFlow"\,->\,"DEHK"} to match \textsc{MadGraph} conventions.
    With the option \mmaInlineCell[]{Input}{FeynmanRules->None} (default), one can also manually provide the Feynman rules rather than having them derived. 
    In this case \mmaInlineCell[]{Input}{None} should be replaced with an association of Feynman rules following the output format of \mmaInlineCell[]{Input}{FeynmanRules[lagrangian]}.
    If the Feynman rules are provided through this option, the consistency of conventions must be ensured manually. 
    The options \mmaInlineCell[]{Input}{\mmaStr{"\(\sigma\)Normalization"}\,->\,}\mmaInlineCell[]{Output}{II/4} (default) and \mmaInlineCell[]{Input}{"IncludeD4Vanishing"\,->\,False} (default) let the user set the normalization of $\sigma^{\mu\nu}$ and whether Dirac structures vanishing in four spacetime dimensions such as $\sigma^{\mu\nu} P_L \otimes \sigma_{\mu\nu} P_R$ should be dropped from the UFO.
    Finally, auxiliary information can be included in the \texttt{\_\_init\_\_.py} file using the option \mmaInlineCell[]{Input}{Information-><|"author"->"Matchete", "model\_version"->"0.0", "arxiv"->"0000.00000"|>}.
\end{itemize}

Example applications of these functions, highlighting the usual workflow, are presented in the following appendix.

\section{Example Workflow: \texorpdfstring{$\tilde{S}_1$}{S1t} and EFT Implementation}
\label{app:examples}

In this Appendix, we provide the details of the implementation of the phenomenological example presented in Sec.~\ref{sec:EFT-convergence} in \matchete. 
A~tutorial notebook with the information provided here is also included in the documentation center and on the website~\cite{MatcheteWebsite} of \matchete.
In a first step, we implement the BSM theory in \matchete and show how the Feynman rules and the UFO files can be generated.
Afterwards, we discuss the matching onto the EFT and the corresponding steps in the EFT scenario.

\subsection{Leptoquark Model}
\label{app:leptoquark}

\subsubsection{Model Definition}
We begin by implementing the $\tilde{S}_1 \sim (\bar{\textbf{3}}, \textbf{1})_{4/3}$ leptoquark model defined in Eq.~\eqref{eq:L}. 
The starting point is given by loading the SM model file with an implementation of EWSB:
\begin{mmaCell}{Input}
  \(\mathcal{L}\)SM = LoadModel["SM+breaking"];
\end{mmaCell}
This command loads the definitions of all SM fields (in the broken and unbroken phases) and all gauge groups and, therefore, provides a convenient starting point for simple SM extensions. 
While the Lagrangian \mmaInlineCell[]{Input}{\(\mathcal{L}\)SM} is given in the unbroken phase, all SM symmetry-breaking definitions are loaded on top to facilitate automatic EWSB, as discussed in Appendix~\ref{app:SSB}.
Next, the BSM fields and interactions can be included in the Lagrangian along with definitions for the $\tilde{S}_1$~field and its couplings:\footnote{It is recommended to enter non-commutative multiplication in the Dirac-algebra space with a \texttt{\textbackslash[CenterDot]} (entered as \texttt{[esc].[esc]}) to ensure future compatibility. This replaces the~\texttt{**} notation used in earlier \matchete versions. }
\begin{mmaCell}{Input}
  DefineField[\mmaUnd{\(\mathcal{S}\)1t}, \mmaDef{Scalar}, Indices -> {Bar[\mmaUnd{SU3c}[fund]]}, Charges -> {\mmaUnd{U1Y}[4/3]},
    Mass\,->\,\{Heavy,\mmaUnd{MS1t}\}, UFO\,->\,<|"mass"->"MS1t"|>, NiceForm\,->\,\{"\mmaStr{\mmaSub{\(\tilde{\mathcal{S}}\)}{1}}","\mmaStr{\mmaSub{M}{\mmaSub{\(\tilde{\text{S}}\)}{1}}}"\}]
\end{mmaCell}
\begin{mmaCell}{Input}
  DefineCoupling[\mmaUnd{\(\kappa\)}, SelfConjugate -> False, Indices -> \{\mmaUnd{Flavor},\mmaUnd{Flavor}\}]
\end{mmaCell}
\begin{mmaCell}{Input}
  \mmaUnd{L\(\mathcal{S}\)1t} =  FreeLag[\mmaUnd{\(\mathcal{S}\)1t}] + PlusHc[\mmaUnd{\(\kappa\)}[r,s] Bar[\mmaDef{CConj}[\mmaUnd{d}[a,r]]] ** \mmaUnd{e}[s] \mmaUnd{\(\mathcal{S}\)1t}[a]];
\end{mmaCell}
This Lagrangian is again written in the unbroken phase.
The mass parameter is designated as \texttt{Heavy}, which subsequently allows the field to be integrated out and the low-energy EFT to be determined with a matching. 
Since the masses of the broken and unbroken phase $\tilde{S}_1$ are identical at tree level, we will use the same coupling for the mass. 
Consequently, all \mmaInlineCell[]{Input}{UFO} information associated with the mass should be added here.
We add a full coupling matrix---providing leptoquark couplings to all fermion generations---and only later restrict to the third-generation coupling case;
the coupling~$\kappa$ from Sec.~\ref{sec:EFT-convergence} corresponds to the \mmaInlineCell[]{Output}{\mmaSup{\(\kappa\)}{33}} entry in this appendix.
The full Lagrangian in the unbroken phase is 
\begin{mmaCell}{Input}
  LBSM = \mmaUnd{LSM} + \mmaUnd{L\(\mathcal{S}\)1t};
\end{mmaCell}

We can now implement the symmetry-breaking pattern for the $\tilde{S}_1$. 
The EWSB relations for the SM are already included when loading the \mmaInlineCell[]{Input}{"SM+breaking"} model and do not have to be re-specified or altered in any way as long as there are no changes due to NP. This is the case in the simple example discussed here.
The details of the SM symmetry-breaking patterns have been discussed in Appendix~\ref{app:SSB}.
We first define the $\tilde{S}_1$ in the broken phase, i.e., with electric charge rather than hypercharge: 
\begin{mmaCell}{Input}
  DefineField[\mmaUnd{S1t}, \mmaDef{Scalar}, Indices -> {Bar[\mmaUnd{SU3c}[fund]]}, Charges -> {\mmaUnd{U1em}[4/3]}, \newline Mass -> \{Heavy, MS1t\}, NiceForm -> \{"\mmaStr{\mmaSub{\(\tilde{\text{S}}\)}{1}}","\mmaStr{\mmaSub{M}{\mmaSub{\(\tilde{\text{S}}\)}{1}}}"\}, \newline UFO -> <|"name"->"S1t","pdg"->50001,"mass"->"MS1t","width"->"WS1t"|>]
\end{mmaCell}
where we included a \mmaInlineCell[]{Input}{"pdg"} code, which is required for all broken-phase fields when exporting UFOs, and further naming information for the export.
Then, we specify the (in this case trivial) field decomposition of the $\tilde{S}_1$ under EWSB (the open color index is shared in both phases):
\begin{mmaCell}{Input}
  FieldDecomposition[\mmaUnd{\(\mathcal{S}\)1t}[a], S1t[a]]
\end{mmaCell}
The Lagrangian in the broken phase can then be obtained using
\begin{mmaCell}{Input}
  LBSMbroken = ImplementVacuumConditions[ToBrokenPhase[LBSM]];
\end{mmaCell}
For illustration, we can inspect the BSM part in the broken phase:
\begin{mmaCell}{Input}
  \mmaUnd{L\(\mathcal{S}\)1tbroken} = ToBrokenPhase[\mmaUnd{L\(\mathcal{S}\)1t}]\,//\,NiceForm
\end{mmaCell}
\begin{mmaCell}{Output}
  \mmaSub{D}{\(\mu\)}\mmaSubSup{\(\overline{\tilde{S}}\)}{1}{a}\,\mmaSub{D}{\(\mu\)}\mmaSub{\(\tilde{S}\)}{1a}\,-\,\mmaDef{\mmaSubSup{\(\mathcal{M}\)}{\mmaSub{\(\tilde{\mathcal{S}}\)}{1}}{2}}\,\mmaSubSup{\(\overline{\tilde{S}}\)}{1}{a}\,\mmaSub{\(\tilde{S}\)}{1a}\,+\,\mmaFrac{16}{9}\mmaFrac{\mmaSup{sw}{2}\,\mmaSup{SCe}{2}}{\mmaSup{cw}{2}}\,\mmaSubSup{\(\overline{\tilde{S}}\)}{1}{a}\,\mmaSub{\(\tilde{S}\)}{1a}\,\mmaSup{\mmaDef{\(\mathcal{Z}\)}}{\(\mu\)2}\,+\,\mmaFrac{4II}{3}\mmaFrac{sw\,SCe}{cw}\,\mmaSub{D}{\(\mu\)}\mmaSubSup{\(\overline{\tilde{S}}\)}{1}{a}\,\mmaSub{\(\tilde{S}\)}{1a}\,\mmaSup{\mmaDef{\(\mathcal{Z}\)}}{\(\mu\)}\,-\,\mmaFrac{4II}{3}\mmaFrac{sw\,SCe}{cw}\,\mmaSubSup{\(\overline{\tilde{S}}\)}{1}{a}\,\mmaSub{D}{\(\mu\)}\mmaSub{\(\tilde{S}\)}{1a}\,\mmaSup{\mmaDef{\(\mathcal{Z}\)}}{\(\mu\)} \newline +\,\mmaSup{\(\overline{\kappa}\)}{pr}\,\mmaSubSup{\(\overline{\tilde{S}}\)}{1}{a}\,(\mmaSubSup{\mmaDef{\(\bar{\mathfrak{d}}\)}}{a}{p}\,\(\cdot\)\,C\,\mmaSub{P}{L}\,\(\cdot\)\,\mmaSup{\mmaDef{\(\bar{\mathfrak{e}}\)}}{r\(\intercal\)})\,+\,\mmaSup{\(\kappa\)}{pr}\,\mmaSub{\(\tilde{S}\)}{1a}\,(\mmaSup{\mmaDef{\(\mathfrak{d}\)}}{ap\(\intercal\)}\,\(\cdot\)\,C\,\mmaSub{P}{R}\,\(\cdot\)\,\mmaSup{\mmaDef{\(\mathfrak{e}\)}}{r})
\end{mmaCell}
where the covariant derivative of $ \tilde{S}_1 $ in the broken phase is $D_\mu \tilde{S}_1 = \left(\partial_\mu + i g_s G^A_\mu T^{A\ast} - ie\tfrac{4}{3} \mathcal{A}_\mu\right) \tilde{S}_1$ in the conventions used for the Feynman rule output.

\subsubsection{Feynman Rules for \slq}
\label{sec: FR slq}
\label{FR BSM}
\begin{figure}[tbp]
    \centering
    \begin{subfigure}[b]{0.3\textwidth}
        \centering
        \includegraphics[width=\textwidth]{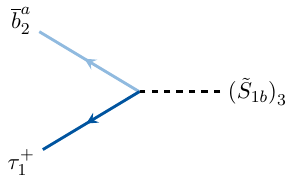}
    \end{subfigure}
    \hfill
    \begin{subfigure}[b]{0.3\textwidth}
        \centering
        \includegraphics[width=\textwidth]{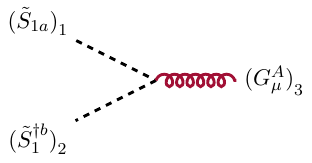}
    \end{subfigure}
    \hfill
    \begin{subfigure}[b]{0.3\textwidth}
        \centering
        \includegraphics[width=\textwidth]{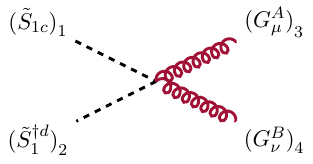}
    \end{subfigure}
    \caption{Feynman diagrams for the relevant leptoquark vertices shown in the \matchete output in Sec.~\ref{sec: FR slq}. From left to right: the Yukawa-type fermion-fermion-scalar vertex, the scalar-scalar-gluon vertex, and the scalar-scalar-gluon-gluon vertex.}
    \label{fig:UV-vertices}
\end{figure}

The Feynman rules of the broken-phase \slq are determined with the call
\begin{mmaCell}{Input}
  FeynmanRules[\mmaUnd{L\(\mathcal{S}\)1tbroken}]\,//\,NiceForm
\end{mmaCell}
\begin{mmaCell}{Output}
  <|\{G,\,S1t,\,\mmaSup{S1t}{\(\ast\)}\} -> (-II\,\mmaSub{g}{s}\,\mmaSubSup{T}{c}{Ab}\,\mmaSubSup{p}{\(\mu\)}{2}\,+\,II\,\mmaSub{g}{s}\,\mmaSubSup{T}{c}{Ab}\,\mmaSubSup{p}{\(\mu\)}{3})\,\mmaStr{\mmaSub{\(\langle\)\mmaSubSup{\(\overline{\tilde{S}}\)}{1}{c}\(\rangle\)}{3}\,\mmaSub{\(\langle\)\mmaSub{\(\tilde{S}\)}{1b}\(\rangle\)}{2}\,\mmaSub{\(\langle\)\mmaSup{G}{A\(\mu\)}\(\rangle\)}{1}},
    \(\ldots\),\,\{S1t,\,goe,\,god\} -> II\,\mmaSup{\(\kappa\)}{rp}\,\mmaSub{\(\delta\)}{bc}\,\mmaStr{\mmaSub{\(\langle\)\mmaSub{\(\tilde{S}\)}{1c}\(\rangle\)}{3}}\,(\mmaSup{\mmaStr{\mmaSub{\(\langle\)\mmaSup{\(\mathfrak{e}\)}{p}\(\rangle\)}{1}}}{\(\intercal\)}\,\(\cdot\)\,C\,\mmaSub{P}{R}\,\(\cdot\)\,\mmaStr{\mmaSub{\(\langle\)\mmaSup{\(\mathfrak{d}\)}{br}\(\rangle\)}{2}}),
    \{\mmaSup{S1t}{\(\ast\)},\,\mmaSup{goe}{\(\ast\)},\,\mmaSup{god}{\(\ast\)}\} -> -II\,\mmaSup{\(\overline{\kappa}\)}{rp}\,\mmaSub{\(\delta\)}{bc}\,\mmaStr{\mmaSub{\(\langle\)\mmaSubSup{\(\overline{\tilde{S}}\)}{1}{c}\(\rangle\)}{3}}\,(\mmaStr{\mmaSub{\(\langle\)\mmaSubSup{\(\overline{\mathfrak{d}}\)}{b}{r}\(\rangle\)}{2}}\,\(\cdot\)\,C\,\mmaSub{P}{L}\,\(\cdot\)\,\mmaSup{\mmaStr{\mmaSub{\(\langle\)\mmaSup{\(\overline{\mathfrak{e}}\)}{p}\(\rangle\)}{1}}}{\(\intercal\)}),
    \{G,\,G,\,S1t,\,\mmaSup{S1t}{\(\ast\)}\} -> (II\,\mmaSubSup{g}{s}{2}\,\mmaSub{g}{\(\mu\nu\)}\,\mmaSubSup{T}{a}{Ac}\,\mmaSubSup{T}{d}{Ba}\,+\,II\,\mmaSubSup{g}{s}{2}\,\mmaSub{g}{\(\mu\nu\)}\,\mmaSubSup{T}{d}{Aa}\,\mmaSubSup{T}{a}{Bc})\,\mmaStr{\mmaSub{\(\langle\)\mmaSubSup{\(\overline{\tilde{S}}\)}{1}{d}\(\rangle\)}{4}\,\mmaSub{\(\langle\)\mmaSub{\(\tilde{S}\)}{1c}\(\rangle\)}{3}\,\mmaSub{\(\langle\)\mmaSup{G}{A\(\mu\)}\(\rangle\)}{1}\,\mmaSub{\(\langle\)\mmaSup{G}{B\(\nu\)}\(\rangle\)}{2}},\,\(\ldots\)|>
  
\end{mmaCell}
where for brevity we do not show the Feynman rules for interactions with the $Z$ boson and photon. 
The Feynman rules are provided in the form of an association where every key is a list of the external fields for one of the vertices.
The association values are the Feynman rules of the corresponding vertices, where the external fields are shown in gray brackets with subscripts which uniquely label the fields within each vertex.
For illustration, we also visualize the relevant vertices from the above output of \matchete in Fig.~\ref{fig:UV-vertices}.

\subsubsection{Coupling Orders, Parameter Card, Flavor Symmetries, and UFO Export}
\label{app:ParamCardBSM}

The parameter card for the $\tilde{S}_1$ model mostly overlaps with that of the~SM. 
Only three additional parameters need to be included, that is, the mass, width, and Yukawa coupling matrix of the~$\tilde{S}_1$.
A template for the parameter card is obtained using
\begin{mmaCell}{Input}
  DefaultParamCard[LBSMbroken, DefaultValues\,->\,True, Gauge\,->\,\mmaDef{R}\(\xi\), \mmaDef{R}\(\xi\)\,->\,1];
\end{mmaCell}
\begin{mmaCell}{Print}
  Gauge fixing using Feynman gauge. Note: this UFO will still be usable in 
  unitary gauge. For MadGraph use 'set gauge unitary' to run in unitary gauge 
  (default) and 'set gauge Feynman' to actually use Feynman gauge.
  
  The following SM input relations have been determined:
  \Big\{\mmaSub{g}{s}\,->\,\mmaSqrt{4\,\(\pi\)\,aS}, SCe\,->\,\mmaSqrt{4\,\(\pi\)\,aEM}, v\,->\,\mmaFrac{1}{\mmaSqrt{Gf\,\mmaSqrt{2}}}, \(\lambda\)\,->\,Gf\,\mmaSqrt{2}\,\mmaSup{Mh}{2},
   sw\,->\,\mmaFrac{1}{2}\,\mmaSqrt{4\,\(\pi\)\,aEM}\,\mmaFrac{1}{MW}\,\mmaFrac{1}{\mmaSqrt{Gf\,\mmaSqrt{2}}}, cw\,->\,\mmaSqrt{1\,-\,\mmaSup{sw}{2}}\Big\}
  
\end{mmaCell}
This generates a \textsc{JSON} file in the current working directory containing the parameter card.
The option \mmaInlineCell[]{Input}{DefaultValues\,->\,True} is used to automatically determine the relation between SM input observables and the model parameters. 
The relations determined are also printed in the notebook.
This feature is limited to BSM theories that do not modify the EWSB pattern of the SM and can, thus, not be applied to the SMEFT and other more complex scenarios.
The option \mmaInlineCell[]{Input}{Gauge\,->\,\mmaDef{R}\(\xi\)} indicates that the parameter card should include the gauge parameters.
Alternatively, one can also gauge fix the Lagrangian in advance by setting \mmaInlineCell[]{Input}{LBSMbroken\,=\,GaugeFixLagrangian[LBSMbroken,\,Gauge\,->\,\mmaDef{R}\(\xi\)];}.\cprotect\footnote{If this is done, one should use the option \mmaInlineCell[]{Input}{Gauge\,->\,Automatic} (which is the default value) when calling \mmaInlineCell[]{Input}{DefaultParamCard}. This will automatically detect the gauge of the provided Lagrangian.} 
Moreover, the option \mmaInlineCell[]{Input}{\mmaDef{R}\(\xi\)\,->\,1} (this is also the default value) is used to restrict the general $R_\xi$~gauge to the Feynman gauge, i.e., all gauge parameters are fixed to unity before generating the parameter card and before exporting the UFO files.
Instead, one can use \mmaInlineCell[]{Input}{\mmaDef{R}\(\xi\)\,->\,General} to keep symbolic gauge-fixing parameters in the UFO. Other values are not supported.
Note that the \mmaInlineCell[]{Input}{Gauge} option is a native \matchete option and hence only supports the general $R_\xi$~gauge and unitary gauge. The restriction to Feynman gauge can be imposed only during the export through the functions \mmaInlineCell[]{Input}{DefaultParamCard} and \mmaInlineCell[]{Input}{ExportUFO}.

\begin{figure}[tbp]
    \centering
    \vspace*{-1cm}
    \begin{minipage}{0.8\textwidth}
    \begin{lstlisting}[language=json,firstnumber=1]
{
    "parameters" : {
      ...
    "MS1t" : {
      "name"     : "MS1t",
      "nature"   : "external",
      "type"     : "real",
      "value"    : 0.0,
      "texname"  : "MS1t",
      "lhablock" : "MASS",
      "lhacode"  : [ 50001 ]
      },
    "WS1t" : {
      "name"     : "WS1t",
      "nature"   : "external",
      "type"     : "real",
      "value"    : 0.0,
      "texname"  : "WS1t",
      "lhablock" : "DECAY",
      "lhacode"  : [ 50001 ]
    },
    "Re_Kappa_33" : {
      "name"     : "Re_Kappa_33",
      "nature"   : "external",
      "type"     : "real",
      "value"    : "0",
      "texname"  : "Re_Kappa_33",
      "lhablock" : "COUPLINGS",
      "lhacode"  : 1025
    },
    "Im_Kappa_33" : {
      "name"     : "Im_Kappa_33",
      "nature"   : "external",
      "type"     : "real",
      "value"    : "0",
      "texname"  : "Im_Kappa_33",
      "lhablock" : "COUPLINGS",
      "lhacode"  : 1026
    },
    "Kappa_33" : {
      "name"     :"Kappa_33",
      "nature"   :"internal",
      "type"     :"complex",
      "value"    :"complex(0,1)*Im_Kappa_33+Re_Kappa_33",
      "texname"  :"Kappa_33"
    },
    ...
  }
}
    \end{lstlisting}
    \end{minipage}
    \cprotect\caption{BSM parameters included in the template parameter card JSON file obtained by \mmaInlineCell[]{Input}{DefaultParamCard} for the $\tilde{S}_1$ leptoquark model. The parameters shown here are the mass (\lstinline[columns=fixed,language=json]{"MS1t"}), width~(\lstinline[columns=fixed,language=json]{"WS1t"}), and the third-generation coupling~(\lstinline[columns=fixed,language=json]{"Kappa_33"}) of $\tilde{S}_1$. The latter is parametrized in terms of its real~(\lstinline[columns=fixed,language=json]{"Re_Kappa_33"}) and imaginary~(\lstinline[columns=fixed,language=json]{"Im_Kappa_33"}) parts. The other flavor combinations for \mmaInlineCell[]{Input}{\mmaUnd{\(\kappa\)}} (if no flavor symmetry is specified) and the SM parameters are also present in the parameter card but not shown here.}
    \label{fig:jsonLQ}
\end{figure}

A~snapshot of the parameter card is displayed in Fig.~\ref{fig:jsonLQ}, showing only the entries for the mass, width, and coupling to third-generation fermions of the $\tilde{S}_1$ leptoquark. 
The couplings of $\tilde{S}_1$ to the other flavors are included in the generated parameter card but omitted here for brevity.
By default all BSM parameters are initialized with a vanishing value, whereas the SM parameters included in the parameter card are assigned the values shown above when the option \mmaInlineCell[]{Input}{DefaultValues\,->\,True} is used (as is the case here). 
Otherwise, they are also initialized with vanishing values.
Complex couplings, such as (\lstinline[columns=fixed,language=json]{"Kappa_33"}), are parametrized in terms of their real~(\lstinline[columns=fixed,language=json]{"Re_Kappa_33"}) and imaginary~(\lstinline[columns=fixed,language=json]{"Im_Kappa_33"}) parts, since external parameters must be real in the UFO format.

In principle one can now proceed to generate the UFO files.
However, it is often convenient to only include the parameters required for a specific analysis in the UFO file and to remove all unused parameters. 
For our EFT validity study in Sec.~\ref{sec:EFT-convergence}, that amounts to dropping all couplings of the $\tilde{S}_1$~leptoquark to fermions other than the third generation. 
This can also be achieved in the parameter card by setting the value of the corresponding coupling (\lstinline[columns=fixed,language=json]{"Kappa_11"}, \lstinline[columns=fixed,language=json]{"Kappa_12"}, \ldots) to \lstinline[columns=fixed,language=json]{"value":"ZERO"}. 
This will entirely remove these parameters from the model during the UFO generation.

An alternative to eliminating the unnecessary parameters is provided by employing flavor symmetries.
In the example at hand, we can make use of a $\mathrm{U}(2)_e \times \mathrm{U}(2)_d$ flavor symmetry under which the two light generations of right-handed leptons and right-handed down-type quarks transform as doublets, whereas the third generation remains a singlet.\footnote{One could also define a full $\mathrm{U}(2)^5$ symmetry here which would produce the same result. We use $\mathrm{U}(2)^2$ only for brevity.}
This effectively forbids all couplings of $\tilde{S}_1$ to fermions except for the third generation.
This flavor symmetry can be defined in \matchete using:
\begin{mmaCell}{Input}
  DefineGlobalGroup[SU2d, SU[2]];    DefineGlobalGroup[U1d, U1];
  DefineGlobalGroup[SU2e, SU[2]];    DefineGlobalGroup[U1e, U1];

  flavorGroup = <|
    e -> \{\{1,2\} -> \{SU2e[fund], U1e[1]\}\},
    d -> \{\{1,2\} -> \{SU2d[fund], U1d[1]\}\}|>;
   
\end{mmaCell}
where we first defined the flavor symmetry as global groups\footnote{\matchete exclusively handles simple or Abelian continuous symmetry groups. Therefore, we loosely identify $\mathrm{U}(N)\sim\mathrm{SU}(N)\times\mathrm{U}(1)$.} before writing the transformation properties for the generations of the fields as an association.
Specifying the third generation to be a singlet (\mmaInlineCell[]{Input}{\{3\}\,->\,Singlet}) is not required.
This flavor symmetry can now be assigned to the Lagrangian, more concretely to the coupling~\mmaInlineCell[]{Input}{\mmaUnd{\(\kappa\)}}, using:
\begin{mmaCell}{Input}
  flavorInvariants\,=\,ImposeFlavorSymmetry[LBSM,\,flavorGroup,\,Except->\{Yu,Yd,Ye\}];
\end{mmaCell}
Excluding the Yukawa couplings from the flavor-symmetry assumptions is not strictly necessary here, since they are in any case replaced by the fermion masses and thus do not show up in the broken-phase Lagrangian.
The invariants under this flavor symmetry can be inspected using 
\begin{mmaCell}{Input}
  flavorInvariants[\mmaUnd{\(\kappa\)}][\!\![1,1]\!\!] // ArrayRules
\end{mmaCell}
\begin{mmaCell}{Output}
  \{\{3,3\}->1, \{_,_\}->0\}
\end{mmaCell}
which returns the \mmaInlineCell[]{Input}{ArrayRules} of a \mmaInlineCell[]{Input}{SparseArray} representing the matrix
\begin{align}
    \kappa^{pr} &= \begin{pmatrix}
        0 & 0 & 0 \\
        0 & 0 & 0 \\
        0 & 0 & 1
    \end{pmatrix}^{\!\!pr}.
\end{align}
The flavor invariant can now be employed for the parametrization of the coupling~\mmaInlineCell[]{Input}{\mmaUnd{\(\kappa\)}} in the parameter card.\cprotect\footnote{\mmaInlineCell[]{Input}{\mmaUnd{\(\kappa\)}} is generally a complex parameter and should be defined as such in \matchete for the flavor-symmetry module to function properly. That is, one cannot directly define~\mmaInlineCell[]{Input}{\mmaUnd{\(\kappa\)}} as a real coupling. This restriction has to be set through the parameter card if desired.}
This is achieved by running \mmaInlineCell[]{Input}{DefaultParamCard} with the additional option \mmaInlineCell[]{Input}{\mmaDef{FlavorInvariants}}:
\begin{mmaCell}{Input}
  DefaultParamCard[LBSMbroken, \mmaDef{FlavorInvariants}\,->\,flavorInvariants, 
    DefaultValues\,->\,True, Gauge\,->\,\mmaDef{R}\(\xi\), \mmaDef{R}\(\xi\)\,->\,1];
\end{mmaCell}
\begin{mmaCell}{Print}
  Gauge fixing using Feynman gauge. Note: this UFO will still be usable in 
  unitary gauge. For MadGraph use 'set gauge unitary' to run in unitary gauge 
  (default) and 'set gauge Feynman' to actually use Feynman gauge.
  
  The following SM input relations have been determined:
  \Big\{\mmaSub{g}{s}\,->\,\mmaSqrt{4\,\(\pi\)\,aS}, SCe\,->\,\mmaSqrt{4\,\(\pi\)\,aEM}, v\,->\,\mmaFrac{1}{\mmaSqrt{Gf\,\mmaSqrt{2}}}, \(\lambda\)\,->\,Gf\,\mmaSqrt{2}\,\mmaSup{Mh}{2},
   sw\,->\,\mmaFrac{1}{2}\,\mmaSqrt{4\,\(\pi\)\,aEM}\,\mmaFrac{1}{MW}\,\mmaFrac{1}{\mmaSqrt{Gf\,\mmaSqrt{2}}}, cw\,->\,\mmaSqrt{1\,-\,\mmaSup{sw}{2}}\Big\}
  
\end{mmaCell}
The resulting parameter card has all components of the $ \kappa $~matrix set to \lstinline[columns=fixed,language=json]{"value":"ZERO"} except for \lstinline[columns=fixed,language=json]{"Kappa_33"} to remove them from the UFO files during the export in the next stage.

Before generating the UFO files it is necessary to define coupling orders to control which contributions should be considered in simulations. 
This can be done by defining
\begin{mmaCell}{Input}
  DefineCouplingOrder["QED", \{\{v,-1\}, \mmaUnd{M\(\mathfrak{d}\)}, \mmaUnd{M\(\mathfrak{u}\)}, \mmaUnd{M\(\mathfrak{e}\)}, \mmaUnd{\(e\)}, \{\mmaUnd{\(\lambda\)},2\}\}, 2]
  DefineCouplingOrder["QCD", gs, 1]
  DefineCouplingOrder["NP", \mmaUnd{\(\kappa\)}, 1]
  
\end{mmaCell}
where fermion masses~(\mmaInlineCell[]{Input}{\mmaUnd{M\(\mathfrak{d}\)}}, \mmaInlineCell[]{Input}{\mmaUnd{M\(\mathfrak{u}\)}}, \mmaInlineCell[]{Input}{\mmaUnd{M\(\mathfrak{e}\)}}) and QED gauge coupling (\mmaInlineCell[]{Input}{\mmaUnd{\(e\)}}) are conventionally assigned to be order one (default) under the \mmaInlineCell[]{Input}{"QED"} counting, while the Higgs VEV~(\mmaInlineCell[]{Input}{v}) is order minus one and the Higgs quartic~(\mmaInlineCell[]{Input}{\mmaUnd{\(\lambda\)}}) is assigned order two.
The strong gauge coupling is given order one under the \mmaInlineCell[]{Input}{"QCD"} counting, while the $\tilde{S}_1$ coupling~(\mmaInlineCell[]{Input}{\mmaUnd{\(\kappa\)}}) to fermions is assigned order one under a new counting \mmaInlineCell[]{Input}{"NP"}.
The hierarchy between the couplings is such that two insertions of the \mmaInlineCell[]{Input}{"QCD"} or \mmaInlineCell[]{Input}{"NP"} couplings are counted as the same order as one insertion of a \mmaInlineCell[]{Input}{"QED"} coupling. 

The UFO files can now be generated using:
\begin{mmaCell}{Input}
  ExportUFO[LBSMbroken, Gauge\,->\,\mmaDef{R}\(\xi\), \mmaDef{R}\(\xi\)\,->\,1,
    InputFile\,->\,FileNameJoin[\{Directory[],\,"model_parameters.json"\}],
    OutputDirectory\,->\,FileNameJoin[\{Directory[],\,"UFO_S1t"\}] 
  ];
\end{mmaCell}
\begin{mmaCell}{Print}
  Gauge fixing using Feynman gauge. Note: this UFO will still be usable in 
  unitary gauge. For MadGraph use 'set gauge unitary' to run in unitary gauge 
  (default) and 'set gauge Feynman' to actually use Feynman gauge.
\end{mmaCell}
which uses the previously generated parameter card file (named \mmaInlineCell[]{Input}{"model_parameters.json"}) placed in the current working directory.
The resulting UFO files are stored in the new directory \mmaInlineCell[formatline=\mmanobreak]{Input}{"UFO_S1t"} under the current working directory, as specified in the input.
We have specified the same gauge (\mmaInlineCell[]{Input}{Gauge\,->\,\mmaDef{R}\(\xi\)} and \mmaInlineCell[]{Input}{\mmaDef{R}\(\xi\)\,->\,1}) as in the call of \mmaInlineCell[]{Input}{DefaultParamCard}.

\subsection{Leptoquark EFT}
\label{app:examples-LQEFT}

After generating the UFO file for the full $\tilde{S}_1$ leptoquark model in the previous section, we proceed to obtain the UFO for its corresponding low-energy EFT description.

\subsubsection{Matching}
Instead of defining the EFT by hand, we can start from the Lagrangian of the full $\tilde{S}_1$ model in the unbroken electroweak phase and use the \matchete core functionality to integrate out the heavy BSM scalar.
This can be achieved with
\begin{mmaCell}{Input}
  LEFT6 = Match[LBSM, EFTOrder\,->\,6, LoopOrder\,->\,0];
  LEFT6 - LSM // NiceForm
\end{mmaCell}
\begin{mmaCell}{Output}
  \mmaFrac{1}{\mmaSubSup{SCM}{\mmaSub{\(\tilde{S}\)}{1}}{2}}\,\mmaSup{\(\overline{\kappa}\)}{pr}\,\mmaSup{\(\kappa\)}{st}\,(\mmaSubSup{\(\overline{d}\)}{a}{p}\,\(\cdot\)\,C\,\mmaSub{P}{L}\,\(\cdot\)\,\mmaSup{\(\overline{e}\)}{r\(\intercal\)})\,(\mmaSup{d}{as\(\intercal\)}\,\(\cdot\)\,C\,\mmaSub{P}{R}\,\(\cdot\)\,\mmaSup{e}{t})
\end{mmaCell}
which gives the dimension-six tree-level EFT Lagrangian. 
Due to the leptoquark structure, the one resulting operator is not in the form of the Warsaw basis. 
For convenience, and in order to remove the charge conjugation matrices, we employ a four-dimensional Fierz identity to bring this operator to the desired form:
\begin{mmaCell}{Input}
  LEFT6 = GreensSimplify[LEFT6, \mmaDef{ReductionIdentities}\,->\,\mmaDef{FourDimensional}];
  LEFT6 - LSM // NiceForm
\end{mmaCell}
\begin{mmaCell}{Output}
  \mmaFrac{1}{\mmaSubSup{2\,SCM}{\mmaSub{\(\tilde{S}\)}{1}}{2}}\,\mmaSup{\(\overline{\kappa}\)}{pr}\,\mmaSup{\(\kappa\)}{st}\,(\mmaSubSup{\(\overline{d}\)}{a}{p}\,\(\cdot\)\,\mmaSub{\(\gamma\)}{\(\mu\)}\,\mmaSub{P}{R}\,\(\cdot\)\,\mmaSup{d}{as})\,(\mmaSup{\(\overline{e}\)}{r}\,\(\cdot\)\,\mmaSub{\(\gamma\)}{\(\mu\)}\,\mmaSub{P}{R}\,\(\cdot\)\,\mmaSup{e}{t})
\end{mmaCell}

For the phenomenological example in Sec.~\ref{sec:EFT-convergence}, we considered also dimension-eight effects. 
The EFT Lagrangian up to mass-dimension eight is obtained as above by changing the option \mmaInlineCell[]{Input}{EFTOrder\,->\,8} for the call of the \mmaInlineCell[]{Input}{Match} function:
\begin{mmaCell}{Input}
  LEFT8 = EOMSimplify[Match[LBSM, EFTOrder\,->\,8, LoopOrder\,->\,0],
    \mmaDef{ReductionIdentities}\,->\,\mmaDef{FourDimensional}]/. Yd|Ye\,->\,0;
\end{mmaCell}
We also apply \mmaInlineCell[]{Input}{EOMSimplify} to remove redundant operator structures at~$d=8$, and we drop the subleading terms proportional to the down-type quark and charged-lepton Yukawas. 
Note that the resulting minimal operator set is not identical to the operators in the Murphy basis~\cite{Murphy:2020rsh}.

Since there are no additional particles or structures incorporating the Higgs doublet in this EFT Lagrangian, the SM EWSB routine from Appendix~\ref{app:SSB}, which is included in the previously loaded model file~\mmaInlineCell[]{Input}{"SM+breaking"}, can be fully applied here without any additions or complications.
Thus, the broken-phase Lagrangian is obtained by:
\begin{mmaCell}{Input}
  LEFT8broken = ImplementVacuumConditions[ToBrokenPhase[LEFT8]];
\end{mmaCell}

\subsubsection{Feynman Rules for the EFT up to Dimension Eight}
\label{FR EFT6}

The Feynman rules of the dimension-eight Lagrangian can then be determined using:
\begin{mmaCell}{Input}
  FeynmanRules[LEFT8broken,\,\{\mmaUnd{\(\mathfrak{e}\)},\,\mmaDef{Conj}[\mmaUnd{\(\mathfrak{e}\)}],\,\mmaUnd{\(\mathfrak{d}\)},\,\mmaDef{Conj}[\mmaUnd{\(\mathfrak{d}\)}]\}]\,//\,FullSimplify\,//\,NiceForm
\end{mmaCell}
\begin{mmaCell}{Output}
  \mmaFrac{II}{2}\,\mmaFrac{1}{\mmaSubSup{SCM}{\mmaSub{\(\tilde{S}\)}{1}}{4}}\,\mmaSup{\(\overline{\kappa}\)}{tr}\,\mmaSup{\(\kappa\)}{sp}\,\mmaSub{\(\delta\)}{cd}\bigg(\big(\mmaStr{\mmaSub{\(\langle\)\mmaSup{\(\overline{\mathfrak{e}}\)}{r}\(\rangle\)}{2}}\,\(\cdot\)\,\mmaSub{\(\gamma\)}{\(\mu\)}\,\mmaSub{P}{R}\,\(\cdot\)\,\mmaStr{\mmaSub{\(\langle\)\mmaSup{\(\mathfrak{e}\)}{p}\(\rangle\)}{1}}\big)
    \Big(\mmaSubSup{SCM}{\mmaSub{\(\tilde{S}\)}{1}}{2}\mmaStr{\mmaSub{\(\langle\)\mmaSubSup{\(\overline{\mathfrak{d}}\)}{d}{t}\(\rangle\)}{4}}\,\(\cdot\)\,\mmaSub{\(\gamma\)}{\(\mu\)}\,\mmaSub{P}{R}\,\(\cdot\)\,\mmaStr{\mmaSub{\(\langle\)\mmaSup{\(\mathfrak{d}\)}{cs}\(\rangle\)}{3}}\,+\,\big(-\,\mmaSubSup{p}{\(\mu\)}{1}\mmaSubSup{p}{\(\nu\)}{4}\,+\,\mmaSubSup{p}{\(\mu\)}{4}\mmaSubSup{p}{\(\nu\)}{2}\big)\big(\mmaStr{\mmaSub{\(\langle\)\mmaSubSup{\(\overline{\mathfrak{d}}\)}{d}{t}\(\rangle\)}{4}}\,\(\cdot\)\,\mmaSub{\(\gamma\)}{\(\nu\)}\,\mmaSub{P}{R}\,\(\cdot\)\,\mmaStr{\mmaSub{\(\langle\)\mmaSup{\(\mathfrak{d}\)}{cs}\(\rangle\)}{3}}\big)\Big)
    -\,\big(\mmaStr{\mmaSub{\(\langle\)\mmaSup{\(\overline{\mathfrak{e}}\)}{r}\(\rangle\)}{2}}\,\(\cdot\)\,\mmaSub{\(\gamma\)}{\(\nu\)}\,\mmaSub{P}{R}\,\(\cdot\)\,\mmaStr{\mmaSub{\(\langle\)\mmaSup{\(\mathfrak{e}\)}{p}\(\rangle\)}{1}}\big)\Big(\big(-\,\mmaSubSup{p}{\(\mu\)}{1}\mmaSubSup{p}{\(\nu\)}{3}\,+\,\mmaSubSup{p}{\(\mu\)}{3}\mmaSubSup{p}{\(\nu\)}{1}\,+\,\mmaSubSup{p}{\(\nu\)}{2}(\mmaSubSup{p}{\(\mu\)}{3}\,+\,\mmaSubSup{p}{\(\mu\)}{4})\big)\big(\mmaStr{\mmaSub{\(\langle\)\mmaSubSup{\(\overline{\mathfrak{d}}\)}{d}{t}\(\rangle\)}{4}}\,\(\cdot\)\,\mmaSub{\(\gamma\)}{\(\mu\)}\,\mmaSub{P}{R}\,\(\cdot\)\,\mmaStr{\mmaSub{\(\langle\)\mmaSup{\(\mathfrak{d}\)}{cs}\(\rangle\)}{3}}\big)
      +\,\big(\mmaSubSup{p}{\(\mu\)}{3}\mmaSubSup{p}{\(\mu\)}{4}\,+\,\mmaSubSup{p}{\(\mu\)}{1}\mmaSubSup{p}{\(\mu\)}{2}\big)\big(\mmaStr{\mmaSub{\(\langle\)\mmaSubSup{\(\overline{\mathfrak{d}}\)}{d}{t}\(\rangle\)}{4}}\,\(\cdot\)\,\mmaSub{\(\gamma\)}{\(\nu\)}\,\mmaSub{P}{R}\,\(\cdot\)\,\mmaStr{\mmaSub{\(\langle\)\mmaSup{\(\mathfrak{d}\)}{cs}\(\rangle\)}{3}}\big)\Big)\bigg)
  
\end{mmaCell}

\begin{mmaCell}{Input}
  FeynmanRules[LEFT8broken,\,\{\mmaUnd{G},\,\mmaUnd{\(\mathfrak{e}\)},\,\mmaDef{Conj}[\mmaUnd{\(\mathfrak{e}\)}],\,\mmaUnd{\(\mathfrak{d}\)},\,\mmaDef{Conj}[\mmaUnd{\(\mathfrak{d}\)}]\}]\,//\,FullSimplify\,//\,NiceForm
\end{mmaCell}
\begin{mmaCell}{Output}
  \mmaFrac{II}{2}\,\mmaFrac{\mmaSub{g}{s}}{\mmaSubSup{SCM}{\mmaSub{\(\tilde{S}\)}{1}}{4}}\,\mmaSup{\(\overline{\kappa}\)}{tr}\,\mmaSup{\(\kappa\)}{sp}\,\mmaSubSup{T}{c}{Ed}\,\mmaStr{\mmaSub{\(\langle\)\mmaSup{G}{E\(\kappa\)}\(\rangle\)}{5}}\big(\mmaStr{\mmaSub{\(\langle\)\mmaSubSup{\(\overline{\mathfrak{d}}\)}{d}{t}\(\rangle\)}{4}}\,\(\cdot\)\,\mmaSub{\(\gamma\)}{\(\mu\)}\,\mmaSub{P}{R}\,\(\cdot\)\,\mmaStr{\mmaSub{\(\langle\)\mmaSup{\(\mathfrak{d}\)}{cs}\(\rangle\)}{3}}\big)\Big(\big(\mmaSubSup{p}{\(\kappa\)}{3}\,-\,\mmaSubSup{p}{\(\kappa\)}{4}\big)\big(\mmaStr{\mmaSub{\(\langle\)\mmaSup{\(\overline{\mathfrak{e}}\)}{r}\(\rangle\)}{2}}\,\(\cdot\)\,\mmaSup{\(\gamma\)}{\(\mu\)}\,\mmaSub{P}{R}\,\(\cdot\)\,\mmaStr{\mmaSub{\(\langle\)\mmaSup{\(\mathfrak{e}\)}{p}\(\rangle\)}{1}}\big)
    +\,\big(\mmaSubSup{p}{\(\mu\)}{1}\,-\,\mmaSubSup{p}{\(\mu\)}{2}\big)\big(\mmaStr{\mmaSub{\(\langle\)\mmaSup{\(\overline{\mathfrak{e}}\)}{r}\(\rangle\)}{2}}\,\(\cdot\)\,\mmaSup{\(\gamma\)}{\(\kappa\)}\,\mmaSub{P}{R}\,\(\cdot\)\,\mmaStr{\mmaSub{\(\langle\)\mmaSup{\(\mathfrak{e}\)}{p}\(\rangle\)}{1}}\big)\Big)
    
\end{mmaCell}

\begin{figure}[tbp]
    \centering
    \begin{subfigure}[c]{0.3\textwidth}
        \centering
        \includegraphics[width=\textwidth]{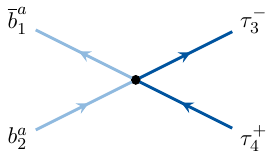}
    \end{subfigure}
    \hfill
    \begin{subfigure}[c]{0.3\textwidth}
        \centering
        \includegraphics[width=\textwidth]{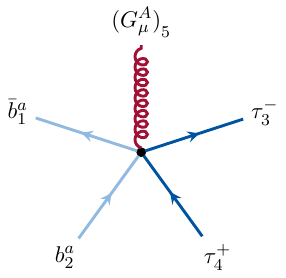}
    \end{subfigure}
    \hfill
    \begin{subfigure}[c]{0.3\textwidth}
        \centering
        \includegraphics[width=\textwidth]{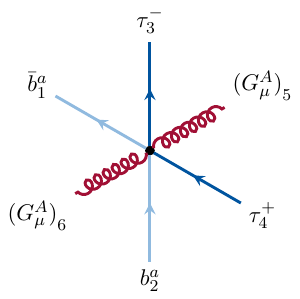}
    \end{subfigure}
    \caption{Feynman diagrams for the relevant vertices generated by the \matchete output above: (left) the dimension-six four-fermion vertices, and (center, right) the dimension-eight four-fermion vertices with one and two additional gluons, respectively.}
    \label{fig:EFT-vertices}
\end{figure}

\begin{mmaCell}{Input}
  FeynmanRules[LEFT8broken,\,\{\mmaUnd{G},\mmaUnd{G},\mmaUnd{\(\mathfrak{e}\)},\mmaDef{Conj}[\mmaUnd{\(\mathfrak{e}\)}],\mmaUnd{\(\mathfrak{d}\)},\mmaDef{Conj}[\mmaUnd{\(\mathfrak{d}\)}]\}]//FullSimplify//NiceForm
\end{mmaCell}
\begin{mmaCell}{Output}
  \mmaFrac{II}{2}\,\mmaFrac{\mmaSubSup{g}{s}{2}}{\mmaSubSup{SCM}{\mmaSub{\(\tilde{S}\)}{1}}{4}}\,\mmaSup{\(\overline{\kappa}\)}{tr}\,\mmaSup{\(\kappa\)}{sp}\,\mmaSub{g}{\(\kappa\)\mmaSub{\(\mu\)}{1}}\big(\mmaSubSup{T}{a}{Ed}\,\mmaSubSup{T}{c}{Fa}\,+\,\mmaSubSup{T}{c}{Ea}\,\mmaSubSup{T}{a}{Fd}\big)\mmaStr{\mmaSub{\(\langle\)\mmaSup{G}{E\(\kappa\)}\(\rangle\)}{5}}\,\mmaStr{\mmaSub{\(\langle\)\mmaSup{G}{F\mmaSub{\(\mu\)}{1}}\(\rangle\)}{6}}{\big(\mmaStr{\mmaSub{\(\langle\)\mmaSubSup{\(\overline{\mathfrak{d}}\)}{d}{t}\(\rangle\)}{4}}\,\(\cdot\)\,\mmaSub{\(\gamma\)}{\(\mu\)}\,\mmaSub{P}{R}\,\(\cdot\)\,\mmaStr{\mmaSub{\(\langle\)\mmaSup{\(\mathfrak{d}\)}{cs}\(\rangle\)}{3}}\big)}{\big(\mmaStr{\mmaSub{\(\langle\)\mmaSup{\(\overline{\mathfrak{e}}\)}{r}\(\rangle\)}{2}}\,\(\cdot\)\,\mmaSup{\(\gamma\)}{\(\mu\)}\,\mmaSub{P}{R}\,\(\cdot\)\,\mmaStr{\mmaSub{\(\langle\)\mmaSup{\(\mathfrak{e}\)}{p}\(\rangle\)}{1}}\big)}
    
\end{mmaCell}
Instead of \mmaInlineCell[]{Input}{\mmaDef{Conj}} one can also use \mmaInlineCell[]{Input}{Bar} to indicate the conjugate external legs.
We show only the three Feynman rules relevant for the phenomenological example of Sec.~\ref{sec:EFT-convergence} for brevity.
A graphical visualization of the three corresponding interaction vertices in the dimension-eight EFT is found in Fig.~\ref{fig:EFT-vertices}.

\subsubsection{Parameter Card and UFO Export}
Obtaining the parameter card works completely analogously to the case of the full BSM theory discussed in Appendix~\ref{app:ParamCardBSM}.
Since we express the Wilson coefficients of the EFT operators still in terms of the leptoquark mass $\mathcal{M}_{\tilde{S}_1}$ and coupling $\kappa$, we can actually reuse the very same parameter card from the full model scenario (see Fig.~\ref{fig:jsonLQ}). 
However, the default value for the leptoquark mass (\lstinline[columns=fixed,language=json]{"MS1t"}) has to be set to a non-vanishing value, e.g., \lstinline[columns=fixed,language=json]{"value":1.0}, to avoid divergences.
The width of the $\tilde{S}_1$ is no longer required and will be ignored when reading in the \textsc{JSON} file.\cprotect\footnote{Another difference is that the leptoquark mass is assigned \lstinline[columns=fixed,language=json]{"lhablock":"MASS"} when running \mmaInlineCell[]{Input}{DefaultParamCard} on the BSM Lagrangian and \lstinline[columns=fixed,language=json]{"lhablock":"COUPLINGS"} when deriving the EFT parameter card.}

Moreover, the definition of flavor symmetries (using \mmaInlineCell[]{Input}{ImposeFlavorSymmetry}) and the generation of the UFO files (with \mmaInlineCell[]{Input}{ExportUFO}) work analogously to the full model scenario in Appendix~\ref{app:ParamCardBSM}.

\section{SMEFT Operator Definitions}
\label{app:operator-defs}
The various SMEFT implementations in \matchete provided with this work contain all operators of the Warsaw, Mainz, or Murphy bases and we refer to Refs.~\cite{Grzadkowski:2010es,MainzBasis,Murphy:2020rsh} (see Sec.~\ref{sec:d=8_UFO} for modifications of the $d=8$ operator basis) for the definitions of the operators.
For convenience, we also collected all operators explicitly mentioned in this manuscript in Tables~\ref{tab:Warsaw-ops} and~\ref{tab:Murphy-ops}.




\begin{table}[tb]
    \renewcommand{\arraystretch}{1.5}
    \centering
    \scalebox{0.85}{
        $
        \begin{array}[t]{|lc||lc||lc|}
            \hline
            \multicolumn{2}{|c||}{\text{Class $\boldsymbol{H^6}$}} &
            \multicolumn{2}{c||}{\text{Class $\boldsymbol{X^2H^2}$}} &
            \multicolumn{2}{c|}{\text{Class $\boldsymbol{\psi^2H^2D}$}}
            \\ \hline
            Q_H & (H^\dagger H)^3 &
            Q_{HG} & (H^\dagger H) G_{\mu\nu}^A G^{A\mu\nu} &
            Q_{H\ell}^{(1)} & (H^\dagger i \overleftrightarrow D_\mu H) (\overline \ell_p \gamma^\mu\ell_r)
            \\ \cline{1-2}
            \multicolumn{2}{|c||}{\text{Class $\boldsymbol{H^4D^2}$}} &
            Q_{HW} & (H^\dagger H) W_{\mu\nu}^I W^{I\mu\nu} &
            Q_{H\ell}^{(3)} & (H^\dagger i \overleftrightarrow D^I_\mu H) (\overline \ell_p \tau^I \gamma^\mu\ell_r)
            \\ \cline{1-2}
            Q_{H\square} & (H^\dagger H) \square (H^\dagger H) &
            Q_{HB} & (H^\dagger H) B_{\mu\nu} B^{\mu\nu} &
            Q_{He} & (H^\dagger i \overleftrightarrow D_\mu H) (\overline e_p \gamma^\mu e_r)
            \\
            Q_{HD} & ( H^\dagger D^\mu H )^2 &
            Q_{HWB} & (H^\dagger \tau^I H) W_{\mu\nu}^I B^{\mu\nu} &
            Q_{Hd} & (H^\dagger i \overleftrightarrow D_\mu H) (\overline d_p \gamma^\mu d_r)
            \\ \hline
            \multicolumn{2}{|c||}{\text{Class $\boldsymbol{\psi^2H^3} + \text{H.c.}$}} &
            \multicolumn{2}{c||}{\text{Class $\boldsymbol{\psi^4}$}} &
            \multicolumn{2}{c}{}
            \\ \cline{1-4}
            Q_{eH} & (H^\dagger H) ( \overline \ell_p e_r H) &
            Q_{ll} & (\overline \ell_p \gamma_\mu \ell_r) (\overline\ell_s \gamma^\mu \ell_t) &
            \multicolumn{2}{c}{}
            \\
            Q_{uH} & (H^\dagger H) ( \overline q_p u_r \widetilde H) &
            Q_{\ell e} & (\overline\ell_p \gamma_\mu \ell_r) (\overline e_s \gamma^\mu e_t) &
            \multicolumn{2}{c}{}
            \\
            Q_{dH} & (H^\dagger H) ( \overline q_p d_r \widetilde H) &
            Q_{qqq} & \varepsilon^{abc} \varepsilon^{il} \varepsilon^{jk} ({q}^\intercal_{aip} C q_{bjr})({q}^\intercal_{cks} C \ell_{lt}) &
            \multicolumn{2}{c}{}
            \\ \cline{1-4}
        \end{array}
        $
    }
    \caption{Operators from the $d=6$ Warsaw basis~\cite{Grzadkowski:2010es} used in this article.}
    \label{tab:Warsaw-ops}
\end{table}
\begin{table}[tb]
    \renewcommand{\arraystretch}{1.5}
    \centering
    \scalebox{0.85}{
        $
        \begin{array}[t]{|lc || lc|}
            \hline
            \multicolumn{2}{|c||}{\text{Class $\boldsymbol{X^2H^4}$}} &
            \multicolumn{2}{c|}{\text{Class $\boldsymbol{\psi^2H^4D}$}}
            \\ \hline
            Q_{G^2H^4}^{(1)} & (H^\dagger H)^2 G_{\mu\nu}^A G^{A\mu\nu} &
            Q_{\ell^2H^4D}^{(1)} & i (\overline\ell_p \gamma^\mu \ell_r) (H^\dagger \overleftrightarrow D_\mu H) (H^\dagger H)
            \\
            Q_{W^2H^4}^{(1)} & (H^\dagger H)^2 W_{\mu\nu}^I W^{I\mu\nu} &
            Q_{\ell^2H^4D}^{(2)} & i (\overline\ell_p \gamma^\mu \tau^I \ell_r) [(H^\dagger \overleftrightarrow D_\mu^I H) (H^\dagger H) + (H^\dagger \overleftrightarrow D_\mu H) (H^\dagger \tau^I H)]
            \\
            Q_{W^2H^4}^{(3)} & (H^\dagger \tau^I H) (H^\dagger \tau^J H) W_{\mu\nu}^I W^{J\mu\nu} &
            Q_{\ell^2H^4D}^{(3)} & i \varepsilon^{IJK} (\overline\ell_p \gamma^\mu \tau^I \ell_r) (H^\dagger \overleftrightarrow D_\mu^J H) (H^\dagger \tau^K H)
            \\
            Q_{WBH^4}^{(1)} & (H^\dagger H) (H^\dagger \tau^I H) W_{\mu\nu}^I B^{\mu\nu} &
            Q_{\ell^2H^4D}^{(4)} & \varepsilon^{IJK} (\overline\ell_p \gamma^\mu \tau^I \ell_r) (H^\dagger \tau^J H) D_\mu (H^\dagger \tau^K H)
            \\
            Q_{B^2H^4}^{(1)} & (H^\dagger H)^2 B_{\mu\nu} B^{\mu\nu} &
            Q_{e^2H^4D} & i (\overline e_p \gamma^\mu e_r) (H^\dagger \overleftrightarrow D_\mu H) (H^\dagger H)
            \\ \hline
            \multicolumn{2}{|c||}{\text{Class $\boldsymbol{\psi^2H^5} + \text{H.c.}$}} &
            \multicolumn{2}{c|}{\text{Class $\boldsymbol{\psi^4HD} + \text{H.c.}$}}
            \\ \hline
            Q_{\ell eH^5} & (H^\dagger H)^2 (\overline \ell_p e_r H) &
            Q_{\ell q^2uHD}^{(1)} & i\varepsilon^{abc}\varepsilon^{km}D_\mu {H^{n}}^\ast (q^\intercal_{amp}C\gamma^\mu u_{br})(q^\intercal_{cns}C \ell_{kt})
            \\
            Q_{quH^5} & (H^\dagger H)^2 (\overline q_p u_r \widetilde H) &
            Q_{\ell q^2uHD}^{(2)} & i\varepsilon^{abc}\varepsilon^{jm}D_\mu {H^{n}}^\ast(q^\intercal_{amp}C\gamma^\mu u_{br})(q^\intercal_{cjs}C \ell_{nt})
            \\
            Q_{qdH^5} & (H^\dagger H)^2 (\overline q_p d_r H) &
            Q_{\ell q^2uHD}^{(3)} & i\varepsilon^{abc} \varepsilon^{km}H^{n\ast}(q^\intercal_{amp}C\gamma^\mu u_{br})(D_\mu q^\intercal_{cns}C \ell_{kt})
            \\ \hline
            \multicolumn{2}{|c||}{\text{Class $\boldsymbol{H^8}$}} &
            \multicolumn{2}{c|}{\text{Class $\boldsymbol{\psi^4D^2} + \text{H.c.}$}}
            \\ \hline
            Q_{H^8} & (H^\dagger H)^4 &
            Q_{\ell qudD^2}^{(2)} & \varepsilon^{abc} \varepsilon^{ij} (\ell^\intercal_{ip} C \sigma^{\mu\nu} q_{ajr}) (D_\mu d^\intercal_{bs} C D_\nu u_{ct})
            \\ \hline
            \multicolumn{2}{|c||}{\text{Class $\boldsymbol{\psi^4H^2}$}} &
            \multicolumn{2}{c|}{\text{Class $\boldsymbol{\psi^4H^2} + \text{H.c.}$}}
            \\ \hline
            Q_{\ell^4H^2} & (\overline\ell_p \gamma^\mu \ell_r) (\overline\ell_s \gamma_\mu \ell_t) (H^\dagger H) &
            Q_{\ell qd^2H^2} & \varepsilon^{abc}\varepsilon^{jk}\varepsilon^{mn}(\ell^\intercal_{jp} C q_{amr})(d^\intercal_{bs} C d_{ct})H_k H_n
            \\
            Q_{\ell^4H^2}^{(2)} & (\overline\ell_p \gamma^\mu \ell_r) (\overline\ell_s \gamma_\mu \tau^I \ell_t) (H^\dagger H) &
            Q_{eq^2dH^2} & \varepsilon^{abc}\varepsilon^{jk}\varepsilon^{mn}(e^\intercal_p C d_{ar})(q^\intercal_{bjs}C q_{cmt})H_k H_n
            \\ \cline{3-4}
            Q_{\ell^2e^2H^2}^{(1)} & (\overline\ell_p \gamma^\mu \ell_r) (\overline e_s \gamma_\mu e_t) (H^\dagger H) &
            \multicolumn{2}{c|}{\text{Class $\boldsymbol{H^6D^2}$}}
            \\ \cline{3-4}
            Q_{\ell^2e^2H^2}^{(2)} & (\overline\ell_p \gamma^\mu \tau^I \ell_r) (\overline e_s \gamma_\mu e_t) (H^\dagger \tau^I H) &
            Q_{H^6}^{(1)} & (H^\dagger H)^2 (D_\mu H^\dagger D^\mu H)
            \\
            \multicolumn{2}{|c||}{} &
            Q_{H^6}^{(2)} & (H^\dagger H) (H^\dagger \tau^I H) (D_\mu H^\dagger \tau^I D^\mu H)
            \\ \hline
        \end{array}
        $
    }
    \caption{Subset of $d=8$ operators from the Murphy basis~\cite{Murphy:2020rsh} with the changes listed in Sec.~\ref{sec:d=8_UFO}.}
    \label{tab:Murphy-ops}
\end{table}

\end{appendix}

\bibliography{references}  

\end{document}